\documentclass[12pt]{article}

\usepackage{latexsym}
\usepackage{amsthm}
\usepackage{amsmath,amssymb,amsfonts,bm,amsbsy,bbm,mathabx}
\usepackage{epsfig}
\usepackage{graphicx,rotating,lscape}
\usepackage{verbatim}
\usepackage{multirow,color}
\usepackage{geometry}
\usepackage{setspace}
\usepackage{float}
\usepackage{physics}
\usepackage{mathtools}
\usepackage{enumitem}
\usepackage{caption, subcaption}
\usepackage{makecell}
\usepackage{mathtools}

\usepackage{algorithm}
\usepackage{algorithmic}
\usepackage[algo2e]{algorithm2e}

\usepackage{natbib}
\usepackage{authblk}

\DeclareMathOperator*{\argmin}{arg\,min}

\newcommand{\E}{E}

\newtheorem{lemma}{\underline{\bf Lemma}}

\newtheorem{pro}{\underline{\bf Proposition}}
\newtheorem{Th}{\underline{\bf Theorem}}
\newtheorem{Rem}{\underline{\bf Remark}}

\newtheorem{definition}{\underline{\bf Definition}}
\newtheorem{assumption}{\underline{\bf Assumption}}

\def\bse{\begin{eqnarray*}}
	\def\ese{\end{eqnarray*}}
\def\be{\begin{eqnarray}}
	\def\ee{\end{eqnarray}}
\def\bsq{\begin{equation*}}
	\def\esq{\end{equation*}}
\def\bq{\begin{equation}}
	\def\eq{\end{equation}}
\def\bi{\begin{itemize}}
	\def\ei{\end{itemize}}

\def\wh{\widehat}
\def\wt{\widetilde}

\def\trans{^{\rm T}}

\def\bbeta{{\boldsymbol\beta}}
\def\bt{{\boldsymbol\theta}}
\def\bgamma{{\boldsymbol\gamma}}

\def\bmu{{\boldsymbol\mu}}
\def\bxi{\boldsymbol\xi}

\def\bGamma{{\boldsymbol\Gamma}}
\def\bSigma{{\boldsymbol\Sigma}}

\def\bphi{{\boldsymbol\phi}}

\def\bOmega{{\boldsymbol\Omega}}
\def\bDelta{{\boldsymbol\Delta}}

\def\0{{\bf 0}}
\def\X{{\bf X}}
\def\bX{\mathbb{X}}
\def\x{{\bf x}}
\def\bZ{{\bf Z}}

\def\bB{\mathbb{B}}
\def\bE{{\bf E}}
\def\fb{{\bf b}}
\def\A{{\bf A}}

\def\Fb{\mathbb{F}}

\def\I{{\bf I}}

\def\bY{{\bf Y}}

\def\bM{\mathbb{M}}
\def\Kb{\mathbb{K}}

\def\bI{\mathbb{I}}
\def\bC{\mathbb{C}}
\def\bT{\mathbb{T}}
\def\T{{\bf T}}
\def\bW{{\bf W}}
\def\bU{{\bf U}}
\def\RR{{\mathbb{R}}}
\def\PP{{\mathbb{P}}}
\def\E{{\hbox{E}}}
\def\bv{{\boldsymbol{v}}}
\def\be{{\boldsymbol{e}}}
\def\cG{{\mathcal{G}}}
\def\trans{^{\rm T}}
\def\J{{\bf J}}
\def\Z{{\bf Z}}
\def\bm{{\bf m}}

\def\boxit#1{\vbox{\hrule\hbox{\vrule\kern6pt\vbox{\kern6pt#1\kern6pt}\kern6pt\vrule}\hrule}}

\def\boxit#1{\vbox{\hrule\hbox{\vrule\kern6pt\vbox{\kern6pt#1\kern6pt}\kern6pt\vrule}\hrule}}

\makeatletter
\def\widebreve{\mathpalette\wide@breve}
\def\wide@breve#1#2{\sbox\z@{$#1#2$}%
	\mathop{\vbox{\m@th\ialign{##\crcr
				\kern0.08em\brevefill#1{0.8\wd\z@}\crcr\noalign{\nointerlineskip}%
				$\hss#1#2\hss$\crcr}}}\limits}
\def\brevefill#1#2{$\m@th\sbox\tw@{$#1($}%
	\hss\resizebox{#2}{\wd\tw@}{\rotatebox[origin=c]{90}{\upshape(}}\hss$}
\makeatletter

\usepackage{xr}
\usepackage{mathrsfs}

\makeatletter
\newcommand*{\addFileDependency}[1]{
	\typeout{(#1)}
	\@addtofilelist{#1}
	\IfFileExists{#1}{}{\typeout{No file #1.}}
}
\makeatother

\newcommand*{\myexternaldocument}[1]{
	\externaldocument{#1}
	\addFileDependency{#1.tex}
	\addFileDependency{#1.aux}
}

\myexternaldocument{supp_revision2}

\usepackage{geometry}
\usepackage{setspace}
\usepackage{tikz}
\usetikzlibrary{decorations.pathreplacing}
\newcommand{\customunderbrace}[2]{%
	\begin{tikzpicture}[baseline=(a.base)]
		\node[inner sep=0] (a) {$#1$};
		\draw[thick, decorate, decoration={brace, amplitude=5pt, mirror}]
		([yshift=-3pt]a.south west) -- ([yshift=-3pt]a.south east)
		node[midway, below=6pt] {#2};
	\end{tikzpicture}%
}

\begin{document}
	
\title{Dependable Exploitation of High-Dimensional Unlabeled Data in an Assumption-Lean Framework}

\author[a]{Chao Ying\thanks{This work was completed while the first author was affiliated with the University of Wisconsin-Madison.}}
\author[b]{Siyi Deng}
\author[b]{Yang Ning}
\author[a]{Jiwei Zhao}
\author[c]{Heping Zhang}
\affil[a]{University of Wisconsin-Madison}
\affil[b]{Cornell University}
\affil[c]{Yale University}

\date{\today}

\maketitle

\thispagestyle{empty}
	
	\begin{abstract}
		\noindent
		Semi-supervised learning has attracted significant attention due to the proliferation of applications featuring limited labeled data but abundant unlabeled data. 
		In this paper, we examine the statistical inference problem in an assumption-lean framework which involves a high-dimensional regression parameter, defined by minimizing the least squares, within the context of semi-supervised learning.
		We investigate when and how unlabeled data can enhance the estimation efficiency of a regression parameter functional.
		First, we demonstrate that a straightforward debiased estimator can only be more efficient than its supervised counterpart if the unknown conditional mean function can be consistently estimated at an appropriate rate. 
		Otherwise, incorporating unlabeled data can actually be counterproductive. 
		To address this vulnerability, we propose a novel estimator guaranteed to be at least as efficient as the supervised baseline, even when the conditional mean function is misspecified. 
		This ensures the dependable use of unlabeled data for statistical inference. 
		Finally, we extend our approach to the general M-estimation framework, and demonstrate the effectiveness of our methodology through comprehensive simulation studies and a real data application.
	\end{abstract}
	{\bf Key Words:}
	Semi-supervised learning, model misspecification, high-dimensional inference, assumption-lean framework, efficiency gain, semiparametric efficiency.
	
	\newpage
	\setcounter{page}{1} 
	
	\setcounter{equation}{0}
	
	\section{Introduction}\label{sec_intro}
	
	\subsection{Motivation: abundant high-dimensional unlabeled data}
	
	Unlabeled data are typically more abundant than labeled data, as collecting raw information is often far easier and less costly than manual annotation.
	In many real-world settings, data are continuously generated through sensors, transactions, medical records, and user interactions, while labeling requires human expertise, time, and resources. 
	For example, in healthcare, vast amounts of electronic health records (EHRs) are generated daily, but annotating them for specific diseases or conditions requires expert clinicians. 
	Similarly, in computer vision, millions of images and videos are uploaded online, yet labeling them with precise object categories demands significant manual effort. 
	
	Unlabeled data are often high-dimensional because they are collected from complex real-world sources that capture a vast array of features without prior filtering or annotation. 
	For instance, in EHRs, a single patient's history can include numerous lab results, diagnoses, prescriptions, and physiological measurements, each contributing to a high-dimensional feature space. 
	In computer vision, images and videos consist of millions of pixels, each representing a separate dimension. 
	Likewise, in natural language processing, raw text data encompass extensive vocabulary and contextual dependencies, making them inherently high-dimensional.
	
	\subsection{Our approach: dependable exploitation in an assumption-lean framework}\label{sec:lean}
	
	Throughout, suppose we observe $n$ independent and identically distributed (i.i.d.) samples $(\X_1, Y_1),\cdots,(\X_n, Y_n)\sim (\X, Y)$ from the labeled data, and $N$ i.i.d. samples, $\X_{n+1},\cdots,$\\$\X_{N+n}\sim \X$, from the unlabeled data.
	For notation simplicity, we denote by $\bX=(\X_1,\cdots,\X_n)\trans$\\$\in\RR^{n\times p}$, $\bY=(Y_1,\cdots,Y_n)\trans\in\RR^n$, and $\wt \bX=(\X_1,\cdots,\X_{N+n})\trans \in\RR^{(N+n)\times p}$.
	Note that $p$ can be much larger than $n$, and there is no requirement between $N$ and $n$.
	
	
	Due to the complex and heterogeneous nature of abundant unlabeled data, estimation problems can be casted in the general
	assumption-lean framework \citep{Buja,Assumption}, where the parameter of interest $\bt^*$ is defined via minimizing a loss function such as the least squares, while the true conditional mean might be hard or even infeasible to be consistently estimated under high-dimensionality.
	To be more specific, we consider
	\begin{equation} \label{eq_model}
		Y=f(\X)+\epsilon,
	\end{equation}
	where $f(\X)$ is the conditional mean function $\E(Y|\X)$, $\epsilon$ is independent of $\X\in \RR^p$ with $\E(\epsilon)=0$, $\E(\epsilon^2)=\sigma^2$, and $\sigma^2$ is an unknown parameter.
	For simplicity, we assume $\E(\X)=\0$ and $\E\{f(\X)\}=0$.
	Additionally, we consider linear regression as the working model which leads to the intercept term as $0$.
	Since $\E\{(Y-\X\trans\bt)^2\}=\E[\{f(\X)-\X\trans\bt\}^2]+\sigma^2$, the regression coefficients in a linear model correspond to the $L_2(\PP)$ projection of $f(\X)$ onto the linear space spanned by $\X$, i.e.,
	$\bt^*=\arg\min_{\bt\in \mathbb{R}^p} \E[\{f(\X)-\X\trans\bt\}^2]\in\RR^p$,
	which describes the linear dependence between $Y$ and $\X$.
	Our goal is to construct an asymptotically normal estimator of the linear functional of $\bt^*$ that, no matter whether the linear model is correctly specified or $f(\X)$ is consistently estimated, is guaranteed to be no less efficient than the supervised estimator (the one that uses labeled data only; e.g., debiased lasso estimator).
	
	As noted above, the parameter $\bt^*$ admits a clear interpretation under misspecification of model~\eqref{eq_model}. The proposed method naturally extends to more sophisticated settings such as the M-estimation; we defer details to Section~\ref{sec:ext}.
	
	\subsection{Our novel contributions}\label{sec:contri}
	
	Adopting the assumption-lean framework, 
	\cite{deng2023optimal} proposed estimators of $\bt^*$ that can have a faster convergence rate than the supervised estimators and can achieve the optimal rate under certain conditions, but they did not study the asymptotic variance of the proposed estimator.
	Indeed, the limiting distribution of their proposed estimators are intractable due to the regularization.
	
	In this paper, distinct from \cite{deng2023optimal},
	our ultimate goal is to answer the question:
	\emph{
		How to propose an estimator of the linear functional of $\bt^*$, that is asymptotically normal, and is guaranteed to be no less efficient than the supervised estimator, no matter whether the linear model is correctly specified or the conditional mean function is consistently estimated?}
	
	To answer this question, a natural choice is the debiased estimator; e.g., one can construct a one-step debiased estimator $\wh\bt^d$, using $\wh\bt_{SD}$ proposed in \cite{deng2023optimal} as the initial (details are given in Section~\ref{background}).
	Indeed, one can show that the corresponding estimator $\bv\trans\wh\bt^d$, with any $\bv\in\RR^p$, is asymptotically normal and can attain the semiparametric efficiency bound under certain conditions.
	However, its performance depends on the consistency of the estimate of $f(\X)$.
	If $f(\X)$ cannot be consistently estimated, $\bv\trans\wh\bt^d$ is not guaranteed to be more efficient than the supervised debiased lasso estimator.
	Therefore, this debiased approach does not answer the question above well.
	
	In this paper,
	we propose a dependable semi-supervised estimator which does not require the estimate of $f(\X)$. The main idea is to construct a set of unbiased estimating functions and decorrelate the score function to reduce the variability.
	In Theorem \ref{thm_inf2} below, we show that the proposed estimator $\bv\trans\wh\bt_{S,\psi}^d$ is asymptotically normal, where $\psi$ is a tuning parameter.
	When the linear model is misspecified and $\lim_{n\rightarrow\infty}\frac{n}{n+N}=\rho$ for some $0\leq \rho<1$, the estimator $\bv\trans\wh\bt_{S,\psi}^d$ with $0<\psi<2$ is strictly more efficient than the debiased lasso, leading to more powerful hypothesis tests and shorter confidence intervals.
	We attain the maximum variance reduction by choosing $\psi=1$.
	In addition, if either the linear model is correctly specified $f(\X)=\X\trans\bt^*$ or the size of the unlabeled data is small in that $\lim_{n\rightarrow\infty}\frac{n}{n+N}=1$ (i.e., $N\ll n$), the estimator $\bv\trans\wh\bt_{S,\psi}^d$ is asymptotically equivalent to the debiased lasso estimator.
	In summary, the estimator $\bv\trans\wh\bt_{S,\psi}^d$ provides a dependable use of the unlabeled data, since it is always no worse than the supervised estimators, no matter whether the linear model is correctly specified or the conditional mean function is consistently estimated.

	\subsection{Relevant literature}\label{sec:review}
	
	The benefits of using abundant unlabeled data have been popularly investigated by both computer scientists and statisticians.
	In problems with discrete labels, researchers have proposed a variety of classification algorithms under common assumptions such as manifold assumption and cluster assumption; see, e.g., \cite{cluster, wang2022usb} and comprehensive survey articles \citep{zhu2005semi,chapelle2009semi}.
	In nonparametric regression, \cite{wasserman} developed an estimator that can improve the rate of the mean squared error under the semi-supervised smoothness assumption; also see \cite{kostopoulos2018semi} for an extensive review of semi-supervised regression.

	In recent years, significant progress has been made on utilizing unlabeled data for parameter estimation or empirical risk minimization \citep{yuval2022semi}. However, the question we posed in Section~\ref{sec:contri}, along with the goal we aim to achieve in this paper, has not been thoroughly addressed in the literature. 
	The relevant studies reviewed in this section either impose restrictive conditions or yield limited results.
	
	In the low and fixed dimensionality setting, for parameter estimation, \cite{chakrabortty2018} and \cite{azriel} proposed estimators that are more efficient than the least square estimator that uses labeled data only, for each component of $\bt^*$.
	However, the efficiency improvement has no guarantee for the linear combination of $\bt^*$ such as $\theta^*_1+\theta^*_2$.

	Allowing the dimensionality to grow with $n$, the sample size of the labeled data, \cite{zhang2019} proposed a general semi-supervised inference framework to improve the estimation of the population mean $\E(Y)$ without specific distributional assumptions relating the outcome $Y$ and the covariate $\X$.
	However, \cite{zhang2019} can only allow the dimensionality to grow with a no faster than $n^{1/2}$ rate.
	Similar conclusions could also be found under the general M-estimation framework \citep{song2023general}.
	
	With high-dimensional data, \cite{2019jelena} proposed semi-supervised estimators of population mean and variance and established their asymptotic distributions; however, they required $\lim_{n\rightarrow\infty} \frac{n}{n+N}<1/2$ to guarantee the dependable inference on $\E(Y)$ (i.e., more efficient than the sample mean of $Y$ in the labeled data).
	In the high dimensional regime, \cite{caiexplained} considered how to estimate the explained variance ${\bt^{*}}\trans\bSigma\bt^*$ in the semi-supervised setting.
	Their estimator achieved the optimal rate of convergence and was asymptotically normal; however, their results were established under the assumption that the working linear model is correctly specified, which differed from the assumption-lean framework.
	In a high dimensional linear regression setting, the parameter of interest studied in \cite{chen2023enhancing} is the regression coefficient associated with one particular covariate, and this particular covariate plays the critical role in defining the discrepancy between the distributions of the labeled and unlabeled data.
	\cite{chen2023enhancing} studied the benefits of the unlabeled data in terms of enhancing efficiency and robustness, but their proposed methods heavily rely on the model among the covariates under different sparse or dense structures.
	Additionally, \cite{chakrabortty2022semi} concentrated on quantile estimation under high dimensionality, adopting the similar idea as \cite{song2023general}.
	In the absence of sparsity, \cite{livne2022improved} studied the problem of estimating the conditional variance of $\X$ given $Y$ in a linear regression model.
	While \cite{hou2023surrogate} proposed a notable approach termed surrogate assisted semi-supervised inference, their estimator remained vulnerable because its consistency depended entirely on a correctly specified imputation model. 
	Also, their framework lacked guarantees for the dependable use of unlabeled data.

	Our work is related to a growing literature that strengthens statistical inference by leveraging AI/ML predictions \citep{WangMcCormickLeek2020, MotwaniWitten2023}.
	For example, prediction-powered inference (PPI), introduced by \citet{angelopoulos2023prediction}, is a semi-supervised framework that uses predictions from an AI/ML model to enable valid statistical inference; however, PPI can be less efficient than the naive labeled-only estimator so may fail to reliably leverage unlabeled data.
	In response, several methods were proposed over the past three years to address this efficiency gap.
	For scalar parameters, \citet{angelopoulos2023ppi++} proposed a simple tuning procedure that guarantees efficiency gain even under prediction-model misspecification.
	More broadly, \citet{miao2025assumption} developed a post-prediction adaptive inference approach that guarantees valid inference without assumptions on the quality of the ML predictions; \citet{GronsbellGaoShiMcCawEtAl2025} studied inference under squared-error loss; and \citet{shan2025sada} extended these ideas to settings with multiple sets of predictions.
	Despite these advances, this line of work typically defines the target parameter in a standard low-dimensional setting.
	In contrast, in this paper, we study how to reliably exploit unlabeled data in an assumption-lean, high-dimensional regime, where the parameter of interest is defined through the high-dimensional coefficient vector in a linear working model.
	This setting is particularly relevant when the goal is to characterize the association between the response and a high-dimensional covariate (or a functional thereof).
	To attain efficiency dominance, our proposed estimator does not require estimating $f(\X)$; it is therefore prediction-free and agnostic to the outcome mean model.
	In addition, accounting for high dimensionality is central to our theoretical analysis and introduces additional technical challenges.

	\subsection{Structure of the paper and notation}
	
	\paragraph{Structure}
	In what follows, Section~\ref{sec:prelim} provides preliminary materials, with Section~\ref{background} reviewing the straightforward debiased estimator which relies on a consistent estimate of $f(\X)$.
	In Section~\ref{sec_dependable}, we propose a novel dependable procedure which does not rely on the estimation of $f(\X)$ and is guaranteed to be no worse than the one using the labeled data only.
	Our proposal can be naturally extended to more sophisticated settings such as the M-estimation, with details in Section~\ref{sec:ext}.
	Numerical experiments and a real data application are in Sections~\ref{sec_simu} and \ref{sec_data}, respectively.
	The paper is concluded with a discussion in Section~\ref{sec_disc}.
	All the technical proofs are contained in the Supplement.
	
	
	\paragraph{Notation}
	For $\bv=(v^{(1)},\cdots,v^{(p)})\trans \in \mathbb{R}^p$, and $1 \leq q \leq \infty$, we define $\norm{\bv}_q=(\sum_{i=1}^p |v^{(i)}|^q)^{1/q}$, $\norm{\bv}_0=|\{i: v^{(i)}\neq 0\}|$, where $|A|$ is the cardinality of a set $A$. Denote $\norm{\bv}_{\infty}=\max_{1\leq i \leq p} |v^{(i)}|$ and $\bv^{\otimes 2}=\bv \bv\trans$.
	For a matrix $\bM=[\bM_{ij}]$,
	define $\norm{\bM}_{\max}=\max_{ij}|\bM_{ij}|$, $\norm{\bM}_1=\max_{j}\sum_{i}|\bM_{ij}|$, $\norm{\bM}_{\infty}=\max_{i}\sum_{j}|\bM_{ij}|$. 
	For $S\subseteq \{1,\cdots,p\}$, let $\bv_S=\{v^{(k)}: k\in S\}$ and $S^c$ be the complement of $S$.
	For matrix $\bX\in\RR^{n\times p}$ and index set $L\subseteq \{1,\cdots,n\}$, $\bX_L=\{\bX_{i\cdot}:i\in L\}\trans\in \RR^{|L|\times p}$.
	For two positive sequences $a_n$ and $b_n$, we write $a_n\asymp b_n$ if $C\leq a_n/b_n\leq C'$ for some $C,C'>0$. Similarly, we use $a\lesssim b$ to denote $a\leq Cb$ for some constant $C>0$.
	
	%
	
	\section{Preliminaries}\label{sec:prelim}
	
	\subsection{Review of estimation and inference with labeled data only}
	
	In supervised learning that only uses the labeled data, there are a large number of penalized methods for estimating $\bt^*$, such as lasso \citep{lasso} and Dantzig selector \citep{candes2007}.
	The supervised Dantzig selector is defined as
	\begin{equation}
		\label{dantini}
		\wh\bt_D=\arg\min \|\bt\|_1,~~\textrm{s.t.}~~ \Big\|\frac{1}{n}\sum_{i=1}^n (Y_i-\X_i\trans\bt)\X_i\Big\|_\infty \leq \lambda_D,
	\end{equation}
	where $\lambda_D$ is a tuning parameter.
	Similarly, the supervised lasso estimator is defined as
	$\wh\bt_L=\argmin_{\bt \in \RR^p} \sum_{i=1}^n (Y_i-\X_i\trans\bt)^2/(2n)+\lambda_L\norm{\bt}_1$.
	The Dantzig selector and lasso are theoretically equivalent \citep{bickel2009}.
	For statistical inference, there has been some recent research on debiased lasso estimators for hypothesis tests and confidence intervals, for example, \cite{zhang2011confidence, vandegeer2014, javanmard2018, cai2015confidence, ning2017, neykov2018unified}, a list that is far from exhaustive.
	Under certain regularity conditions such as $s\log p/\sqrt{n}=o(1)$, following the proof in \cite{mann2015}, one can show that, the debiased lasso estimator for $\bv\trans\bt^*$ is $\sqrt{n}$-consistent and the corresponding asymptotic variance equals $\bv\trans\bOmega \Kb\bOmega\bv$, where $\bOmega=\bSigma^{-1}$ is the precision matrix and
	$\Kb=\E(\T_{i1}^{\otimes 2})$; see Remark~\ref{rem_efficiency2} in Section~\ref{sec:theory}.
	
	While considerable progress has been made towards understanding estimation and inference in the fully supervised setting, research in the semi-supervised setting remains limited.
	Notably, under model (\ref{eq_model}),
	where linear regression serves as the working model, the covariate $\X$ no longer functions as an ancillary statistic for the regression parameter $\boldsymbol{\theta}^*$. Therefore, the information of $\X$ in the unlabeled data may improve the estimation and inference of $\bt^*$.
	
	\subsection{Straightforward debiased estimator}\label{background}
	
	In an earlier work, \cite{deng2023optimal} proposed a semi-supervised estimator $\wh\bt_{SD}$ in their Section 3.
	As long as the conditional mean function is consistently estimated by $\wh f(\cdot)$ and some regularity conditions are satisfied, Theorem 3.2 in \cite{deng2023optimal} showed that $\wh\bt_{SD}$ is minimax optimal.
	To be more specific,
	\begin{equation} \label{est}
		\wh\bt_{SD}=\arg\min \|\bt\|_1,~~\textrm{s.t.}~~ \|\wh\bSigma_{n+N}\bt-\wh\bxi\|_\infty \leq \lambda_{SD},
	\end{equation}
	where the cross-fitting technique (with details provided in Algorithm~\ref{alg}) is used to compute $\wh\bxi$; i.e., 
	$\wh\bxi=(\wh \bxi_1+\wh \bxi_2)/2$,
	$\wh \bxi_j= \frac{1}{n_j}\sum_{i\in D_j^*} \X_iY_i-\frac{1}{n_j}\sum_{i\in D_j^*} \X_i\wh f^{-j}(\X_i)+\frac{1}{n_j+N_j}\sum_{i\in D_j}\X_i\wh f^{-j}(\X_i)$.
	
	
	
	Motivated by the formulation of the regularized estimator $\wh \bt_{SD}$ in (\ref{est}), one can view $h(\wt\bX,\bY;\bt)=\wh\bSigma_{n+N}\bt-\wh\bxi$ as an estimating function for $\bt$.
	Borrowing the idea from the classical one-step estimator and the debiased lasso, one can easily construct 
	\begin{equation}\label{eq_debias1}
		\wh \bt^d =\wh \bt_{SD}-\wh \bOmega h(\wt\bX,\bY;\wh\bt_{SD})=\wh \bt_{SD}+\wh \bOmega(\wh\bxi-\wh\bSigma_{n+N}\wh \bt_{SD}),
	\end{equation}
	where $\wh \bOmega$ is an estimator of $\bOmega=\bSigma^{-1}$
	with details provided in Supplement~\ref{app:A1}.
	
	In Supplement~\ref{app:A}, we detail the theoretical properties of the debiased estimator $\wh \bt^d$.
	More specifically, under some assumptions and certain regularity conditions, the debiased estimator $\bv\trans\wh \bt^d$ is $\sqrt{n}$-consistent, and the corresponding asymptotic variance is $\bv\trans(\sigma^2\bOmega+\frac{n}{n+N}\bGamma)\bv$, where $\bGamma=\E[\bW^{\otimes 2}\{f(\X)-\X\trans\bt^*\}^2]$ and $\bW=\bOmega \X$.
	Moreover, one can show that
	$n^{1/2}\bv\trans(\wh\bt^d-\bt^*) / \{\bv\trans(\wh\sigma^2\wh\bOmega+\frac{n}{n+N}\wh\bGamma)\bv\}^{1/2} \stackrel{d}{\longrightarrow} \mathcal{N}(0,1)$,
	where the specific form of the estimators $\wh\sigma^2$, $\wh\bOmega$ and $\wh\bGamma$ are provided in Supplement~\ref{app:A1}.
	
	However, the debiased estimator has a serious drawback.
	The consistency and asymptotic normality of $\wh \bt^d$ relies on the consistent estimation of the conditional mean function $f(\X)$, say, $\wh f(\X)$.
	If, unfortunately, $f(\X)$ cannot be consistently estimated, the estimator $\bv\trans\wh\bt^d$ is not guaranteed to be more efficient than the supervised debiased lasso estimators \citep{vandegeer2014}.
	This is not an ideal phenomenon.
	The main goal of this paper, with the novel method presented below in Section~\ref{sec_dependable}, is to propose an estimator that is guaranteed to be always more efficient than the supervised debiased lasso estimators, thus provides the dependable use of the unlabeled data.
	
	\section{Proposed Estimator towards Dependable \\
		Semi-Supervised Inference}\label{sec_dependable}
	
	To exploit high-dimensional unlabeled data, the most difficult step is to estimate
	the conditional mean function $f(\X)$ correctly.
	In this section we propose a novel dependable semi-supervised inference approach, which does not rely on the estimation of the conditional mean function but guarantees the efficiency gain compared to the supervised approach, thus provides the dependable use of the unlabeled data.

	\subsection{Motivation}
	
	Given any $p$-dimensional function $m(\X) \colon \RR^p\to \RR$, define $\bmu=\E\{\X m(\X)\}$, then $\X m(\X)-\bmu$ is an unbiased estimating function for zero.
	While it does not directly involve the unknown parameter $\bt^*$, it does play an important role in the dependable semi-supervised inference approach.
	Using $\X m(\X)-\bmu$ as the covariate, we postulate a $p$-variate working regression model with response variable $\X(Y-\X\trans\bt^*)$; i.e.,
	\begin{equation}\label{eq_regression2}
		\X(Y-\X\trans\bt^*)=\bB\trans\{\X m(\X)-\bmu\}+\bE,
	\end{equation}
	where $\bE\in\RR^p$ is the error vector and the coefficient matrix $\bB\in\RR^{p\times p}$ is
	$\bB=(\E[\{\X m(\X)-\bmu\}^{\otimes 2}])^{-1}\E\{\X^{\otimes 2}m(\X)(Y-\X\trans\bt^*)\}$.
	Since (\ref{eq_regression2}) is only a working model, the error $\bE$ and the covariate $\X m(\X)-\bmu$ are not necessarily independent. Recall that the response variable $\X(Y-\X\trans\bt^*)$ corresponds to the score function of $\bt^*$ in the linear regression model, and can be rewritten as $\X\{\epsilon+\eta(\X)\}$, where $\epsilon=Y-f(\X)$ and $\eta(\X)=f(\X)-\X\trans\bt^*$ is the nonlinear effect. Since $\epsilon$ and $\X m(\X)-\bmu$ are independent, the goal of model (\ref{eq_regression2}) is to explain the nonlinear effect $\X\eta(\X)$ by the covariate $\X m(\X)-\bmu$. Indeed, we show in Remark \ref{rem_compare} that the optimal choice of $m(\X)$ is $\eta(\X)$ and in this case the nonlinear effect $\X\eta(\X)$ can be perfectly explained by $\X m(\X)-\bmu$.

	Given $\X m(\X)-\bmu$ and the coefficient matrix $\bB$, we define a class of unbiased estimating functions for $\bt^*$ as $h_{\psi}(\X,Y; \bt)=\X(Y-\X\trans\bt)-\psi \bB\trans\{\X m(\X)-\bmu\}=\bar\bxi_\psi-\X\X\trans\bt$, where
	\begin{equation}\label{eq_xibar}
		\bar\bxi_\psi=\X Y-\psi \bB\trans\{\X m(\X)-\bmu\},
	\end{equation}
	with $\psi\in\RR$ being a tuning parameter that balances two unbiased functions $\X m(\X)-\bmu$ and $\X(Y-\X\trans\bt^*)$. In particular, we have $\E\{h_{\psi}(\X,Y; \bt^*)\}=\0$ for any $\psi$. Indeed, we show in Remark \ref{rem_efficiency2} that the optimal choice of $\psi$ is $\psi=1$, which implies $h_{\psi}(\X,Y; \bt^*)=\bE$ in view of (\ref{eq_regression2}). Thus, from a geometric perspective, $h_{\psi}(\X,Y; \bt)$ is the residual by projecting the score function $\X(Y-\X\trans\bt)$ onto the set of unbiased estimating functions $\X m(\X)-\bmu$ in the $L_2(\PP)$ norm. Following the insight from the above geometric interpretation, we now propose the dependable semi-supervised inference approach.
	
	\subsection{Dependable semi-supervised inference}\label{sec:main}
	
	To formulate the inference procedure, we first consider how to estimate the coefficient matrix $\bB$. In view of (\ref{eq_regression2}) and the followup discussion, to estimate $\bB$, we can either pre-specify a nonlinear function $m(\X)$ or perhaps use a more flexible approach to estimate $m(\X)$ from the data. 
	To see this, we can define $m(\X)=\argmin_{g\in \cG} \E\{Y-g(\X)\}^2$, where $\cG$ is a pre-specified class of functions of $\X$; e.g., class of linear functions, additive functions, functions correspond to interaction model \citep{zhao2016analysis}, single-index model \citep{radchenko2015high, yang2017high, eftekhari2021inference} or multi-index model \citep{yang2017learning}.
	Consider a concrete example where the choice of $\cG=\{\sum_{j=1}^p \alpha_j X_j+\sum_{1\leq k<\ell\leq p} \beta_{k\ell}X_\ell X_k: \alpha_j, \beta_{k\ell}\in\RR\}$ corresponds to the class of functions with main effects and the second-order interactions. By fitting a penalized interaction model \citep{zhao2016analysis},  we can construct an estimator $\wh m(\X)$. 
	In the rest of the paper, we assume an estimator $\wh m(\X)$ of $m(\X)$ is available. 
	The detailed technical conditions on $\wh m(\X)$ are shown in Assumption \ref{assumptionE} and Theorem \ref{thm_inf2}.
	More discussions about other choices of class $\cG$ can be found in the Supplement~\ref{sec:functional_class}.
	We apply a cross-fitting approach to estimate $\bB$. 
	Given the estimator $\wh m^{-j}(\cdot)$ obtained from the labeled data $D^*\backslash D_j^*$ for $j=1,2$, we can estimate the $k$th column of $\bB$ by
	\begin{eqnarray} \label{eq_B}
		\wh \bB^j_{\cdot k}=\arg\min_{\bbeta\in \mathbb{R}^p} \frac{1}{n_j}\sum_{i\in D_j^*} \Big[X_{ik} (Y_i- \X_i\trans \wh\bt_D)-\bbeta\trans \{\X_i\wh m^{-j}(\X_i) -\wh \bmu^j\}\Big]^2+\wt\lambda_k\norm{\bbeta}_1,
	\end{eqnarray}
	where $\wh\bt_D$ is the supervised Dantzig estimator in \eqref{dantini}, $\wh\bmu^j=\frac{1}{n_j}\sum_{i\in D^*_j}\wh m^{-j}(\X_i)\X_i$, $\lambda_D$ and $\wt\lambda_k$ are two tuning parameters. We note that it is possible to estimate $\bmu=\E\{\X m(\X)\}$ by using both labeled and unlabeled data $D_j$. However, the rate of the estimator $\wh \bB^j_{\cdot k}$ remains the same. The final estimator of $\bB_{\cdot k}$ is $\wh \bB_{\cdot k}=(\wh\bB^1_{\cdot k}+\wh\bB^2_{\cdot k})/2$, and this leads to $\wh\bB=(\wh \bB_{\cdot 1},...,\wh \bB_{\cdot p})$.

	Motivated by the form of $\bar\bxi_\psi$ in (\ref{eq_xibar}), we construct the following estimate of $\bxi=\E(\X Y)$,
	\begin{equation} \label{newxi}
		\wh \bxi_{S,\psi}= \frac{\sum_{i=1}^n\X_i Y_i}{n}-\frac{\psi}{2}\wh \bB\trans\sum_{j=1}^2\left\{\frac{\sum_{i\in D^*_{j}}\X_i \wh m^{-j}(\X_i)}{n_j}-\frac{\sum_{i\in D_j} \X_i \wh m^{-j}(\X_i)}{n_j+N_j}\right\},
	\end{equation}
	where we apply the cross-fitting technique again. 
	We note that different from the estimator $\wh \bmu^j$ used in $\wh \bB^j_{\cdot k}$, we estimate $\bmu$ by $\frac{1}{n_j+N_j}\sum_{i\in D_j} \X_i \wh m^{-j}(\X_i)$ in (\ref{newxi}), which incorporates the information from the unlabeled data.

	Similar to $\wh\bt^d$ in (\ref{eq_debias1}), we propose the following dependable semi-supervised estimator
	\begin{equation}\label{eq_debias2}
		\wh \bt^d_{S,\psi}=\wh\bt_D+\wh \bOmega(\wh\bxi_{S,\psi}-\wh \bSigma_n\wh \bt_D),
	\end{equation}
	where $\wh\bt_D$ is the supervised Dantzig estimator in \eqref{dantini},
	$\wh \bxi_{S,\psi}$ is defined in (\ref{newxi}), $\wh \bSigma_n=\frac{1}{n}\sum_{i=1}^n\X_i^{\otimes 2}$ and $\wh \bOmega$ is the node-wise lasso  estimator in (\ref{hatomega}). It is worthwhile to note that we estimate $\bSigma$ by $\wh \bSigma_n$ in (\ref{eq_debias2}), whereas we use $\wh \bSigma_{n+N}=\frac{1}{n+N}\sum_{i=1}^{n+N}\X_i^{\otimes 2}$ in the estimator $\wh\bt^d$. Indeed, this is a critical difference as replacing $\wh \bSigma_n$ with $\wh \bSigma_{n+N}$ in  (\ref{eq_debias2}) corresponds to an estimating function different from $h_{\psi}(\X,Y; \bt)$ and therefore no longer leads to a more efficient estimator.

	\subsection{Theory}\label{sec:theory}
	
	To show the theoretical properties of $\wh \bt^d_{S,\psi}$, we require the following assumptions.
	
	\begin{assumption}\label{assumptionE}
		\begin{enumerate}
			\item[(E1)] The smallest eigenvalue of $\E[\{\X m(\X)-\bmu\}^{\otimes 2}]$ is lower bounded by a positive constant. 
			The 2nd moment of $E_j$ in (\ref{eq_regression2}) is less than $C$ and $|m(\X_i)|\leq C$, for some constant $C$. 
			The estimator $\wh m^{-j}(\cdot)$ satisfies $\norm{\wh m^{-j}-m}_2=O_p(c_n)$ for a deterministic sequence $c_n$. 
			We require $s_BK_1^2\Big(c_n+\sqrt{\frac{\log p}{n}}\Big)=o(1)$ and $\sqrt{\frac{s\log p}{n}}=O(1)$.
			\item[(E2)] The columns of the matrix $\bB$ are sparse with $\max_{1\leq k\leq p}\norm{\bB_{\cdot k}}_0 = s_B$ and $\max_{1\leq k\leq p}\norm{\bB_{\cdot k}}_1$\\$\leq L_B$ for some $L_B$ that may grow with $n$.
			\item[(E3)] $\norm{\bOmega}_\infty\leq L_{\Omega}$ and $(K_1L_\Omega)^2s_{\Omega}\sqrt{\log p/(n+N)}=o(1)$, where $s_{\Omega}$ is the maximum rowwise sparsity of $\bOmega$ defined in Assumption \ref{assumption_inf} (presented in Supplement~\ref{app:A}).
			\item[(E4)]  $\E|\epsilon|^{2+\delta}=O(1)$ and $\E|\eta|^{2+\delta}=O(1)$, where $\epsilon=Y-f(\X)$ and $\eta=f(\X)-\X\trans\bt^*$.
		\end{enumerate}
	\end{assumption}

	Assumption (E1) guarantees that the restricted eigenvalue (RE) condition holds for the estimation of $\bB_{\cdot k}$ in (\ref{eq_B}). 
	Assumption (E2) is the sparsity assumption of $\bB_{\cdot k}$. For example, in the ideal case that we choose $m(\X)=f(\X)-\X\trans\bt^*$ and then $\bB=\bI_p$ is sparse. 
	We note that there are other practical settings that the sparsity assumption of $\bB_{\cdot k}$ is reasonable (e.g., $\X$ is blockwise independent). We defer the detailed discussion to Supplement~\ref{app_sparseB}.
	In Assumption (E2), we further require that the matrix $L_\infty$ norm of $\bB$ is bounded by $L_B$, which is used to establish the rate of $\wh\bB$.
	Assumption (E3) and (A2) in Assumption \ref{assumption_est} (presented in Supplement~\ref{app:A}) together imply the strong boundedness condition and $\max_{1\leq i\leq n+N}\max_{1\leq k\leq p}|\X_{i,-k}\trans\bgamma_k|=O(K)$ in \cite{vandegeer2014} which further guarantees the rate of $\wh\bOmega$ in the matrix $L_{\infty}$ norm.
	In particular, to control the remainder term in the asymptotic expansion of $\wh\bt^d_{S,\psi}$, we need $\norm{\bOmega}_\infty\leq L_{\Omega}$. Together with the boundedness assumption $\|\X_i\|_\infty\leq K_1$ in Assumption \ref{assumption_est} (presented in Supplement~\ref{app:A}), it implies $\|\bOmega \X_i\|_\infty\leq \norm{\bOmega}_\infty\|\X_i\|_\infty\leq K_1 L_{\Omega}$.
	Note that (E3) is equivalent to Assumption \ref{assumption_inf} (presented in Supplement~\ref{app:A}) by replacing $K_1 L_{\Omega}$ with $K$.
	Assumption (E4) assumes that $\epsilon$ and $\eta$ have bounded $(2+\delta)$ moment, which is used to simplify the Lyapunov condition.
	Denote
	\begin{equation} \label{eq_bGamma_psi}
		\bGamma_{\psi}=\E(\T_{i1}^{\otimes 2})-\frac{N(2\psi-\psi^2)}{n+N}\{\E(\T_{i2}\T_{i1}\trans)\}\trans\{\E(\T_{i2}^{\otimes 2})\}^{-1}\E(\T_{i2}\T_{i1}\trans),
	\end{equation}
	where $\T_{i1}=\X_i(Y_i-\X_i\trans\bt^*)$ and $\T_{i2}=\X_i m(\X_i)-\bmu$.
	
	\begin{Th}\label{thm_inf2}
		Suppose Assumptions \ref{assumptionE} and \ref{assumption_est} (presented in Supplement~\ref{app:A}) hold. We choose $\lambda_D\asymp K_1\sqrt{\frac{(\sigma^2+\Phi^2)\log p}{n}}$ in (\ref{dantini}), $\lambda_k\asymp K\sqrt{\frac{\log p}{n+N}}$ in (\ref{eq_hatgamma}) and $\wt\lambda_k=\wt\lambda_{opt}$ in (\ref{eq_B}), where $\wt\lambda_{opt}$ is defined in Proposition \ref{newB}.  Then for any $\bv\neq \0 \in\RR^p$, $\bv\trans (\wh\bt^d_{S,\psi}-\bt^*)$ equals
		\begin{align}\label{eq_thm_inf12_1}
			\bv\trans\bOmega\left[\frac{\bX\trans (\bY-\bX\bt^*)}{n}-\psi \bB\trans\left\{\frac{\sum_{i=1}^{n} \X_i m(\X_i)}{n} -\frac{\sum_{i=1}^{n+N} \X_i m(\X_i)}{n+N}\right\}\right]+O_p(\bar\delta_n),
		\end{align}
		where
		$\bar\delta_n=\|\bv\|_1 L_{\Omega} K_1^2\sqrt{\frac{\log p}{n}} \left\{L_Bs_\Omega\sqrt{\frac{\log p}{n+N}}+s_B^{1/2}L_B\left(\sqrt{\frac{\log p}{n}}+c_n\right)+K_1(s\vee s_B)\sqrt{\frac{\log p}{n}}\right\}$
		with $c_n$ defined in Assumption \ref{assumptionE} (E1) and $\E[\{f(\X)-\X\trans\bt^*\}^2]\leq\Phi^2$. In addition, if $\bv\trans\bOmega\bGamma_\psi\bOmega\bv\geq C\|\bv\|_2^2$ for some constant $C$, $\bar\delta_n/\|\bv\|_2=o(n^{-1/2})$ and
		\begin{align}\label{eq_thm_inf2_1}
			\Big(\frac{\|\bv\|_1L_\Omega K_1}{\|\bv\|_2}\Big)^{2+\delta} \frac{1}{n^{\delta/2}}\Big(1+\frac{L_B^{2+\delta}N}{n+N}\Big)=o(1),
		\end{align}
		then
		$n^{1/2}\bv\trans(\wh\bt^d_{S,\psi}-\bt^*) / (\bv\trans\bOmega\bGamma_\psi\bOmega\bv)^{1/2} \stackrel{d}{\longrightarrow} \mathcal{N}(0,1)$.
	\end{Th}
	
	We now elaborate on the technical conditions used in Theorem~\ref{thm_inf2}.
	In
	\eqref{eq_thm_inf12_1}, the remainder term $\bar\delta_n$ characterizes the effect of the plug-in estimators $\wh\bOmega$, $\wh\bB$ and $\wh\bt_D$.
	To establish the asymptotic normality of $\bv\trans (\wh\bt^d_{S,\psi}-\bt^*)$, we further need to assume that $\delta_n$ is sufficiently small and the Lyapunov condition (\ref{eq_thm_inf2_1}) holds so that one can apply the central limit theorem to the leading terms in \eqref{eq_thm_inf12_1}.
	To further simplify the conditions in Theorem \ref{thm_inf2}, one can assume $\|\bv\|_1/\|\bv\|_2$, $K_1, L_{\Omega}, L_B$ are all of order $O(1)$.
	As a result, (\ref{eq_thm_inf2_1}) always holds and the condition $\bar\delta_n/\|\bv\|_2=o(n^{-1/2})$ is implied by
	\begin{align}\label{eq_thm_inf2_3}
		s_{\Omega}{\frac{\log p}{\sqrt{n+N}}}+(s\vee s_B){\frac{\log p}{\sqrt{n}}}+s_B^{1/2}c_n\sqrt{\log p}=o(1).
	\end{align}
	Note that the condition $s({\log p}/{\sqrt{n}})=o(1)$ implied by (\ref{eq_thm_inf2_3}) is actually slightly stronger than the condition (\ref{inf1_5}) required by the debiased estimator $\bv\trans \wh\bt^d$, with details presented in Theorem \ref{inf1} in Supplement~\ref{app:A}.
	This is because $\wh\bt_D$ is not only used as an initial estimator when constructing the one-step estimator $\wh\bt^d_{S,\psi}$ but also used as a plug-in estimator to estimate $\bB$ in (\ref{eq_B}). The error of $\wh\bt_D$ accumulates in the asymptotic expansion of $\bv\trans (\wh\bt^d_{S,\psi}-\bt^*)$, leading to the slow order $s({\log p}/{\sqrt{n}})$ in (\ref{eq_thm_inf2_3}). In terms of  the rate of $\wh m^{-j}$ in the $L_2(\PP)$ norm, (\ref{eq_thm_inf2_3}) requires $c_n=o\{(s_B\log p)^{-1/2}\}$, slightly stronger than the condition for $\wh f^{-j}$ in (\ref{inf1_5}).
	Notably, 
	the condition $\|\bv\|_1/\|\bv\|_2=O(1)$ is not always satisfied; in this case, \eqref{eq_thm_inf2_3} becomes
	\begin{align}\label{eq_thm_inf22_3}
		\frac{\|\bv\|_1}{\|\bv\|_2}\left\{s_{\Omega}{\frac{\log p}{\sqrt{n+N}}}+(s\vee s_B){\frac{\log p}{\sqrt{n}}}+s_B^{1/2}c_n\sqrt{\log p}+n^{-\frac{\delta}{2(2+\delta)}}\right\}=o(1).
	\end{align}
	
	\begin{Rem}
		The bound in (\ref{eq_thm_inf22_3}) requires that the ratio $\|\bv\|_1/\|\bv\|_2$ not be too large, which excludes cases where $\bv$ contains many large entries (e.g., $\bv=(1,1,...,1)\trans$). This observation is consistent with \cite{cai2015confidence}, which shows that the debiased estimator does not yield optimal confidence intervals for $\bv\trans\wh \bt^d_{S,\psi}$ when $\bv$ is dense. 
		
		To illustrate situations where our results apply, first note that if $\bv=\be_j$, the $j$th basis vector in $\RR^p$, then $\bv\trans\wh\bt^d_{S,\psi}=\wh\theta^d_{S,\psi,j}$ reduces to the estimator of $\theta^d_{S,\psi,j}$. 
		In this case, condition (\ref{eq_thm_inf22_3}) reduces to \eqref{eq_thm_inf2_3}. 
		Similar observations apply when $\bv\trans\bt^*$ is a linear combination of $\bt^*$ with $\|\bv\|_0$ fixed.
		More generally, the set of vectors $\bv$ in $\RR^p$ satisfying (\ref{eq_thm_inf22_3}) forms a cone $\Big\{\frac{\|\bv\|_1}{\|\bv\|_2}\leq t_n[s_{\Omega}{\frac{\log p}{\sqrt{n+N}}}+(s\vee s_B){\frac{\log p}{\sqrt{n}}}+s_B^{1/2}c_n\sqrt{\log p}+n^{-\frac{\delta}{2(2+\delta)}}]^{-1}\Big\}$ for some $t_n=o(1)$. 
		Compared with  \cite{cai2015confidence}, which considered the debiased estimator for $\bv\trans\bt^*$ with sparse $\bv$, condition (\ref{eq_thm_inf22_3}) may still hold when $\bv$ is approximately sparse, i.e., when it contains many small but nonzero entries. 
		Our results therefore remain applicable in such settings.
		
	\end{Rem}
	
	\begin{Rem}[Efficiency improvement and optimality]\label{rem_efficiency2}
		When the linear model is correctly specified, i.e., $f(\X)=\X\trans\bt^*$, we have $\E(\T_{i2}\T_{i1}\trans)=\0$ and $\bGamma_\psi=\sigma^2\bSigma$.
		Thus, the asymptotic variance of $\bv\trans\wh\bt^d_{S,\psi}$ reduces to $\sigma^2\bv\trans\bOmega\bv$, which agrees with that of the supervised debiased estimator.
		In the following, we assume $\E(\T_{i2}\T_{i1}\trans)=\E\{\X^{\otimes 2}m(\X)\eta\}$ is of full rank. Since $\E(\T_{i2}^{\otimes 2})$ is strictly positive definite by Assumption (E1),  it implies that $\{\E(\T_{i2}\T_{i1}\trans)\}\trans\{\E(\T_{i2}^{\otimes 2})\}^{-1}\E(\T_{i2}\T_{i1}\trans)$ is strictly positive definite. We consider the following two cases.
		\begin{itemize}
			\item[(1)] $\lim_{n\rightarrow\infty}\frac{n}{n+N}=1$. Recall that the asymptotic variance of the supervised debiased estimator is $\bv\trans\bOmega \Kb\bOmega\bv$, where $\Kb=\E(\T_{i1}^{\otimes 2})$ with $\T_{i1}=\X_i(Y_i-\X_i\trans\bt^*)$; see \cite{mann2015}. In this case, the asymptotic variance of $\bv\trans\wh\bt^d_{S,\psi}$ is identical to $\bv\trans\bOmega \Kb\bOmega\bv$. Thus there is no efficiency improvement when $n\gg N$.
			
			\item[(2)]  $\lim_{n\rightarrow\infty}\frac{n}{n+N}=\rho$ for some $0\leq \rho<1$. 
			In this case, the asymptotic variance $\bv\trans\bOmega\bGamma_\psi\bOmega\bv$ is strictly smaller than $\bv\trans\bOmega \Kb\bOmega\bv$ if and only if $2\psi-\psi^2>0$, i.e. $0<\psi<2$. 
			Thus, our estimator $\bv\trans\wh\bt^d_{S,\psi}$ with $0<\psi<2$ is more efficient than the supervised estimator.  
			Interestingly, the asymptotic variance $\bv\trans\bOmega\bGamma_\psi\bOmega\bv$ is minimized when taking $\psi=1$. Thus, the estimator $\bv\trans\wh\bt^d_{S,\psi=1}$ is optimal within the following class of estimators $\{\bv\trans\wh\bt^d_{S,\psi}: \psi\in\RR\}$ in terms of asymptotic efficiency.
			In view of the form of $\bGamma_\psi$, the variance reduction becomes more evident as $\rho$ goes to $0$ (i.e., $N$ increases).
			In the interest of space, we defer more detailed discussions regarding the efficiency gain in Supplement~\ref{sec:supp_efficiency}.
		\end{itemize}
		
	\end{Rem}
	
	\begin{Rem}[Comparison with the estimator $\wh\bt^d$]\label{rem_compare}
		To see the connection of the two estimators $\wh\bt^d$ and $\wh\bt^d_{S,\psi}$ in (\ref{eq_debias1}) and (\ref{eq_debias2}), consider the ideal case with $m(\X)=f(\X)-\X\trans\bt^*$. From (\ref{eq_bGamma_psi}), we can show that with $\psi=1$,
		$\bGamma_{\psi}=\sigma^2\bSigma+\E(\X^{\otimes 2}\eta^2)-\frac{N}{n+N}\E(\X^{\otimes 2}\eta^2)=\sigma^2\bSigma+\frac{n}{n+N}\E(\X^{\otimes 2}\eta^2)$,
		where $\eta=f(\X)-\X\trans\bt^*$. The asymptotic variances of $\bv\trans\wh\bt^d$ and $\bv\trans\wh\bt^d_{S,\psi}$ are identical, since $\bv\trans\bOmega\bGamma_\psi\bOmega\bv=\bv\trans(\sigma^2\bOmega+\frac{n}{n+N}\bGamma)\bv$, where $\bGamma=\bOmega\E(\X^{\otimes 2}\eta^2)\bOmega$. Thus, in the ideal case when $f(\X)$ is known, using $\wh\bt^d_{S,\psi}$ with $m(\X)=f(\X)-\X\trans\bt^*$ would not suffer efficiency loss compared to $\wh\bt^d$ and both estimators improve the efficiency of the debiased estimator (and attains the semi-parametric efficiency bound under certain conditions); see Remark \ref{rem_efficiency2} above and Remark \ref{rem_efficiency} (presented in Supplement~\ref{app:A}). However, if there is no sufficient information for us to estimate $f(\X)$ consistently, the estimator $\bv\trans\wh\bt^d$ may not improve the efficiency of the debiased estimator, whereas $\bv\trans\wh\bt^d_{S,\psi}$ does not rely on the estimation of  $f(\X)$ and guarantees the efficiency improvement for any $m(\X)$ that satisfies the conditions in Theorem \ref{thm_inf2}. We note that the amount of efficiency improvement of $\bv\trans\wh\bt^d_{S,\psi}$ depends on the choice of $m(\X)$. Unless we choose $m(\X)=f(\X)-\X\trans\bt^*$, the estimator $\bv\trans\wh\bt^d_{S,\psi}$ in general would not attain the semi-parametric efficiency bound.
	\end{Rem}

	From Remarks \ref{rem_efficiency2} and \ref{rem_compare}, we can see that the estimator $\bv\trans\wh\bt_{S,\psi}^d$ provides a dependable use of the unlabeled data, since it is no worse than the supervised approach, no matter whether the linear model is correctly specified or the conditional mean function is consistently estimated.

	As mentioned in the introduction, when the dimension $p$ is fixed, \cite{azriel} and \cite{chakrabortty2018} investigated how to  incorporate the unlabeled data to improve the estimation efficiency for $\theta_j^*$. In addition to the technical challenges arise from the high dimensionality, the way we construct our estimator
	$\wh\bt^d_{S,\psi}$ is different from theirs. Unlike $\wh\bt^d_{S,\psi}$, their estimators cannot guarantee the efficiency improvement if the parameter of interest is the linear combination of $\bt^*$ (e.g., $\theta^*_1+\theta^*_2$). We refer to Supplement \ref{sec_compare} for more details.

	\subsection{Variance estimation}
	We now consider how to estimate the asymptotic variance of $\bv\trans\wh\bt^d_{S,\psi}$.
	To estimate $\bGamma_\psi$ in (\ref{eq_bGamma_psi}), we note that $\wh\bB\trans$ is an estimate of $\E(\T_{i2}\T_{i1}\trans)\trans\{\E(\T_{i2}^{\otimes 2})\}^{-1}$. We can further estimate $\bM_1=\E(\T_{i1}^{\otimes 2})$ and $\bM_2=\E(\T_{i2}\T_{i1}\trans)$ by
	$\wh\bM_1=\frac{1}{n}\sum_{i=1}^n (Y_i-\X_i\trans\wh\bt_D)^2\X_i^{\otimes 2}$
	and $\wh\bM_2=(\wh\bM^1_2+\wh\bM^2_2)/2$, where
	$\wh\bM_2^j=\frac{1}{n_j}\sum_{i\in D^*_j} (Y_i-\X_i\trans\wh\bt_D)\wh m^{-j}(\X_i)\X_i^{\otimes 2}$.
	Given these estimates, an estimator of $\bGamma_\psi$ is defined as
	$\wh\bGamma_\psi=\wh\bM_1-\frac{N(2\psi-\psi^2)}{n+N}\wh\bB\trans\wh\bM_2$.

	\begin{pro}\label{prop_variance2}
		Assume conditions in Theorem \ref{thm_inf2}, $\E(\epsilon^4)=O(1), \E(\eta^4)=O(1)$ and $Rem=o(1)$, where
		$Rem=K_1(s_B+s_B^{1/2}L_B)\left(\sqrt{\frac{\log p}{n}}+c_n\right)+K_1^2\sqrt{\frac{ss_B\log p}{n}}+K_1L_B\sqrt{\frac{s\log p}{n}}$.
		Then,
		\begin{equation}\label{eq_prop_variance2_1}
			\Big|\bv\trans\wh\bOmega\wh\bGamma_\psi\wh\bOmega\bv-\bv\trans\bOmega\bGamma_\psi\bOmega\bv\Big|=O_p\Big\{\|\bv\|_1^2(R_1+R_2+R_3)\Big\},
		\end{equation}
		where $R_1=K_1L_{\Omega}^2s_{\Omega}\sqrt{\frac{\log p}{n+N}} \|\bGamma_\psi\|_{\max}$, $R_2=K_1^3L_{\Omega}^2\sqrt{\frac{s\log p}{n}}$, $R_3=\frac{NK_1^2L_{\Omega}^2}{n+N}Rem$.
		Thus, if $\|\bv\|_1^2(R_1+R_2+R_3)/\|\bv\|_2^2=o(1)$, we have
		\begin{align}\label{eq_prop_variance2_2}
			n^{1/2}\bv\trans(\wh\bt^d_{S,\psi}-\bt^*) / (\bv\trans\wh\bOmega\wh\bGamma_\psi\wh\bOmega\bv)^{1/2} \stackrel{d}{\longrightarrow} \mathcal{N}(0,1).
		\end{align}
	\end{pro}
	
	
	We note that the three terms $R_1, R_2$ and $R_3$ in (\ref{eq_prop_variance2_1}) stem from the estimation errors of $\wh\bOmega$, $\wh\bt_D$ and $\wh\bB$, respectively. To further simplify the conditions in Proposition \ref{prop_variance2}, let us consider the case that $\|\bv\|_1/\|\bv\|_2$, $K_1, L_{\Omega}, L_B$ and $\|\bGamma_\psi\|_{\max}$ are all of order $O(1)$.
	Then, the asymptotic normality in (\ref{eq_prop_variance2_2}) is valid provided $(s\vee s_B)\sqrt{\frac{\log p}{n}}=o(1)$, $s_{\Omega}\sqrt{\frac{\log p}{n+N}}=o(1)$ and $s_Bc_n=o(1)$.
	
	\begin{algorithm}[htbp]
		\renewcommand{\algorithmicrequire}{\textbf{Input:}}
		\renewcommand{\algorithmicensure}{\textbf{Output:}}
		\caption{The algorithm to compute the estimators $\wh \bt_{SD}$, $\wh \bt^d$ and $\wh\bt^d_{S,\psi}$ via cross-fitting.}\label{alg}
		\begin{algorithmic}[1]
			\REQUIRE $D^*=\{(\X_i, Y_i): i=1,\cdots, n\}$ from the label data, $U=\{\X_i: i=n+1,\cdots, n+N\}$ from the unlabel data. Let $D=D^* \cup U$, $D^*=D_1^*\cup D^*_2$, $U=U_1\cup U_2$, $D_1=D_1^*\cup U_1$ and $D_2=D_2^*\cup U_2$, with $|D_j^*|=n_j$ and $|U_j|=N_j$, $j=1,2$.\\
			\ENSURE Estimators $\wh \bt_{SD}$, $\wh \bt^d$ and $\wh\bt^d_{S,\psi}$.
			\STATE Estimate $\wh f^{-j}(\cdot)$ and $\wh m^{-j}(\cdot)$ using data $D^*\backslash D^*_j$ for $j=\{1,2\}$, respectively;
			\STATE Compute the estimator $\wh \bt_{SD}$ proposed in \cite{deng2023optimal} via equation \eqref{est}.
			\STATE Compute the straightforward debiased estimator $\wh \bt^d$ via equation \eqref{eq_debias1},
			and obtain the confident intervals via equation \eqref{eq_prop_variance3}. 
			Note that the details of the estimator $\wh \bt^d$ are presented in Supplement~\ref{app:A}.
			\STATE Compute the proposed estimator $\wh\bt^d_{S,\psi}$ via equation \eqref{eq_debias2}, and obtain the confident intervals via equation \eqref{eq:interval}.
		\end{algorithmic}
	\end{algorithm}
	
	Lastly, from (\ref{eq_prop_variance2_2}), we can construct the $(1-\alpha)$ confidence interval for $\bv\trans\bt^*$ as
	\begin{align}\label{eq:interval}
		[\bv\trans\wh\bt^d_{S,\psi}-z_{1-\alpha/2}n^{-1/2}\wh{sd}, \bv\trans\wh\bt^d_{S,\psi}+z_{1-\alpha/2}n^{-1/2}\wh{sd}],
	\end{align}
	where $z_{1-\alpha/2}$ is the $1-\alpha/2$ quantile of a standard normal distribution and $\wh{sd}=(\bv\trans\wh\bOmega\wh\bGamma_\psi\wh\bOmega\bv)^{1/2}$. 
	Similarly, if the interest is in testing the hypothesis $H_0: \bv\trans\bt^*=0$, we can construct the test statistic $n^{1/2}\bv\trans\wh\bt^d_{S,\psi}/(\bv\trans\wh\bOmega\wh\bGamma_\psi\wh\bOmega\bv)^{1/2}$ based on (\ref{eq_prop_variance2_2}).

	\subsection{Algorithm}
	
	For clarity, we summarize the algorithm to compute the estimators $\wh \bt_{SD}$, $\wh \bt^d$ and $\wh\bt^d_{S,\psi}$ via cross-fitting in Algorithm~\ref{alg}.

	\section{Extension}\label{sec:ext}
	
	So far, we have only considered the situation that the parameter of interest $\bt^*$ is defined as $\bt^{*}=\arg\min_{\bt\in\mathbb{R}^{p}}\E\{(Y-\X{\trans}\bt)^2\}$.
	In this section, we extend our proposed methodology to the more general M-estimation framework; i.e.,
	$\bt^{*}=\arg\min_{\bt\in\mathbb{R}^{p}}\E\{L(\X,Y;\bt)\}$,
	with $L(\X,Y;\bt)$ twice continuously differentiable in $\bt$, 
	$\E\{\nabla_{\bt\bt{\trans}}L(\X,Y;\bt^{*})\}$ positive definite, and that $\E\{\nabla_{\bt\bt{\trans}}L(\X,Y;\bt)\}$ being nonsingular for all $\bt$ in a neighborhood of $\bt^{*}$.
	Clearly, the previously defined $\bt^*$ corresponds to the situation that $\nabla_\bt L(\X, Y; \bt)=\X(Y-\X\trans\bt)$, and the parameter of interest $\bbeta$ studied in \cite{hou2023surrogate} corresponds to the situation that $\nabla_\bt L(\X, Y; \bt)=\X\{Y-g(\bbeta\trans\X)\}$ with a known function $g(\cdot)$.
	
	In this general M-estimation framework, the supervised Dantzig selector is defined as $\wh\bt_{M,D}=\arg\min_{\bt}\|\bt\|_1$ s.t. $\Big\|\frac{1}{n}\sum_{i=1}^n \nabla_\bt L(\X_i,Y_i;\bt)\Big\|_\infty \le \lambda_{M,D}$
	with $\lambda_{M,D}$ a tuning parameter. 
	In our proposed methodology, we project the gradient $\nabla_\bt L(\X,Y;\bt)$ as
	$\nabla_\bt L(\X,Y;\bt)=\bB\trans\{\bm(\X)-\bmu\}+\bE$,
	where $\bm(\x)=\{m_1(\x),\ldots,m_p(\x)\}\trans$, $\bmu=\E\{\bm(\X)\}$, and the projection coefficient is
	$\bB=\left[\E\{\bm(\X)-\bmu\}^{\otimes 2}\right]^{-1}
	\E\left[\{\bm(\X)-\bmu\}\,\nabla_\bt\trans L(\X,Y;\bt)\right]$.
	Next, we define the estimating function $h_{\psi}(\X,Y; \bt)
	=\nabla_\bt L(\X,Y;\bt)-\psi\,\bB\trans\{\bm(\X)-\bmu\}$
	which has mean zero for any $\psi\in\RR$.
	
	Analogous to $\wh \bt^d_{S,\psi}$ in \eqref{eq_debias2}, we define the proposed estimator here as
	$\wh \bt^d_{M,S,\psi}=\wh\bt_{M,D}
	-\left\{\frac{1}{n}\sum_{i=1}^n\nabla_{\bt\bt\trans}L(\X_i,Y_i;\wh\bt_{M,D})\right\}^{-1}
	\left\{\frac{1}{n}\sum_{i=1}^n h_\psi(\X_i,Y_i;\wh\bt_{M,D})\right\}$, 
	where the last term 
	can be written as 
	$\frac{1}{n}\sum_{i=1}^n\nabla_\bt L(\X_i,Y_i;\wh\bt_{M,D})
	-\psi\,\wh\bB\trans 
	\sum_{j=1}^2\left\{
	\frac{1}{n_j}\sum_{i\in D_j^*}\wh\bm^{-j}(\X_i)
	-\frac{1}{n_j+N_j}\sum_{i\in D_j}\wh\bm^{-j}(\X_i)\right\}$.
	Similarly defining $\T_{i1}=\nabla_{\bt}L(\X_i,Y_i;\bt^*)$, $\T_{i2}=\bm(\X_i)-\bmu$ and $\bOmega=\left\{\E\nabla_{\bt\bt\trans} L(\X,Y;\bt^*)\right\}^{-1}$, one can easily develop the analogous result as Theorem~\ref{thm_inf2}.
	In the interest of space, we only provide a heuristic description of the extension here without presenting the full details.

	\section{Simulation Studies}\label{sec_simu}
	
	\subsection{Data generating models and practical implementation}
	
	We first generate a $p$-dimensional multivariate normal random vector $\bU \sim \mathcal{N}(0,\bSigma)$ with  $\Sigma_{jk}=0.3^{|j-k|}$. We set the covariate $\X=(X_1,...,X_p)\trans$ to be $X_1=|U_1|$ and $X_j=U_j ~\mathrm{for}~ 1<j\leq p$. The reason we take $X_1=|U_1|$ is that this transformation implies $\E(X_1^k X_j)=0$ for $j\neq 1$ but the parameter $\theta^*_1$ for centered $X_1$ is nonzero. 
	
	We first consider a non-additive model and call it \underline{Model 1}, that
	$Y=0.6(X_1+X_2)^2+0.4X_4^3-X_5+2X_6+\epsilon$,
	where $\epsilon \sim \mathcal{N}(0,1)$. 
	To calculate the corresponding regression parameter $\bt^*$ under the working linear model, we first center $Y$ and $X_1$ so that their means are 0. By Proposition 4 in \cite{mann2015}, the support of $\bt^*$ is $S=\{1,2,4,5,6\}$ and the corresponding regression parameter $\bt^*$ is $(1.48,1.04,0,1.2,-1,2,0,...,0)\trans$.
	
	Before we proceed to illustrate the results, we discuss several practical implementation issues for the proposed methods.
	To compute our optimal semi-supervised estimator $\wh\bt_{SD}$ in (\ref{est}), we apply the group lasso with spline basis to estimate a sparse additive regression function $\wh f$ \citep{huang2010}. To be specific, we use the cubic spline basis with degree of freedom $df=5$. To select the penalty parameter in group lasso and make computation easier, the BIC criterion is used; see Section 4 in \cite{huang2010} for the definition. After we derive the estimator $\wh f$ and subsequently $\wh \bxi$, we modify the source code in the \textsf{flare} package to compute the Dantzig type estimator $\wh\bt_{SD}$, where the tuning parameter $\lambda_{SD}$ is selected by 5 fold cross-validation. Given the estimator $\wh\bt_{SD}$, we can compute the one-step estimator $\wh\bt^d$ in (\ref{eq_debias1}) for inference, where $\wh\bOmega$ is obtained by the node-wise lasso using the \textsf{glmnet} package with tuning parameter selected by 5 fold cross-validation.
	
	To implement the dependable semi-supervised method, we choose $\wh m(\cdot)=\wh f(\cdot)$ the estimated sparse additive function obtained previously. We estimate each column of the coefficient matrix $\bB$ by (\ref{eq_B}) using lasso with tuning parameters selected by  cross-validation. With the optimal choice $\psi=1$ (see Remark \ref{rem_efficiency2}), we can compute the dependable semi-supervised estimator $\wh \bt^d_{S,\psi=1}$ in (\ref{eq_debias2}), where $\wh\bOmega$ is obtained previously and the Dantzig selector $\wh\bt_D$ is computed using the \textsf{flare} package.
	
	To compare the inference results, we consider two versions of debiased lasso estimators,
	\begin{equation} \label{est_Ld}
		\wh \bt_1^d =\wh \bt_{lasso}+ \bar\bOmega\Big(\frac{1}{n}\sum_{i=1}^{n}\X_iY_i-\wh\bSigma_{n}\wh \bt_{lasso}\Big),~~\wh \bt_2^d =\wh \bt_{lasso}+\wh \bOmega\Big(\frac{1}{n}\sum_{i=1}^{n}\X_iY_i-\wh\bSigma_{n}\wh \bt_{lasso}\Big),
	\end{equation}
	where $\wh \bt_{lasso}$ and $\bar \bOmega$ are the standard lasso and node-wise lasso estimator applied to the labeled data. The only difference between $\wh \bt_1^d$ and $\wh \bt_2^d$ is the way of estimating the precision matrix $\bOmega$. The two estimators $\wh \bt_1^d, \wh\bt_2^d$ and the associated confidence intervals can be computed using the \textsf{hdi} package with Robust option.
	
	\subsection{Numerical results}
	
	With sample size $n\in\{100,300,500\}$, the ratio $N/n\in\{1,4,8\}$ and the dimension $p\in\{200,500\}$, we compare the performance of the four methods: $\wh\bt^d_1$ (D-Lasso1, that only uses labeled data with sample size $n$), $\wh \bt^d_2$ (D-Lasso2), both defined in (\ref{est_Ld}), the straightforward debiased estimator D-SSL $\wh\bt^d$ defined in (\ref{eq_debias1}), and the proposed dependable semi-supervised estimator S-SSL $\wh \bt^d_{S,\psi=1}$ defined in (\ref{eq_debias2}).
	
\begin{figure}[!htbp] 
		\centering
		\includegraphics[width=\textwidth,height=1.15\textwidth]{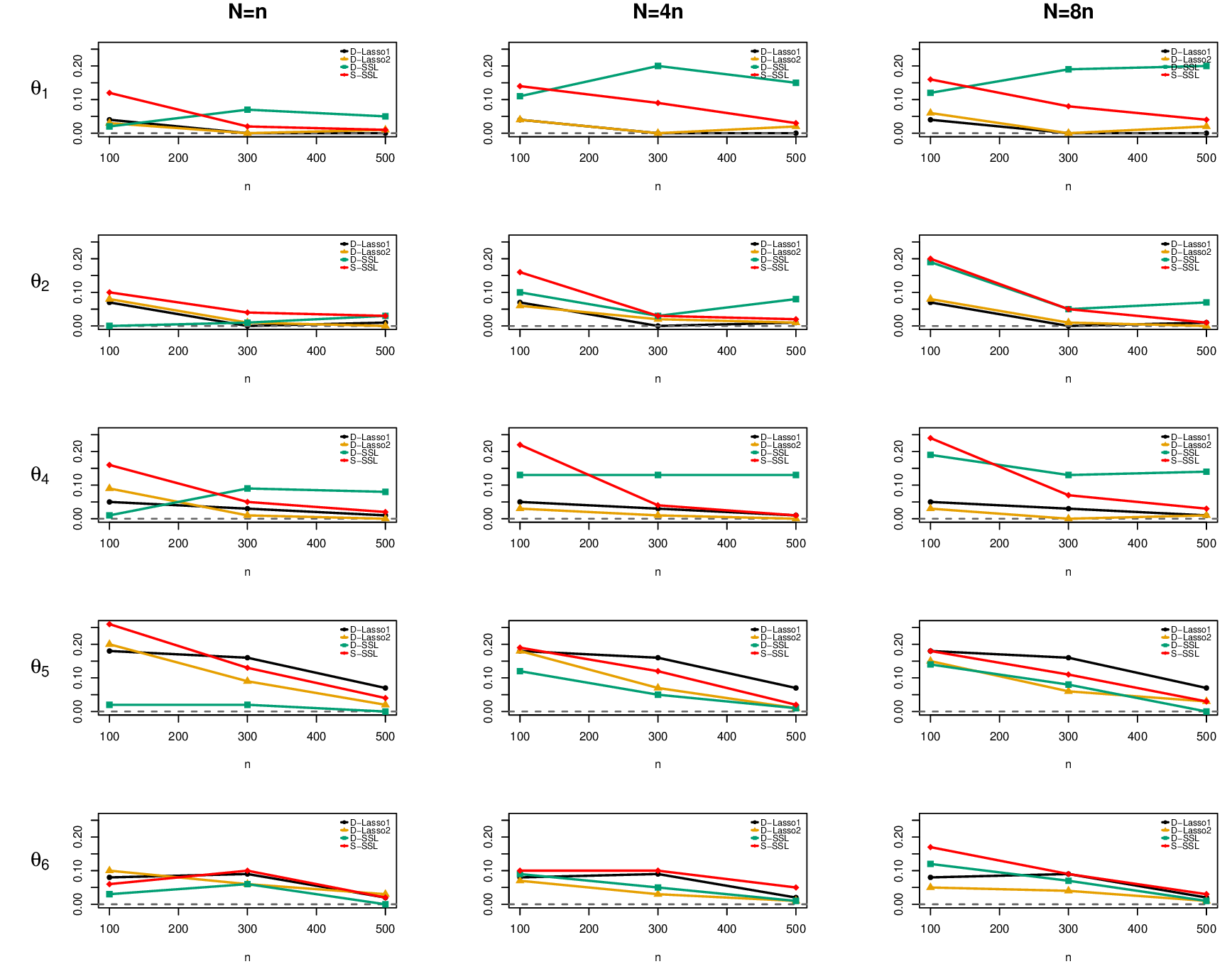}
		\caption{
			Simulation results for \underline{Model 1} with $p=200$: absolute difference between the empirical 95\% coverage probability and the nominal level 0.95.
			In all panels, rows represent different parameters, columns represent different $N/n$ ratios, and each panel plots the trend over the sample size $n$.
		}
		\label{fig:200}
	\end{figure}

	For the $p=200$ case, we report the empirical bias (Bias), standard deviation (SD), root mean squared error (RMSE) and the half length of 95\% confidence interval (len/2) for each of the single parameters $\theta_1$, $\theta_2$, $\theta_4$, $\theta_5$ and $\theta_6$ in Table~\ref{sim:200main}, and plot the absolute difference between the empirical 95\% coverage probability and the nominal level 0.95 in Figure~\ref{fig:200}.
	In the interest of space, the corresponding results for the $p=500$ case are placed in Table~\ref{sim:500main} and Figure~\ref{fig:500} in the Supplement.
	These results are based on 100 simulation replications.
	In Table~\ref{sim:time}, we also report the computation time (in seconds) of one simulation replication of these four methods for both $p=200$ and $p=500$.
	
	From these results, in the majority of the scenarios we consider, the proposed method S-SSL has the smallest SD and RMSE, compared to the methods D-Lasso1 and D-Lasso2.
	This shows, even with a misspecified conditional mean function modeling strategy, S-SSL can still achieve efficiency gain, indicating the dependable use of the unlabeled data.
	The coverage rate of the method S-SSL is close to the nominal level, especially when the sample size increases to $n=500$.
	However, the method D-SSL has a low coverage rate in some cases.
	This results from a poor estimation of conditional mean as the true model is no longer additive.
	Computing the proposed method S-SSL takes a slightly longer time than all other methods, due to its sophisticated nature.
	An interesting phenomenon we observe in this comparison is that computation takes longer when $n$ is approximately equal to $p$, but decreases once $n$ becomes larger than $p$.
	In practice, estimation of the conditional mean function can be difficult especially under high-dimensionality, therefore we recommend  S-SSL as it provides a dependable use of unlabeled data even if the imposed conditional mean model is incorrect.
	
	We also conduct similar numerical investigations when the data generating model is additive and we call it \underline{Model 2}.
	In the interest of space, we defer the detailed results to Supplement~\ref{sec:sim_supp}.
	Besides the parallel results in Table~\ref{sim:t1}, Table~\ref{sim:t3} and Table~\ref{sim:time_supp}, we also conduct different sensitivity analyses with different condition mean function estimation methods, different tuning parameter selection methods, and different estimands; see Table~\ref{sim:t9}, Table~\ref{sim:t12}, Table~\ref{sim:t10} and Table~\ref{sim:t11} in the Supplement for details.

	\begin{table}[!htbp] 
		\caption{Simulation results for \underline{Model 1} with $p=200$:
			Bias, SD and RMSE stand for empirical bias, standard deviation, and root mean squared error, respectively, len represents the length of 95\% confidence interval.
			The estimators D-Lasso1 (that only uses labeled data with sample size $n$) and D-Lasso2 are $\wh\bt^d_1$ and $\wh \bt^d_2$, defined in (24).
			The straightforward debiased estimator D-SSL is defined in (5).
			The proposed dependable semi-supervised estimator S-SSL is defined in (14).
			The best performance is \textbf{bolded} during the comparison.
		}\label{sim:200main}
		\center
		\scalebox{0.70}{
			\begin{tabular}{cccrrrrrrrrrrrr}
				\hline
				&		&	&\multicolumn{4}{c}{$n=100$}  &\multicolumn{4}{c}{$n=300$}&\multicolumn{4}{c}{$n=500$}		\\
				&	$N$	& &\text{Bias}& \text{SD}& \text{RMSE}  &\text{len/2}  &\text{Bias}&\text{SD}  & \text{RMSE}& \text{len/2} &\text{Bias}&\text{SD}  & \text{RMSE}& \text{len/2}\\
				\hline
				\multirow{10}{*}{$\theta_1$}&&	{D-Lasso1}
				& \bf{0.017}& 1.076     & 1.071     & 1.437     &  \bf{-0.056}& 0.257     & 0.262     & 0.542     & -0.014     & 0.215     & 0.214     & 0.426     \\
				\cline{3-15}
				&\multirow{3}{*}{$n$}&	{D-Lasso2}
				&-0.057     & 0.514     & 0.515     & 0.944     &	 -0.057     & 0.259     & 0.264     & 0.541     & \bf{-0.011}& 0.215     & 0.214     & 0.425     \\
				&		&	{D-SSL}
				&-0.162     & 0.465     & 0.490     & 1.112     &	 -0.098     & 0.237     & 0.256     & \bf{0.417}& -0.041     & 0.202     & 0.206     & \bf{0.306}\\
				&		&	{S-SSL}
				&-0.180     & \bf{0.449}& \bf{0.482}& \bf{0.812}&	 -0.102     & \bf{0.228}& \bf{0.249}& 0.463     & -0.045     & \bf{0.177}& \bf{0.182}& 0.370     \\
				\cline{3-15}
				&\multirow{3}{*}{$4n$}&	{D-Lasso2}
				&-0.055     & 0.506     & 0.506     & 0.944     &	 \bf{-0.054}& 0.260     & 0.265     & 0.543     & \bf{-0.013}& 0.217     & 0.216     & 0.426     \\
				&		&	{D-SSL}
				&-0.166     & 0.479     & 0.505     & 0.807     &	 -0.114     & 0.248     & 0.272     & \bf{0.353}& -0.050     & 0.199     & 0.204     & \bf{0.266}\\
				&		&	{S-SSL}
				&-0.217     & \bf{0.414}& \bf{0.465}& \bf{0.727}&	 -0.126     & \bf{0.205}& \bf{0.240}& 0.405     & -0.063     & \bf{0.168}& \bf{0.178}& 0.325     \\
				\cline{3-15}
				&\multirow{3}{*}{$8n$}&	{D-Lasso2}
				&-0.065     & 0.514     & 0.516     & 0.932     &	 \bf{-0.054}& 0.262     & 0.267     & 0.543     & \bf{-0.013}& 0.216     & 0.216     & 0.424     \\
				&		&	{D-SSL}
				&-0.173     & 0.492     & 0.519     & 0.717     &	 -0.118     & 0.242     & 0.268     & \bf{0.334}& -0.056     & 0.197     & 0.204     & \bf{0.254}\\
				&		&	{S-SSL}
				&-0.250     & \bf{0.399}& \bf{0.469}& \bf{0.700}&	 -0.134     & \bf{0.197}& \bf{0.237}& 0.386     & -0.066     & \bf{0.160}& \bf{0.172}& 0.309     \\
				\hline
				\multirow{10}{*}{$\theta_2$}&&	{D-Lasso1}
				&-0.102     & 0.708     & 0.712     & 0.856     &   0.011     & 0.167     & 0.166     & 0.327     &  0.016    & 0.125      & 0.126     & 0.260     \\
				\cline{3-15}
				&\multirow{3}{*}{$n$}&	{D-Lasso2}
				&\bf{-0.045}& 0.317     & 0.319     & 0.560     &	  0.011     & 0.166     & 0.165     & 0.327     &  0.013     & 0.125     & 0.125     & 0.260     \\
				&		&	{D-SSL}
				&-0.109     & 0.302     & 0.320     & 0.667     &	 -0.008     & \bf{0.138}& \bf{0.137}& \bf{0.257}& -0.004     & 0.113     & 0.112     & \bf{0.190}\\
				&		&	{S-SSL}
				&-0.094     & \bf{0.273}& \bf{0.288}& \bf{0.498}&	 \bf{-0.002}& 0.158     & 0.157     & 0.289     & \bf{-0.001}& \bf{0.108}& \bf{0.107}& 0.231     \\
				\cline{3-15}
				&\multirow{3}{*}{$4n$}&	{D-Lasso2}
				&\bf{-0.052}& 0.314     & 0.317     & 0.565     &	  0.010     & 0.168     & 0.167     & 0.329     &  0.011     & 0.127     & 0.127     & 0.260     \\
				&		&	{D-SSL}
				&-0.138     & 0.292     & 0.322     & 0.491     &	 -0.012     & \bf{0.130}& \bf{0.130}& \bf{0.222}& \bf{-0.006}& \bf{0.104}& \bf{0.103}& \bf{0.168}\\
				&		&	{S-SSL}
				&-0.138     & \bf{0.258}& \bf{0.291}& \bf{0.444}&	 \bf{-0.009}& 0.145     & 0.145     & 0.262     & -0.009     & \bf{0.104}& \bf{0.103}& 0.210     \\
				\cline{3-15}
				&\multirow{3}{*}{$8n$}&	{D-Lasso2}
				&\bf{-0.052}& 0.319     & 0.321     & 0.560     &	  \bf{0.010}& 0.166     & 0.166     & 0.330     &  0.011     & 0.126     & 0.126     & 0.260     \\
				&		&	{D-SSL}
				&-0.141     & 0.289     & 0.320     & 0.442     &	 -0.014     & \bf{0.126}& \bf{0.126}& \bf{0.211}& \bf{-0.007}& \bf{0.101}& \bf{0.101}& \bf{0.161}\\
				&		&	{S-SSL}
				&-0.154     & \bf{0.256}& \bf{0.297}& \bf{0.426}&	 -0.014     & 0.144     & 0.144     & 0.254     & -0.012     & 0.102     & 0.102     & 0.203     \\
				\hline
				\multirow{10}{*}{$\theta_4$}&&	{D-Lasso1}
				&-0.171     & 0.495     & 0.521     & 0.786     &  -0.050     & 0.189     & 0.194     & 0.382     & -0.024     & 0.137     & 0.139     & 0.280     \\
				\cline{3-15}
				&\multirow{3}{*}{$n$}&	{D-Lasso2}
				&\bf{-0.103}& 0.328     & \bf{0.342}& 0.653     &	 \bf{-0.040}& 0.186     & 0.190     & 0.383     & \bf{-0.017}& 0.136     & 0.137     & 0.280     \\
				&		&	{D-SSL}
				&-0.186     & 0.314     & 0.364     & 0.706     &	 -0.060     & 0.169     & 0.179     & \bf{0.270}& -0.020     & 0.124     & \bf{0.125}& \bf{0.198}\\
				&		&	{S-SSL}
				&-0.191     & \bf{0.293}& 0.349     & \bf{0.548}&	 -0.078     & \bf{0.151}& \bf{0.169}& 0.321     & -0.035     & \bf{0.122}& 0.127     & 0.238     \\
				\cline{3-15}
				&\multirow{3}{*}{$4n$}&	{D-Lasso2}
				&\bf{-0.073}& 0.322     & \bf{0.328}& 0.665     &	 \bf{-0.027}& 0.188     & 0.189     & 0.385     & \bf{-0.010}& 0.134     & 0.134     & 0.282     \\
				&		&	{D-SSL}
				&-0.200     & 0.312     & 0.369     & 0.513     &	 -0.070     & 0.157     & 0.171     & \bf{0.233}& -0.021     & 0.123     & 0.125     & \bf{0.175}\\
				&		&	{S-SSL}
				&-0.225     & \bf{0.275}& 0.354     & \bf{0.476}&	 -0.088     & \bf{0.127}& \bf{0.154}& 0.270     & -0.038     & \bf{0.106}& \bf{0.112}& 0.207     \\
				\cline{3-15}
				&\multirow{3}{*}{$8n$}&	{D-Lasso2}
				&\bf{-0.068}& 0.322     & \bf{0.328}& 0.655     &	 \bf{-0.020}& 0.189     & 0.189     & 0.385     & \bf{-0.006}& 0.136     & 0.136     & 0.281     \\
				&		&	{D-SSL}
				&-0.196     & 0.310     & 0.366     & 0.460     &	 -0.070     & 0.154     & 0.169     & \bf{0.220}& -0.022     & 0.123     & 0.124     & \bf{0.167}\\
				&		&	{S-SSL}
				&-0.228     & \bf{0.271}& 0.353     & \bf{0.450}&	 -0.088     & \bf{0.124}& \bf{0.151}& 0.253     & -0.035     & \bf{0.107}& \bf{0.112}& 0.196     \\
				\hline
				\multirow{10}{*}{$\theta_5$}&&	{D-Lasso1}
				& \bf{0.219}& 0.894     & 0.916     & 0.651     &   0.147     & 0.131     & 0.196     & 0.248     &  0.069     & 0.093     & 0.116     & 0.188     \\
				\cline{3-15}
				&\multirow{3}{*}{$n$}&	{D-Lasso2}
				& 0.267     & 0.280     & 0.386     & 0.452     &	  0.116     & 0.124     & 0.170     & 0.250     &  0.050     & 0.091     & 0.103     & 0.189     \\
				&		&	{D-SSL}
				& 0.271     & 0.250     & \bf{0.368}& 0.643     &	  \bf{0.086}& \bf{0.117}& \bf{0.145}& 0.260     &  \bf{0.034}& \bf{0.087}& \bf{0.093}& 0.191     \\
				&		&	{S-SSL}
				& 0.295     & \bf{0.244}& 0.382     & \bf{0.413}&	  0.124     & 0.127     & 0.177     & \bf{0.233}&  0.057     & 0.089     & 0.106     & \bf{0.176}\\
				\cline{3-15}
				&\multirow{3}{*}{$4n$}&	{D-Lasso2}
				& \bf{0.204}& 0.289     & 0.353     & 0.461     &	  0.089     & 0.126     & 0.154     & 0.253     &  0.032     & 0.091     & 0.096     & 0.190     \\
				&		&	{D-SSL}
				& 0.232     & 0.256     & 0.344     & 0.502     &	  \bf{0.081}& \bf{0.108}& \bf{0.135}& 0.227     &  \bf{0.028}& 0.084     & \bf{0.088}& 0.172     \\
				&		&	{S-SSL}
				& 0.234     & \bf{0.241}& \bf{0.335}& \bf{0.390}&	  0.103     & 0.126     & 0.163     & \bf{0.217}&  0.038     & \bf{0.083}& 0.091     & \bf{0.166}\\
				\cline{3-15}
				&\multirow{3}{*}{$8n$}&	{D-Lasso2}
				& \bf{0.161}& 0.284     & 0.326     & 0.456     &	  \bf{0.074}& 0.128     & 0.148     & 0.254     &  \bf{0.021}& 0.094     & 0.096     & 0.190     \\
				&		&	{D-SSL}
				& 0.220     & 0.257     & 0.337     & 0.455     &	  0.077     & \bf{0.112}& \bf{0.135}& 0.218     &  0.022     & 0.086     & \bf{0.088}& 0.166     \\
				&		&	{S-SSL}
				& 0.215     & \bf{0.233}& \bf{0.316}& \bf{0.379}&	  0.089     & 0.121     & 0.149     & \bf{0.214}&  0.031     & \bf{0.083}& \bf{0.088}& \bf{0.163}\\
				\hline
				\multirow{10}{*}{$\theta_6$}&&	{D-Lasso1}
				&-0.169     & 0.311     & 0.353     & 0.578     &  -0.096     & 0.126     & 0.158     & 0.245     & -0.055     & 0.090     & 0.105     & 0.187     \\
				\cline{3-15}
				&\multirow{3}{*}{$n$}&	{D-Lasso2}
				&\bf{-0.138}& 0.232     & 0.269     & 0.465     &	 -0.080     & 0.125     & 0.148     & 0.247     & -0.042     & 0.088     & 0.097     & 0.189     \\
				&		&	{D-SSL}
				&-0.216     & 0.227     & 0.313     & 0.677     &	 \bf{-0.073}& 0.127     & \bf{0.146}& 0.265     & \bf{-0.039}& 0.084     & \bf{0.092}& 0.194     \\
				&		&	{S-SSL}
				&-0.181     & \bf{0.199}& \bf{0.268}& \bf{0.434}&	 -0.092     & \bf{0.119}& 0.150     & \bf{0.229}& -0.046     & \bf{0.080}& \bf{0.092}& \bf{0.173}\\
				\cline{3-15}
				&\multirow{3}{*}{$4n$}&	{D-Lasso2}
				&\bf{-0.112}& 0.235     & \bf{0.259}& 0.477     &	 \bf{-0.065}& 0.130     & 0.145     & 0.251     & \bf{-0.033}& 0.088     & 0.093     & 0.190     \\
				&		&	{D-SSL}
				&-0.219     & 0.239     & 0.323     & 0.509     &	 -0.071     & 0.125     & 0.143     & 0.230     & \bf{-0.033}& 0.079     & \bf{0.086}& 0.173     \\
				&		&	{S-SSL}
				&-0.179     & \bf{0.221}& 0.283     & \bf{0.404}&	 -0.077     & \bf{0.117}& \bf{0.139}& \bf{0.215}& -0.040     & \bf{0.078}& 0.087     & \bf{0.163}\\
				\cline{3-15}
				&\multirow{3}{*}{$8n$}&	{D-Lasso2}
				&\bf{-0.106}& 0.238     & \bf{0.259}& 0.475     &	 \bf{-0.059}& 0.131     & 0.143     & 0.253     & \bf{-0.027}& 0.088     & 0.091     & 0.191     \\
				&		&	{D-SSL}
				&-0.219     & 0.246     & 0.328     & 0.459     &	 -0.069     & 0.132     & 0.148     & 0.218     & -0.030     & 0.084     & 0.089     & 0.166     \\
				&		&	{S-SSL}
				&-0.174     & \bf{0.221}& 0.281     & \bf{0.393}&	 -0.072     & \bf{0.117}& \bf{0.137}& \bf{0.211}& -0.034     & \bf{0.077}& \bf{0.084}& \bf{0.160}\\
				\hline
		\end{tabular}}
	\end{table}
	
	\begin{table}[!htbp] 
		\caption{Simulation results for \underline{Model 1}: computational time (in seconds) of one simulation replication. 
			The estimates of $\bB$ and $\bOmega$ are implemented in parallel, with each utilizing 11 cores.}\label{sim:time}
		\center
		\begin{tabular}{cccrrrrrr}
			\hline
			& & & \multicolumn{3}{c}{$p=200$} & \multicolumn{3}{c}{$p=500$} \\
			\cline{4-9}
			&		&	&$n=100$&$n=300$&$n=500$	&$n=100$&$n=300$&$n=500$	\\
			\cline{3-9}
			&&	{D-Lasso1}
			& 3.914&   6.527 & 5.198   &10.336&29.015    &65.451    \\
			\hline
			\multirow{9}{*}{$N$}&\multirow{3}{*}{$n$}&	{D-Lasso2}
			& 8.095&   5.205 & 5.636	&19.083&138.068	  &57.667     \\
			&		&	{D-SSL}
			& 8.024&   5.191 & 5.841   &18.356&137.662   &57.512   \\
			&		&	{S-SSL}
			& 9.405&  14.368 & 9.163  &21.995&147.846   &89.433 \\
			\cline{2-9}
			&\multirow{3}{*}{$4n$}&	{D-Lasso2}
			& 5.367&   6.522 & 8.253	&65.645&63.110	  &105.88	     \\
			&		&	{D-SSL}
			& 5.352&   6.743 & 8.972	&65.182&63.507    &107.359	     \\
			&		&	{S-SSL}
			& 6.691&  15.802 & 12.103  &68.588&73.173    &138.447 \\
			\cline{2-9}
			&\multirow{3}{*}{$8n$}&	{D-Lasso2}
			& 5.611&   8.660 & 12.959  &58.896&115.243	  &204.105     \\
			&		&	{D-SSL}
			& 5.687&   9.356 & 14.863   &58.809&117.126   &208.926       \\
			&		&	{S-SSL}
			& 6.969&  18.249 & 17.670  &61.927&126.081   &238.811 \\
			\hline
		\end{tabular}
	\end{table}

	\section{Real Data Application}\label{sec_data}
	
	In this section, we apply our proposed method to a real-world dataset 
	from the Medical Information Mart for Intensive Care III (MIMIC-III) database \citep{johnson2016mimic}.
	MIMIC-III is an openly available electronic health records system developed by the MIT Lab for Computational Physiology.
	It comprises deidentified health-related data associated with intensive care unit patients with rich information including demographics, vital signs, laboratory test, medications, and so on.
	Our initial motivation for this data analysis is the association study for the albumin level in the blood sample, a very indicative biomarker correlated with the phenotypes of different types of diseases \citep{phillips1989association}.
	We focus on a subset with 4784 patients of the whole database that the albumin level is available.
	
	Some data cleaning strategy is inevitable for handling electronic health records database. In our situation, around 54\% covariates contain missing values. Among these covariates with missing values, the missingness proportions are 9.4\% on average and the range is from 0.2\% to 30.8\%.
	For those missing values, we simply impute them using the mean of observed samples, the so-called mean imputation. For many clinical markers with continuous scale, the database collects the minimum, the maximum, as well as the mean, values across a certain period of time. To alleviate the potential collinearity among these variables but also to maintain as much information as possible, we decide to only include the maximum and the mean values in our analysis. Additionally, we convert the categorical variables, such as gender and marital status, to dummy variables.
	The number of features after data pre-processing is $p=162$. We randomly sample 4500 observations out of 4784 patients and divide them into $n=500$ labeled data and $N=4000$ unlabeled data, where the value of the outcome variable \emph{albumin} is removed for the 4000 unlabeled data.
	From the set of 500 labeled instances, a subset of 100 is chosen as the observed labeled data. We then construct unlabeled sample sets by selecting the top 1,000, 2,000, and 3,000 instances from the unlabeled pool, in order to study the effects of varying the unlabeled data size.

	Firstly, following the simulation setup, we use the \textsf{hdi} package with the \textsf{robust} option to obtain the debiased Lasso estimator (D‑Lasso1) from the labeled data. Due to multiple testing, the p-value for each covariate is corrected using the default \textsf{holm} approach. We then implemented the proposed D-SSL and S-SSL procedures using the same configuration as in the simulation design and computed the Holm-adjusted $p$-values. For reference, we also obtained a supervised debiased Lasso estimator based on all entire dataset, which is called ``Oracle" and also computed the Holm-adjusted $p$-values.

\begin{figure}[!htbp]
		\centering	\includegraphics[width=1\linewidth,height=0.5\textheight]{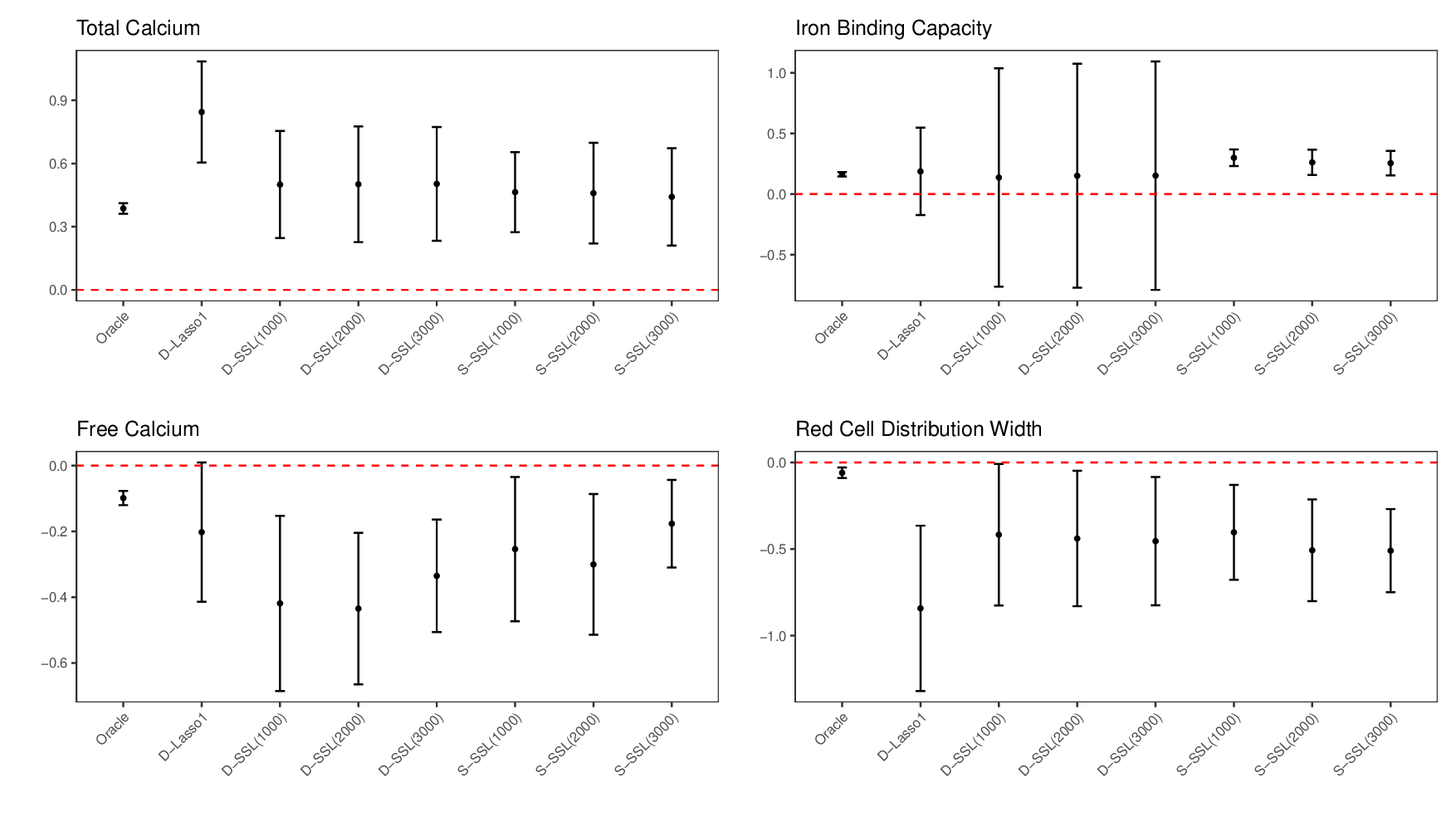}
		\caption{Real Data Application Results: the point estimates and the corresponding confidence intervals of the methods, Oracle, D-Lasso1, D-SSL, and S-SSL, with sample size $n=100$ and $N\in\{1000,2000,3000\}$.}
		\label{fig:realdata}
	\end{figure}

	We focused on four clinically and biologically relevant biomarkers, 
	\emph{Total Calcium} (TC), \emph{Free Calcium} (FC), \emph{Iron Binding Capacity} (IBC) and \emph{Red Cell Distribution Width}
	(RDW), and evaluated three sample-size configurations, $(n,N)\in\{(100,1000),(100,2000),(100,3000)\}$. Figure~\ref{fig:realdata} summarizes the results. All three proposed methods consistently identified TC, a well-established biochemical marker reflecting overall calcium status \citep{payne1973interpretation}. For FC, D-SSL and S-SSL correctly recovered the signal, whereas D-Lasso1 failed to detect it. 
	Free calcium represents the ionized, physiologically active component of serum calcium and therefore carries strong biological relevance \citep{baird2011ionized}. 
	IBC was selected only by S-SSL; this marker reflects transferrin-mediated iron transport and is central to assessing iron metabolism \citep{camaschella2015iron}. 
	Finally, RDW, which quantifies the heterogeneity of red blood cell size \citep{salvagno2015red,patel2009red}, may not be detected by D-SSL at smaller sample sizes but was recovered once more unlabeled data became available. All three methods ultimately selected RDW as $N$ increased. 
	Notably, S-SSL consistently identified all four variables across settings, demonstrating superior detection power.
	
	The confidence intervals exhibit a clear efficiency pattern. For variables selected by S-SSL, their intervals are uniformly shorter than those of D-Lasso1, indicating substantial efficiency gains from incorporating unlabeled data. By contrast, the performance of D-SSL varies across covariates: its intervals can be shorter, comparable, or wider; reflecting the fact that its efficiency improvement depends on correct specification of the working conditional mean model $f(\cdot)$ and is not guaranteed under misspecification.

		\section{Discussion}\label{sec_disc}
		
		In this paper, we propose the semi-supervised estimator
		$\bv\trans\wh\bt^d_{S,\psi}$ for $\bv\trans\bt^*$ with a pre-specified $\bv\in\RR^p$.
		This allows for the development of inference procedures for the general estimand $\bv\trans\bt^*$, such as constructing confidence intervals and conducting hypothesis tests.
		
		A key advantage of the proposed estimator is that it guarantees performance no worse than the supervised approach, ensuring the dependable use of high-dimensional unlabeled data, while avoiding the need to estimate the true conditional mean function $f(\X)$.

		Our framework is more suitable when the goal is to understand the relationship between $Y$ and $\X$ without knowledge of the true regression function, such as when investigating the association between a phenotype and genome-wide SNPs in a genome-wide association study.
		In recent literature, there has been a surge of work defining model-agnostic measures to quantify the importance of covariates for prediction and studying their theoretical properties; these are commonly referred to as variable or feature importance measures \citep{williamson2021nonparametric, williamson2023general, verdinelli2024feature, verdinelli2024decorrelated}. Beyond regression settings, these ideas have also been extended to survival analysis \citep{wolock2025assessing} and causal inference \citep{hines2025variable}.
		Our proposal shares a similar spirit with the variable importance literature in that both aim to understand the role of covariates in a model-agnostic framework. However, the two approaches are fundamentally different. Due to space limitations, we defer a detailed discussion of their similarities and differences to Supplement~\ref{sec:supp_importance}.
		
		Finally, 
		in some applications, labeled and unlabeled data are collected under different conditions or from distinct populations.
		In such cases, assuming that the marginal distribution of $\X$ differs between labeled and unlabeled data \citep{kawakita2013semi} is more appropriate. However, how such distributional differences impact the utility of unlabeled data for estimation and inference remains an open question.
		We plan to explore this issue in future research.
		
		
		


		

		\bibliographystyle{agsm}
		\bibliography{refSSL2}
		
		\clearpage
		\setcounter{page}{1} 
				\renewcommand{\thepage}{S\arabic{page}}
				{\centering
			\section*{SUPPLEMENT}}
		
		\setcounter{equation}{0}\renewcommand{\theequation}{S.\arabic{equation}}
		\setcounter{subsection}{0}\renewcommand{\thesubsection}{S.\arabic{subsection}}
		\setcounter{pro}{0}\renewcommand{\thepro}{S.\arabic{pro}}
		\setcounter{Th}{0}\renewcommand{\theTh}{S.\arabic{Th}}
		\setcounter{lemma}{0}\renewcommand{\thelemma}{S.\arabic{lemma}}
		\setcounter{Rem}{0}\renewcommand{\theRem}{S.\arabic{Rem}}
		\setcounter{assumption}{0}\renewcommand{\theassumption}{S.\arabic{assumption}}
		\setcounter{definition}{0}\renewcommand{\thedefinition}{S.\arabic{definition}}
		
		\setcounter{table}{0}
		\renewcommand{\thetable}{S.\arabic{table}}
		
		\setcounter{figure}{0}
		\renewcommand{\thefigure}{S.\arabic{figure}}
		
		\def\C{{\bf C}}

		\subsection{Preliminary Definition}
		To characterize the tail behavior of random variables, we introduce the following definition.
		
		\begin{definition}[Sub-Gaussian variable and vector]
			A random variable $X$ is called sub-Gaussian if there exists some positive constant $K_2$ such that $\PP(|X|>t)\leq \exp(1-t^2/K^2_2)$ for all $t\geq 0$. The sub-Gaussian norm of $X$ is defined as $\|X\|_{\psi_2}=\sup_{q\geq 1}q^{-1/2}(\E|X|^q)^{1/q}$. A vector $\X\in\RR^p$ is a sub-Gaussian vector if the one-dimensional marginals $\bv\trans\X$ are sub-Gaussian for all $\bv\in\RR^p$, and its sub-Gaussian norm is defined as $\|\X\|_{\psi_2}=\sup_{\|\bv\|_2=1}\|\bv\trans\X\|_{\psi_2}$.
		\end{definition}

		\subsection{Preliminary Lemmas}
		We start with several basic lemmas that we will apply in our proofs.
		
		\begin{lemma}[Lemma B.1 in \cite{victor}] \label{chern}
			Let $\{X_n\},\{Y_n\}$ be sequences of random variables. If for any $c>0$, $\PP(|X_n|>c|Y_n)=o_p(1)$. Then $X_n=o_p(1)$.
		\end{lemma}
		
		\begin{lemma}[Nemirovski moment inequality, Lemma 14.24 in \cite{shd}]\label{repeated_lemma}
			For $m\geq1$ and $p>e^{m-1}$, we have \begin{equation}
				\E\left(\max_{1\leq k\leq p}\left|\sum_{i=1}^n\left[\gamma_k(Z_i)-\E\{\gamma_k(Z_i)\}\right]\right|^m\right)\leq (8\log 2p)^{\frac{m}{2}}\E\left[\left\{\max_{1\leq k\leq p}\sum_{i=1}^n\gamma_k^2(Z_i)\right\}^{m/2}\right]
			\end{equation}
			
		\end{lemma}
		
		\begin{lemma}[Theorem 3.1 in \cite{RE}]\label{RE}
			Assume that $\bX\in \RR^{n\times p}$ has zero mean and covariance $\bSigma$. Furthermore, assume that the rows of $\bX\bSigma^{-1/2}\in \RR^{n\times p}$ are independent sub-gaussian random vector with a bounded sub-gaussian constant and $\Lambda_{\min}(\bSigma)>C_{\min}>0$,  $\max_{1\leq j\leq p}\Sigma_{jj}=O(1)$. Set $0<\delta<1$, $0<s_0<p$, and $L>0$.  Define the following event,
			\begin{equation}
				\mathcal{B}_\delta(n,s_0,L)=\left\{\bX\in \RR^{n\times p}:(1-\delta)\sqrt{C_{\min}} \leq \frac{\norm{\bX \bv}_2}{\sqrt{n}\norm{\bv}_2},\forall \bv\in \mathcal{C}(s_0,L)~\textrm{s.t.}~\bv\neq 0\right\}.
			\end{equation}
			and $\mathcal{C}(s_0,L)=
			\{\bt\in \RR^p:\exists S\subseteq \{1,...,p\},|S|=s_0, \norm{\bt_{S^c}}_1\leq L\norm{\bt_S}_1\}$. Then, there exists a constant $c
			_1=c(L,\delta)$ such that, for sample size $n\geq c_1s_0\log(p/s_0)$, we have
			\begin{equation}
				\PP\left\{\mathcal{B}_\delta(n,s_0,L)\right\}\geq 1-e^{-\delta^2n}.
			\end{equation}
		\end{lemma}

		\subsection{Theoretical Properties of the Debiased Estimator}
		This section is divided into four parts: Section \ref{app:A} presents the central limit theorem for the Debiased Estimator. Section \ref{app:A1} discusses the variance estimation of the Debiased Estimator. Section \ref{app:A2} provides the proof of the central limit theorem for the estimator, and Section \ref{app:A3} examines the convergence rate of the variance estimate for the Debiased Estimator.
		\subsubsection{Asymptotic Properties of Debiased Estimator}\label{app:A}
		
		To show the asymptotic distribution of the proposed estimator $\bv\trans\wh\bt^d$, we require the following assumptions.
		
		\begin{assumption} \label{assumption_est}
			\begin{enumerate}
				\item[(A1)] $\bSigma^{-1/2}\X$ is a zero mean sub-gaussian vector with bounded sub-gaussian norm and ${\rm{Cov}}(\X)=\bSigma$ has smallest eigenvalue $\Lambda_{\min}(\bSigma)\geq C_{\min}>0$ for some positive constant $C_{\min}$. Moreover, $\max_{1\leq j\leq p}\Sigma_{jj}=O(1)$.
				\item[(A2)] $\max_{1\leq i\leq n+N}\norm{\X_i}_\infty \leq K_1$ where we allow $K_1$ to diverge with $(n,N,p)$.
				\item[(A3)] $\E(\epsilon^2)=\sigma^2$ and {$\E[\{f(\X)-\X\trans\bt^*\}^2]\leq\Phi^2$}. 
				
				\item[(A4)]  $\bt^*$ is $s$-sparse with $\norm{\bt^*}_0=s$, and $\frac{s\log p}{n+N}=O(1)$.
			\end{enumerate}
		\end{assumption}

		\begin{assumption}\label{assumption_inf}
			Assume $\max_{1\leq i\leq n+N}\norm{\bOmega \X_i}_\infty \leq K_2$, and  $\max_{1\leq k\leq p}\norm{\bOmega_{k\cdot}}_0\leq s_{\Omega}$ satisfies\\
			$K^2s_{\Omega}\sqrt{\log p/(n+N)}=o(1)$, where $K=K_1\vee K_2$ with $K_1$ defined in Assumption \ref{assumption_est}.
		\end{assumption}
		
		Note that Assumption~\ref{assumption_est} is the same as the Assumption 3.1 in \cite{deng2023optimal}.
		Assumption \ref{assumption_inf} and (A2) in Assumption \ref{assumption_est} together imply the strong boundedness condition and $\max_{1\leq i\leq n+N}\max_{1\leq k\leq p}|\X_{i,-k}\trans\bgamma_k|=O(K)$ in \cite{vandegeer2014} which further guarantees the rate of $\wh\bOmega$ in the matrix $L_{\infty}$ norm. While it is possible to relax the sparsity assumption $\max_{1\leq k\leq p}\norm{\bOmega_{k\cdot}}_0\leq s_{\Omega}$ \citep{javanmard2018}, we make this assumption in order to show the proposed estimator is regular and asymptotically linear, which facilitates the comparison with other competing estimators in terms of asymptotic efficiency.
		Finally, we note that Assumptions \ref{assumption_est} and \ref{assumption_inf} do not impose or imply any upper bound on $\Lambda_{\max}(\bSigma)$. For example,  we allow $\bSigma$ to be an equicorrelation matrix, whose largest eigenvalue is proportional to the dimension $p$. Given these assumptions, the following theorem shows that $\bv\trans\wh\bt^d$ is asymptotically normal for a linear functional $\bv\trans\bt^*$.
		
		\begin{Th} \label{inf1}
			Suppose Assumptions \ref{assumption_est} and \ref{assumption_inf} hold. By choosing $\lambda_{SD}\asymp K_1(\Phi\sqrt{\frac{\log p}{n+N}}+\sigma\sqrt{\frac{\log p}{n}}+b_n\sqrt{\frac{\log p}{n}})$ and $\lambda_k\asymp K\sqrt{\frac{\log p}{n+N}}$ uniformly over $k$, we obtain that for any $\bv\neq \bm{0}\in\RR^p$,
			\begin{align}\label{inf1_1}
				\bv\trans(\wh\bt^d-\bt^*)&=\frac{1}{n}\sum_{i=1}^n\bv\trans\bW_i\{Y_i-f(\X_i)\}+\frac{1}{n+N}\sum_{i=1}^{n+N} \bv\trans\bW_i\{f(\X_i)-\X_i\trans\bt^*\}+O_p(\delta_n),
			\end{align}
			where $\bW_i=\bOmega \X_i$ and $\delta_n=\|\bv\|_1(R_1+R_2)$ with
			$$
			R_1=K_1K(s\vee s_\Omega)\left\{{\frac{\Phi\log p}{n+N}}+{\frac{(\sigma+b_n)\log p}{\sqrt{n(n+N)}}}\right\},~~R_2=K_2b_n\sqrt{\frac{\log p}{n}},
			$$
			and $b_n$ is a deterministic sequence that satisfies $\|\wh f^{-j}-f\|=O_p(b_n)$ for $j=1,2$. In addition, if
			\begin{align}\label{inf1_2}
				\frac{n^{1/2}\delta_n}{\left\{\bv\trans(\sigma^2\bOmega+\frac{n}{n+N}\bGamma)\bv\right\}^{1/2}}=o(1)
			\end{align}
			with $\bGamma=\E[\bW_i^{\otimes 2}\{f(\X_i)-\X_i\trans\bt^*\}^2]$, $\epsilon$ and $\eta(\X)=f(\X)-\X\trans\bt^*$ satisfy
			\begin{align}\label{inf1_3}
				\|\bv\|_1^{2+\delta}K_2^{2+\delta}\left\{\frac{ \E|\epsilon|^{2+\delta}}{n^{\delta/2}(\sigma^2 \bv\trans\bOmega\bv)^{1+\delta/2}}+\frac{\E|\eta(\X)|^{2+\delta}}{(n+N)^{\delta/2}(\bv\trans\bGamma\bv)^{1+\delta/2}}\right\}=o(1),
			\end{align}
			for some $\delta>0$, then
			\begin{align}\label{inf1_4}
				\frac{n^{1/2}\bv\trans(\wh\bt^d-\bt^*)}{\left\{\bv\trans(\sigma^2\bOmega+\frac{n}{n+N}\bGamma)\bv\right\}^{1/2}} \stackrel{d}{\longrightarrow} \mathcal{N}(0,1).
			\end{align}
		\end{Th}
		
		The asymptotic expansion of $\bv\trans(\wh\bt^d-\bt^*)$ is presented in (\ref{inf1_1}), where the remainder term $\delta_n$  consists of two components $R_1$ and $R_2$, which come from the cross product of the estimation errors of $\wh\bOmega$ and $\wh\bt_{SD}$ in Theorem 3.2 of \cite{deng2023optimal} and the plug-in error of $\wh f^{-j}$ in $\wh\bxi$, respectively. To establish the asymptotic normality of $\bv\trans(\wh\bt^d-\bt^*)$, we further need to assume that $\delta_n$ is sufficiently small and the Lyapunov condition holds so that one can apply the central limit theorem to the leading terms in (\ref{inf1_1}). These two conditions  are rigorously formulated in (\ref{inf1_2}) and (\ref{inf1_3}). To further simplify (\ref{inf1_2}) and (\ref{inf1_3}), assume that $\sigma^2 \bv\trans\bOmega\bv\geq C\|\bv\|_2^2$ and $\bv\trans\bGamma\bv\geq C\|\bv\|_2^2$ for some constant $C$, $\E|\epsilon|^{2+\delta}, \E|\eta(\X)|^{2+\delta}, K$ are all $O(1)$ and $b_n=o(1)$. Under these mild conditions, (\ref{inf1_2}) and (\ref{inf1_3}) are implied by
		\begin{align}\label{inf1_5}
			\frac{\|\bv\|_1}{\|\bv\|_2}\left\{\frac{(s\vee s_\Omega)\log p}{\sqrt{n+N}}+b_n\sqrt{\log p}+n^{-\frac{\delta}{2(2+\delta)}}\right\}=o(1).
		\end{align}

		\begin{Rem}
			(1) The bound (\ref{inf1_5}) requires the ratio $\|\bv\|_1/\|\bv\|_2$ cannot be too large which excludes the case that $\bv$ has many large entries (e.g., $\bv=(1,1,...,1)\trans$). This observation agrees with the theoretical results in \cite{cai2015confidence}, as the debiased estimator does not yield optimal confidence intervals for $\bv\trans\bt^*$ when $\bv$ is a dense vector. To see some concrete examples that our results are applicable, we first note that if $\bv=\be_j$ the $j$th basis vector in $\RR^p$, then $\bv\trans\wh\bt^d=\wh\theta^d_j$ reduces to the estimate of $\theta_j$. Our condition (\ref{inf1_5}) becomes $(s\vee s_\Omega)\log p=o(\sqrt{n+N})$ and $b_n=o(1/\sqrt{\log p})$. The former is a standard condition for debiased inference adapted to the semi-supervised setting and the latter is slightly stronger than the consistency of $\wh f^{-j}$ required in Theorem 3.2 of \cite{deng2023optimal}; see Remark 3.3 for details. The same comments are applicable if the parameter of interest $\bv\trans\bt^*$ is a linear combination of $\bt^*$ with $\|\bv\|_0$ fixed.
			
			Indeed, the set of vector $\bv$ in $\RR^p$ satisfying (\ref{inf1_5}) forms a cone $[\frac{\|\bv\|_1}{\|\bv\|_2}\leq t_n\{\frac{(s\vee s_\Omega)\log p}{\sqrt{n+N}}+b_n\sqrt{\log p}+n^{-\frac{\delta}{2(2+\delta)}}\}^{-1}]$ for some $t_n=o(1)$. Compared to  \cite{cai2015confidence} who proposed the debiased estimator for $\bv\trans\bt^*$ with sparse $\bv$, the cone condition (\ref{inf1_5}) may still hold if $\bv$ is approximately sparse with many small but nonzero entries. Our results are still applicable in this case.
			
			(2) Assuming $\|\bv\|_1/\|\bv\|_2$ is a constant and $N\gg n$, we can see from (\ref{inf1_5}) that in the semi-supervised setting we need $(s\vee s_\Omega)\log p=o(\sqrt{n+N})$, which is much weaker than the similar condition $(s\vee s_\Omega)\log p=o(\sqrt{n})$ for the supervised  estimators (up to some logarithmic factors). Thus, with a large amount of unlabeled data, our inference results may still hold for models with large $s$.
		\end{Rem}

		\begin{Rem}[Efficiency improvement and semi-parametric efficiency bound]\label{rem_efficiency}
			We first note that, when the linear model is correctly specified i.e. $f(\X)=\X\trans\bt^*$, we have $\bGamma=\0$ and the asymptotic variance of $\bv\trans\wh\bt^d$ reduces to $\sigma^2\bv\trans\bOmega\bv$, which agrees with the asymptotic variance of the debiased estimator in fully supervised setting and also matches the semi-parametric efficiency bound. In this case, the information of $X$ contained in the unlabeled data is ancillary and does not contribute to the inference on $\bt$; see also \cite{azriel,chakrabortty2018}.
			
			In the following, we assume $\bGamma$ is strictly positive definite. Recall that our asymptotic analysis requires $n,p\rightarrow\infty$ and allows $N$ to be either fixed or grow with $n$. In the following, we discuss the asymptotic variance of $\bv\trans\wh\bt^d$ in (\ref{inf1_5}) according to the magnitude of $N$.
			\begin{itemize}
				\item[(1)] $\lim_{n\rightarrow\infty}\frac{n}{n+N}=1$. Denote $\Kb=\E\left\{\X^{\otimes 2}(Y-X\trans\bt^*)^2\right\}$. It is seen that $\Kb=\sigma^2\bSigma+\bSigma\bGamma\bSigma$. In this case, the asymptotic variance of $\bv\trans\wh\bt^d$ reduces to $\bv\trans(\sigma^2\bOmega+\bGamma)\bv=\bv\trans\bOmega \Kb\bOmega\bv$, which is the asymptotic variance of the debiased estimator in the fully supervised setting; see \cite{mann2015,ning2017}. As expected, when $N\ll n$, the amount of unlabeled data is not sufficiently large to improve the asymptotic efficiency of the estimator.
				\item[(2)]  $\lim_{n\rightarrow\infty}\frac{n}{n+N}=\rho$ for some $0<\rho<1$. In this case, the asymptotic variance $\bv\trans(\sigma^2\bOmega+\rho\bGamma)\bv$ is strictly smaller than $\bv\trans\bOmega \Kb\bOmega\bv=\bv\trans(\sigma^2\bOmega+\bGamma)\bv$. Thus, the unlabeled data can be used to improve the asymptotic efficiency for inference.
				\item[(3)]  $\lim_{n\rightarrow\infty}\frac{n}{n+N}=0$. In the case, the asymptotic variance becomes $\sigma^2\bv\trans\bOmega\bv$. Indeed, if the distribution of $\X$ is known, the semi-parametric efficiency bound for estimating $\bv\trans\bt^*$ is exactly $\sigma^2\bv\trans\bOmega\bv$ as well; see \cite{chakrabortty2018} and the reference therein. Thus, when $N\gg n$, our estimator attains the semi-parametric efficiency bound.
			\end{itemize}
			
		\end{Rem}
		
		\subsubsection{Variance Estimation}\label{app:A1}
		
		In the following, we consider how to estimate the asymptotic variance of $\bv\trans\wh\bt^d$. 
		
		For estimating $\bOmega$,
		one can consider the following node-wise lasso estimator \citep{meinshausen2006high} based on both labeled and unlabeled data $\wt\bX$. For $k \in [p]$, define the vector $\wh{\bgamma}_k=\{\wh\gamma_{k,j}:j\in [p] ~\textrm{and}~ j\neq k\}$ as
		\begin{equation}\label{eq_hatgamma}
			\wh{\bgamma}_k=\argmin_{\gamma\in\mathbb{R}^p}\left\{ \frac{1}{n+N}||\bX_{\cdot k}-\wt \bX_{\cdot -k}\bgamma||_2^2+2\lambda_k\norm{\bgamma}_1\right\}.
		\end{equation}
		Denote by
		$$\wh{\bC}=
		\begin{bmatrix}
			1& -\wh{\gamma}_{1,2}&\dots& -\wh{\gamma}_{1,p} \\
			-\wh{\gamma}_{2,1}& 1 & \dots& -\wh{\gamma}_{2,p} \\
			\vdots & \vdots & \ddots & \vdots\\
			-\wh{\gamma}_{p,1}&-\wh{\gamma}_{p,2} & \dots &1
		\end{bmatrix}
		$$
		and let
		\begin{equation} \label{nodewise lasso}
			\wh \bT^2={\rm{diag}}(\wh{\tau}_1^2,...,\wh{\tau}_p^2), ~\textrm{where}~\wh{\tau}_k^2=\frac{1}{n+N}(\wt \bX_{\cdot k}-\wt \bX_{\cdot -k}\wh\bgamma_k)\trans\wt\bX_{\cdot k}.
		\end{equation}
		Then the node-wise lasso estimator is defined as
		\begin{equation} \label{hatomega}
			\wh\bOmega=\wh \bT^{-2}\wh \bC.
		\end{equation}
		
		To estimate $\sigma^2$, we apply the cross-fitting technique. Specifically,  for $j=\{1,2\}$, define
		$$
		\wh\sigma^2_j=\frac{1}{n_j}\sum_{i\in D^*_j}\left\{Y_i-\wh f^{-j}(\X_i)\right\}^2.
		$$
		We estimate $\sigma^2$ by $\wh\sigma^2=(\wh\sigma^2_1+\wh\sigma^2_2)/2$. Similarly, define
		$$
		\wh\bGamma_j=\frac{1}{n_j+N_j}\sum_{i\in D_j} (\wh\eta_i^{-j})^2\wh\bOmega \X_i\X_i\trans\wh\bOmega,
		$$
		where $\wh\eta_i^{-j}=\wh f^{-j}(\X_i)-\wh\bt_{SD}\trans \X_i$ and $\wh\bOmega$ is defined in (\ref{hatomega}). We then estimate $\bGamma$ by $\wh\bGamma=(\wh\bGamma_1+\wh\bGamma_2)/2$. The following Proposition shows that the asymptotic variance of $\bv\trans\wh\bt^d$ can be consistently estimated by the plug-in estimator $\bv\trans(\wh\sigma^2\wh\bOmega+\frac{n}{n+N}\wh\bGamma)\bv$.

		
		\begin{pro}\label{prop_variance1}
			Suppose Assumptions \ref{assumption_est} and \ref{assumption_inf} hold. To simplify the presentation, we further assume $\E(\epsilon^4)=O(1), \E\left\{\eta^4(\X)\right\}=O(1)$ and $K \sqrt{\frac{s\log p}{n+N}}=o(1)$.
			Then
			\begin{equation}\label{eq_prop_variance1}
				\left|\bv\trans\left(\wh\sigma^2\wh\bOmega+\frac{n}{n+N}\wh\bGamma\right)\bv-\bv\trans\left(\sigma^2\bOmega+\frac{n}{n+N}\bGamma\right)\bv\right|
				=O_p\left\{\|\bv\|_2^2\left(\frac{1}{\sqrt{n}}+b_n^2\right)+\textrm{Rem}_N\right\},
			\end{equation}
			where
			\begin{equation}\label{eq_prop_variance2}
				\textrm{Rem}_N=\frac{n}{n+N}K^2\|\bv\|_1^2b_n+K^3\|\bv\|_1^2(s\vee s_\Omega)\sqrt{\frac{\log p}{n+N}}.
			\end{equation}
			Under the additional assumptions $\sigma^2\bv\trans\bOmega\bv\geq C\|\bv\|_2^2$ and $\textrm{Rem}_N/\|\bv\|_2^2=o(1)$, we have
			\begin{align}\label{eq_prop_variance3}
				\frac{n^{1/2}\bv\trans(\wh\bt^d-\bt^*)}{\left\{\bv\trans(\wh\sigma^2\wh\bOmega+\frac{n}{n+N}\wh\bGamma)\bv\right\}^{1/2}} \stackrel{d}{\longrightarrow} \mathcal{N}(0,1).
			\end{align}
		\end{pro}

		To better understand the convergence rate of the estimated asymptotic variance, we decompose the error in  (\ref{eq_prop_variance1}) into two terms, $\|\bv\|_2^2\left(\frac{1}{\sqrt{n}}+b_n^2\right)$ and $\textrm{Rem}_N$. The former is due to the estimation error of $\wh\sigma^2$ and the latter comes from the error of $\wh\bGamma$ and $\wh\bOmega$. It is of interest to note that, if $N\gg n$, the error term $\textrm{Rem}_N$ may vanish to 0  fast enough, so that the convergence rate of the estimated asymptotic variance in (\ref{eq_prop_variance1}) is dominated by $\|\bv\|_2^2\left(\frac{1}{\sqrt{n}}+b_n^2\right)$. In addition, for many practical estimators $\wh f^{-j}$, such as the group lasso estimator for sparse additive models in Remark 3.3 of \cite{deng2023optimal}, its convergence rate in $L_2(\PP)$ norm is no slower than $n^{-1/4}$, that is $b_n=o(n^{-1/4})$. In this case, the rate in (\ref{eq_prop_variance1}) further reduces to $\|\bv\|^2_2/\sqrt{n}$, which is the best possible rate for estimating the variance even if $\bOmega$, $\bGamma$ and $f(X)$ are known. Thus, the unlabeled data lead to a more accurate estimate of the asymptotic variance.

		Finally, from (\ref{eq_prop_variance3}) we can construct the $(1-\alpha)$ confidence interval for $\bv\trans\bt^*$ as $[\bv\trans\wh\bt^d-z_{1-\alpha/2}n^{-1/2}sd, \bv\trans\wh\bt^d+z_{1-\alpha/2}n^{-1/2}sd]$, where $z_{1-\alpha/2}$ is the $1-\alpha/2$ quantile of a standard normal distribution and $sd=\Big\{\bv\trans(\wh\sigma^2\wh\bOmega+\frac{n}{n+N}\wh\bGamma)\bv\Big\}^{1/2}$. Similarly, if one is interested in testing the hypothesis $H_0: \bv\trans\bt^*=0$, we can construct the test statistic $n^{1/2}\bv\trans\wh\bt^d/\left\{\bv\trans(\wh\sigma^2\wh\bOmega+\frac{n}{n+N}\wh\bGamma)\bv\right\}^{1/2}$ based on (\ref{eq_prop_variance3}).

\subsubsection{Proof of Theorem \ref{inf1}}\label{app:A2}

\begin{proof}
	
	We will first derive some preliminary probability bounds that will be used later in the proof. With (A1)-(A5) in Assumptions (\ref{assumption_est}) and (\ref{assumption_inf}), we can verify the assumptions (B1)-(B4) for strongly bounded case in Theorem 2.4 of \cite{vandegeer2014} holds with $K=K_1\vee K_2$. In particular, we have
	\begin{align*}
		|\{\X^{(-k)}\}\trans\gamma_k|&=|\{\X^{(-k)}\}\trans\bSigma_{-k,-k}^{-1}\bSigma_{-k,k}|=|\{\X^{(-k)}\}\trans\bOmega_{-k,k}||\bOmega_{kk}^{-1}|\\
		&=|\{\X^{(-k)}\}\trans\bOmega_{-k,k}|(\bSigma_{kk}-\bSigma_{k,-k}\bSigma_{-k,-k}^{-1}\bSigma_{-k,k})=O(K_2),
	\end{align*}
	uniformly over $1\leq k\leq p$.
	
	Under the the strongly bounded case with $s_\Omega=o\left(\frac{n+N}{\log p}\right)$, and $\max_k\bSigma_{k,k}=O(1)$, we can apply Theorem 2.4 and Lemma 5.3 in \cite{vandegeer2014}
	and claim that the nodewise lasso estimator satisfies
	\begin{equation}\label{eq_rate_nodewise}
		\norm{\wh \bOmega-\bOmega}_\infty=O_p\left(Ks_{\Omega}\sqrt{\frac{\log p}{n+N}}\right),~~\norm{\bI_p-\wh \bOmega \wh \bSigma_{n+N}}_{\max}=O_p\left(K\sqrt{\frac{\log p}{n+N}}\right).
	\end{equation}
	The first probability bound in (\ref{eq_rate_nodewise}) is directly from Theorem 2.4 of \cite{vandegeer2014}. To see the second probability bound, with the formulation of $\wh \bOmega$ and notation from nodewise Lasso (\ref{nodewise lasso}), we know for each row of $\wh \bOmega$,
	$$
	\norm{\wh \bSigma_{n+N}\bOmega_{k\cdot}\trans-e_k}_\infty\leq \lambda_k/\wh \tau_k^2,
	$$
	where $e_k$ is the unit vector.
	Furthermore, invoking Lemma 5.3 in \cite{vandegeer2014}, we know when we choose a suitable tuning parameter $\lambda_k\asymp K\sqrt{\frac{\log p}{n+N}}$ uniformly over $k$, we have $\max_k 1/\wh \tau_k^2=O_p(1)$. Hence, $\norm{\bI_p-\wh \bOmega \wh \bSigma_{n+N}}_{\max}\leq \max_k \left(\lambda_k/\wh \tau_k^2\right)=O_p\left(K\sqrt{\frac{\log p}{n+N}}\right)$.
	
	In addition, recalling from the derivation of (S.17) in \cite{deng2023optimal}, we have
	\begin{equation}\label{eq_inf1_1}
		\norm{\frac{\bX\trans(\bY- \fb_n)}{n}+\frac{\wt\bX\trans( \fb_{n+N}-\wt\bX\bt^*)}{n+N}}_\infty=O_p\left(K_1\sigma \sqrt{\frac{\log p}{n}}+K_1\Phi\sqrt{\frac{\log p}{n+N}}\right),
	\end{equation}
	where $\fb_n=\left\{f(\X_1),\dots,f(\X_n)\right\}\trans$ and $\fb_{n+N}$ is defined similarly.
	
	Given the above preliminary results, we focus on deriving the limiting distribution of $\bv\trans(\wh\bt^d-\bt^*)$. Recall that we use the following notation $\wh\fb^{-j}_{D^*_j}=\{\wh f^{-j}(\X_i):i\in D_j^*\}$, $\wh\fb^{-j}_{D_j}=\{\wh f^{-j}(\X_i):i\in D_j\}$, and $\fb_{D^*_j}$ and $\fb_{D_j}$ are defined similarly. We decompose the term $\bv\trans(\wh\bt^d-\bt^*)$ as
	\begin{align}
		& \bv\trans(\wh\bt^d-\bt^*)\nonumber\\
		=&\bv\trans\left[(\bI_p-\wh\bOmega\wh\bSigma_{n+N})(\wh\bt_{SD}-\bt^*)+\wh \bOmega\sum_{j=1}^2\left\{\frac{\bX_{D^*_j}\trans(\bY_{D^*_j}-\wh \fb^{-j}_{D^*_j})}{2n_j}+\frac{\wt \bX_{D_j}\trans( \wh \fb^{-j}_{D_j}-\wt\bX_{D_j}\bt^*)}{2n_j+2N_j}\right\}\right]\nonumber\\
		=&\bv\trans\left[(\bI_p-\wh\bOmega\wh\bSigma_{n+N})(\wh\bt_{SD}-\bt^*)+(\wh \bOmega-\bOmega)\left\{\frac{\bX\trans(\bY- \fb_{n})}{n}+\frac{\wt \bX\trans( \fb_{n+N}-\wt \bX\bt^*)}{n+N}\right\}\right.\nonumber\\
		&\left.+\bOmega\left\{\frac{\bX\trans(\bY-\fb_{n})}{n}+\frac{\wt \bX\trans(\fb_{n+N}-\wt \bX\bt^*)}{n+N}\right\}-\bOmega\sum_{j=1}^2\left\{\frac{\bX_{D^*_j}\trans(\wh \fb^{-j}_{D^*_j}- \fb_{D^*_j})}{2n_j}+\frac{\wt \bX_{D_j}\trans( \fb_{D_j}-\wh \fb^{-j}_{D_j})}{2n_j+2N_j}\right\}\right.\nonumber\\
		&\left.+(\bOmega-\wh\bOmega) \sum_{j=1}^2\left\{\frac{\bX_{D^*_j}\trans(\wh \fb^{-j}_{D^*_j}- \fb_{D^*_j})}{2n_j}+\frac{\wt \bX_{D_j}\trans( \fb_{D_j}-\wh \fb^{-j}_{D_j})}{2n_j+2N_j}\right\}\right].\label{eq_pf_debias}
	\end{align}
	Therefore, with the preliminary results above in hand, we can show that
	\begin{align*}
		\norm{(\bI_p-\wh\bOmega\wh\bSigma_{n+N})(\wh\bt_{SD}-\bt^*)}_\infty&\leq \norm{\bI_p-\wh\bOmega\wh\bSigma_{n+N}}_{\max}\norm{\wh\bt_{SD}-\bt^*}_1\\
		&=O_p\left[K_1Ks\left\{\Phi{\frac{\log p}{n+N}}+\sigma{\frac{\log p}{\sqrt{n(n+N)}}}+b_n{\frac{\log p}{\sqrt{n(n+N)}}}\right\}\right],
	\end{align*}
	from (\ref{eq_rate_nodewise}) and Theorem 3.2 of \cite{deng2023optimal}. Similarly,
	\begin{align*}
		&\norm{ (\wh \bOmega-\bOmega)\left\{\frac{\bX\trans(\bY- \fb_{n})}{n}+\frac{\wt\bX\trans( \fb_{n+N}-\wt\bX\bt^*)}{n+N}\right\}}_\infty\\
		&\leq \norm{ \wh \bOmega-\bOmega}_\infty\norm{\frac{\bX\trans(\bY- \fb_{n})}{n}+\frac{\wt\bX\trans( \fb_{n+N}-\wt\bX\bt^*)}{n+N}}_\infty \nonumber\\
		&=O_p\left[K_1Ks_\Omega\left\{\Phi{\frac{\log p}{n+N}}+\sigma{\frac{\log p}{\sqrt{n(n+N)}}}\right\}\right],
	\end{align*}
	from (\ref{eq_inf1_1}) and (\ref{eq_rate_nodewise}). Following the similar argument in the analysis of $I_1$ in (S.14) of \cite{deng2023optimal} together with the assumption that $\norm{\bW}_\infty=\norm{\bOmega X}_\infty \leq K_2$, we obtain
	\begin{align*}
		\norm{\bOmega\sum_{j=1}^2\left\{\frac{\bX_{D^*_j}\trans(\wh \fb^{-j}_{D^*_j}- \fb_{D^*_j})}{2n_j}+\frac{\wt \bX_{D_j}\trans( \fb_{D_j}-\wh \fb^{-j}_{D_j})}{2n_j+2N_j}\right\}}_\infty=O_p\left(K_2b_n\sqrt{\frac{\log p}{n}}\right).
	\end{align*}
	Similarly, we have
	\begin{align*}
		&\norm{(\bOmega-\wh\bOmega) \sum_{j=1}^2\left\{\frac{\bX_{D^*_j}\trans(\wh \fb^{-j}_{D^*_j}- \fb_{D^*_j})}{2n_j}+\frac{\wt \bX_{D_j}\trans( \fb_{D_j}-\wh \fb^{-j}_{D_j})}{2n_j+2N_j}\right\}}_\infty\\
		\leq& \norm{\bOmega-\wh\bOmega}_{\infty} \norm{\sum_{j=1}^2\left\{\frac{\bX_{D^*_j}\trans(\wh \fb^{-j}_{D^*_j}- \fb_{D^*_j})}{2n_j}+\frac{\wt \bX_{D_j}\trans( \fb_{D_j}-\wh \fb^{-j}_{D_j})}{2n_j+2N_j}\right\}}_\infty\\
		=&O_p\left\{K_1Kb_ns_\Omega\frac{\log p}{\sqrt{n(n+N)}}\right\}.
	\end{align*}
	Collecting the above probability bounds and plugging into (\ref{eq_pf_debias}), we obtain
	\begin{align}
		\bv\trans(\wh\bt^d-\bt^*)&=\bv\trans\bOmega\left\{\frac{\bX\trans(\bY-\fb_n)}{n}+\frac{\wt\bX\trans(\fb_{n+N}-\wt\bX\bt^*)}{n+N}\right\}+O_p(\delta_n)\nonumber\\
		&=\frac{1}{n}\sum_{i=1}^n\bv\trans\bW_i\{Y_i-f(\X_i)\}+\frac{1}{n+N}\sum_{i=1}^{n+N} \bv\trans\bW_i\{f(\X_i)-\X_i\trans\bt^*\}+O_p(\delta_n)\nonumber\\
		&=\sum_{i=1}^{n+N} \xi_i+O_p(\delta_n).\label{eq_pf_debias2}
	\end{align}
	where
	$$
	\xi_i=\begin{cases}
		\frac{1}{n}\bv\trans\bW_i\left[Y_i-f(\X_i)+\frac{n}{n+N}\{f(\X_i)-\X_i\trans\bt^*\}\right] & \mbox{for}~ 1\leq i\leq n, \\
		\frac{1}{n+N}\bv\trans\bW_i\{f(\X_i)-\X_i\trans\bt^*\} & \mbox{for}~ n+1\leq i\leq n+N
	\end{cases}
	$$
	and
	$$
	\delta_n=\|\bv\|_1\left[K_1K(s\vee s_\Omega)\left\{\Phi{\frac{\log p}{n+N}}+\sigma{\frac{\log p}{\sqrt{n(n+N)}}}+b_n{\frac{\log p}{\sqrt{n(n+N)}}}\right\}+K_2b_n\sqrt{\frac{\log p}{n}}\right].
	$$
	In the following, we will apply the Lindeberg-Feller Central Limit Theorem to  (\ref{eq_pf_debias2}). First, we note that $\E[\bW_i\{Y_i-f(\X_i)\}]=\0$ and $\E [\bW_i\{f(\X_i)-\X_i\trans\bt^*\}]=\0$. Denote $\eta_i=f(\X_i)-\X_i\trans\bt^*$. We have
	\begin{align*}
		\sum_{i=1}^{n+N}\E(\xi_i^2)&=\sum_{i=1}^n\frac{1}{n^2}\E\left\{\bv\trans\bW_i\left(\epsilon_i+\frac{n}{n+N}\eta_i\right)\right\}^2+\sum_{i=n+1}^{n+N}\frac{1}{(n+N)^2}\E(\bv\trans\bW_i\eta_i)^2\\
		&=\frac{1}{n}\bv\trans\left(\sigma^2\bOmega+\frac{n}{n+N}\bGamma\right)\bv:=t_n^2,
	\end{align*}
	where $\bGamma={\rm{Cov}}[\bW\{f(\X)-\X\trans\bt^*\}]$. The Lyapunov condition holds as follows
	\begin{align*}
		\frac{\sum_{i=1}^{n+N}\E|\xi_i|^{2+\delta}}{t_n^{2+\delta}}&\leq \frac{\|\bv\|_1^{2+\delta} K_2^{2+\delta}\{\sum_{i=1}^n\E(\epsilon_i+\frac{n}{n+N}\eta_i)^{2+\delta}/n^{2+\delta}+\sum_{i=n+1}^{n+N}\E\eta_i^{2+\delta}/(n+N)^{2+\delta}\}}{t_n^{2+\delta}}\\
		&\leq \frac{\|\bv\|_1^{2+\delta} K_2^{2+\delta}[2^{1+\delta}\{\E\epsilon_i^{2+\delta}+(\frac{n}{n+N})^{2+\delta}\E\eta_i^{2+\delta}\}/n^{1+\delta}+\E\eta_i^{2+\delta}N/(n+N)^{2+\delta}]}{t_n^{2+\delta}}\\
		&\leq \frac{2^{1+\delta} \|\bv\|_1^{2+\delta} K_2^{2+\delta}\E\epsilon_i^{2+\delta}}{n^{\delta/2}(\sigma^2 \bv\trans\bOmega\bv)^{1+\delta/2}}+\frac{ \|\bv\|_1^{2+\delta} K_2^{2+\delta}(2^{1+\delta}n+N)\E\eta_i^{2+\delta}}{(n+N)^{1+\delta/2}(\bv\trans\bGamma\bv)^{1+\delta/2}}\rightarrow 0,
	\end{align*}
	where the first inequality follows from $\norm{\bW}_\infty=\norm{\bOmega X}_\infty \leq K_2$ and the second one is due to the convexity of the function $x^{2+\delta}$ for $x>0$. Therefore, the Lindeberg-Feller Central Limit Theorem leads to
	$$
	\sum_{i=1}^{n+N} \xi_i/t_n\stackrel{d}{\longrightarrow} \mathcal{N}(0,1).
	$$
	From (\ref{eq_pf_debias2}) we obtain
	\begin{align*}
		\frac{\bv\trans(\wh\bt^d-\bt^*)}{t_n}&=\sum_{i=1}^{n+N} \xi_i/t_n+O_p(\delta_n/t_n)\stackrel{d}{\longrightarrow} \mathcal{N}(0,1),
	\end{align*}
	as $\delta_n/t_n=o(1)$. This completes the proof.
	
\end{proof}

\subsubsection{Proof of Proposition \ref{prop_variance1}}\label{app:A3}
\begin{lemma}\label{lem_sig}
	Under the same conditions in Proposition \ref{prop_variance1}, we have
	\begin{equation}\label{eq_lem_sig1}
		|\wh\sigma^2-\sigma^2|=O_p(n^{-1/2}+b^2_n),
	\end{equation}
	and
	\begin{equation}\label{eq_lem_gamma}
		\Big|\bv\trans(\wh\bGamma-\bGamma)\bv\Big|=O_p\left\{K^2\|\bv\|_1^2\left(b_n+K\sqrt{\frac{s\log p}{n}}+\frac{K^2s\log p}{n}+Ks_{\Omega}\sqrt{\frac{\log p}{n+N}}\right)\right\}.
	\end{equation}
\end{lemma}

\begin{proof}[Proof of Lemma \ref{lem_sig}]
	To show (\ref{eq_lem_sig1}), it suffices to upper bound $\wh\sigma^2_j-\sigma^2$, that is
	\begin{align}
		\wh\sigma^2_j-\sigma^2&=\frac{1}{n_j}\sum_{i\in D^*_j}\left\{Y_i-f(\X_i)+f(\X_i)-\wh f^{-j}(\X_i)\right\}^2-\sigma^2\nonumber\\
		&=\frac{1}{n_j}\sum_{i\in D^*_j} (\epsilon_i^2-\sigma^2)+\frac{2}{n_j}\sum_{i\in D^*_j}\epsilon_i\left\{f(\X_i)-\wh f^{-j}(\X_i)\right\}+\frac{1}{n_j}\sum_{i\in D^*_j}\left\{f(\X_i)-\wh f^{-j}(\X_i)\right\}^2.\label{eq_pf_prop_variance1}
	\end{align}
	Chebyshev's inequality together with the assumption $\E(\epsilon^4)\leq C$ implies $\frac{1}{n_j}\sum_{i\in D^*_j} (\epsilon_i^2-\sigma^2)=O_p(n^{-1/2})$. As in the derivation of (S.16) in \cite{deng2023optimal} we have
	$$
	\PP\left[\frac{1}{n_j}\sum_{i\in D^*_j}\epsilon_i\left\{f(\X_i)-\wh f^{-j}(\X_i)\right\}>c\sigma b_n/n_j^{1/2}|D_j^*\right]\leq \frac{\|\wh f^{-j}-f\|_2^2}{c^2b_n^2}\wedge 1.
	$$
	As a result, we have
	$$
	\frac{1}{n_j}\sum_{i\in D^*_j}\epsilon_i\left\{f(\X_i)-\wh f^{-j}(\X_i)\right\}=O_p\left(\frac{b_n}{n^{1/2}}\right).
	$$
	Similarly, $\frac{1}{n_j}\sum_{i\in D^*_j}\{f(\X_i)-\wh f^{-j}(\X_i)\}^2\lesssim \|\wh f^{-j}-f\|_2^2=O_p(b^2_n)$.  Plugging into (\ref{eq_pf_prop_variance1}), we have $|\wh\sigma^2_j-\sigma^2|=O_p(n^{-1/2}+b^2_n)$, which further implies (\ref{eq_lem_sig1}).
	
	To show (\ref{eq_lem_gamma}), we decompose $\bv\trans(\wh\bGamma_j-\bGamma)\bv$ as follows
	\begin{align}
		&\bv\trans(\wh\bGamma_j-\bGamma)\bv\nonumber\\
		&=\customunderbrace{\bv\trans\wh\bOmega\frac{1}{n_i+N_j}\sum_{i\in D_j}\X_i\X_i\trans \{(\wh\eta_i^{-j})^2-\eta_i^2\}\wh\bOmega\bv}{$T_1$}+\customunderbrace{\bv\trans\wh\bOmega\frac{1}{n_i+N_j}\sum_{i\in D_j}\{\X_i\X_i\trans \eta_i^2-\E(\X_i\X_i\trans \eta_i^2)\}\wh\bOmega\bv}{$T_2$}\nonumber\\
		&~~~~~+\customunderbrace{\bv\trans(\wh\bOmega-\bOmega) \E(\X_i\X_i\trans \eta_i^2)\wh\bOmega\bv}{$T_3$}+\customunderbrace{\bv\trans\bOmega \E(\X_i\X_i\trans \eta_i^2)(\wh\bOmega-\bOmega)\bv}{$T_3$}\label{eq_pf_prop_variance2}
	\end{align}
	Let us first consider $T_1$, which can be rewritten as
	\begin{align*}
		T_1&=\frac{1}{n_i+N_j}\sum_{i\in D_j}(\bv\trans\wh\bOmega \X_i)^2\{\wh\eta_i^{-j}-\eta_i\}^2+\frac{2}{n_i+N_j}\sum_{i\in D_j}(\bv\trans\wh\bOmega \X_i)^2\eta_i\{\wh\eta_i^{-j}-\eta_i\}.
	\end{align*}
	We know that $|\bv\trans\wh\bOmega \X_i|\leq \|\bv\|_1\|\wh\bOmega \X_i\|_\infty\leq  \|\bv\|_1(\|\bOmega \X_i\|_\infty+\|\wh\bOmega-\bOmega\|_\infty \|\X_i\|_\infty)\lesssim K\|\bv\|_1$, since $K s_{\Omega}\{\log p/(n+N)\}^{1/2}=o(1)$. In addition, $(\wh\eta_i^{-j}-\eta_i)^2\leq 2\{\wh f^{-j}(\X_i)-f(\X_i)\}^2+2\{\X_i\trans(\wh\bt_{SD}-\bt^*)\}^2$. Combining these results, Theorem 3.2 of \cite{deng2023optimal} and $b_n=o(1)$, we derive
	$$
	\Big|\frac{1}{n_i+N_j}\sum_{i\in D_j}(\bv\trans\wh\bOmega \X_i)^2(\wh\eta_i^{-j}-\eta_i)^2\Big|=O_p\left\{K^2\|\bv\|_1^2\left(b_n^2+\frac{K_1^2s\log p}{n}\right)\right\}
	$$
	and
	\begin{align*}
		&\Big|\frac{1}{n_i+N_j}\sum_{i\in D_j}(\bv\trans\wh\bOmega \X_i)^2\eta_i(\wh\eta_i^{-j}-\eta_i)\Big|\\
		&\leq \Big|\frac{1}{n_i+N_j}\sum_{i\in D_j}(\bv\trans\wh\bOmega \X_i)^2\eta_i^2\Big|^{1/2} \Big|\frac{1}{n_i+N_j}\sum_{i\in D_j}(\bv\trans\wh\bOmega \X_i)^2(\wh\eta_i^{-j}-\eta_i)^2\Big|^{1/2}\\
		&=O_p\left\{K^2\|\bv\|_1^2\left(b_n+K\sqrt{\frac{s\log p}{n}}\right)\right\},\\
	\end{align*}
	where the first step holds by Cauchy–Schwarz inequality. This implies the following rate for $T_1$:
	$$
	|T_1|=O_p\left\{K^2\|\bv\|_1^2\left(b_n+K\sqrt{\frac{s\log p}{n}}+\frac{K^2s\log p}{n}\right)\right\}.
	$$
	For $T_2$, we can show that
	\begin{align*}
		T_2&= \bv\trans(\wh\bOmega-\bOmega) \bZ_n(\wh\bOmega-\bOmega)\bv+2 \bv\trans(\wh\bOmega-\bOmega)\bZ_n\bOmega\bv+\bv\trans\bOmega\bZ_n\bOmega\bv
	\end{align*}
	where $\bZ_n=\frac{1}{n_i+N_j}\sum_{i\in D_j}\{\X_i\X_i\trans \eta_i^2-\E(\X_i\X_i\trans \eta_i^2)\}$. We can bound the three terms in the right hand side of the above equation separately. As an illustration, we have
	$$
	\bv\trans(\wh\bOmega-\bOmega) \bZ_n(\wh\bOmega-\bOmega)\bv\leq \|\bv\|_1^2\|\wh\bOmega-\bOmega\|^2_\infty\|\bZ_n\|_{\max}
	=O_p\left(\|\bv\|_1^2\frac{K^2s^2_{\Omega}\log p}{n+N}K^2\sqrt{\frac{\log p}{n+N}}\right),
	$$
	where in the last step holds we plugin the rate of $\wh\bOmega$ in (\ref{eq_rate_nodewise}) and apply Lemma \ref{repeated_lemma} to upper bound $\|\bZ_n\|_{\max}$ (as $X_{i}$ is uniformly bounded by $K_1\leq K$ and $\E(\eta_i^2)$ is bounded as well). Using a similar argument, one can derive
	$$
	| \bv\trans(\wh\bOmega-\bOmega)\bZ_n\bOmega\bv|\leq \|\bv\|_1\|\wh\bOmega-\bOmega\|_\infty\|\bZ_n\bOmega\bv\|_\infty
	=O_p\left(\|\bv\|_1^2Ks_{\Omega}\sqrt{\frac{\log p}{n+N}}K^2\sqrt{\frac{\log p}{n+N}}\right),
	$$
	and
	$$
	\bv\trans\bOmega\bZ_n\bOmega\bv=O_p\left(\|\bv\|_1^2K^2\sqrt{\frac{\log p}{n+N}}\right).
	$$
	Under the additional assumption that $K s_{\Omega}\{\log p/(n+N)\}^{1/2}=o(1)$, we can simplify the rate of $T_2$ as
	$$
	|T_2|=O_p\left(\|\bv\|_1^2K^2\sqrt{\frac{\log p}{n+N}}\right).
	$$
	Finally, let us consider $T_3$ and $T_4$. For $T_3$, we have
	$$
	T_3=\bv\trans(\wh\bOmega-\bOmega) \E(\X_i\X_i\trans \eta_i^2\bOmega\bv)+\bv\trans(\wh\bOmega-\bOmega) \E(\X_i\X_i\trans \eta_i^2)(\wh\bOmega-\bOmega)\bv,
	$$
	where the first term is identical to $T_4$ and therefore it suffices to only consider the rate of $T_3$. Since $\|\E(\X_i\X_i\trans \eta_i^2\bOmega\bv)\|_\infty\leq \|\bv\|_1K^2\E(\eta_i^2)\lesssim \|\bv\|_1K^2$ and $\|\E(\X_i\X_i\trans \eta_i^2)\|_{\max}\lesssim K^2$, we have
	$$
	|T_3|\lesssim \|\bv\|_1Ks_{\Omega}\sqrt{\frac{\log p}{n+N}} \|\bv\|_1K^2+\|\bv\|_1^2\frac{K^2s^2_{\Omega}\log p}{n+N}K^2
	=O_p\left(\|\bv\|_1^2K^3s_{\Omega}\sqrt{\frac{\log p}{n+N}}\right).
	$$
	Collecting the upper bounds for $T_1,\dots, T_4$, from (\ref{eq_pf_prop_variance2}) we obtain the rate in (\ref{eq_lem_gamma}).
\end{proof}

\begin{proof}[Proof of Proposition \ref{prop_variance1}]
	
	Note that $\bv\trans\bOmega\bv\leq \|\bv\|_2^2\lambda_{\max}(\bOmega)\lesssim \|\bv\|_2^2$ since $\lambda_{\max}(\bOmega)=1/\lambda_{\min}(\bSigma)\leq 1/C$. From Lemma \ref{lem_sig}, we can show that
	\begin{align*}
		&\bv\trans\wh\sigma^2\wh\bOmega\bv-\bv\trans\sigma^2\bOmega\bv\\
		&=(\wh\sigma^2-\sigma^2)\bv\trans\bOmega\bv+(\wh\sigma^2-\sigma^2)\bv\trans(\wh\bOmega-\bOmega)\bv+\sigma^2\bv\trans(\wh\bOmega-\bOmega)\bv\\
		&\lesssim \|\bv\|_2^2(n^{-1/2}+b_n^2)+ (n^{-1/2}+b_n^2)\|\bv\|_1\|\bv\|_\infty Ks_{\Omega}\sqrt{\frac{\log p}{n+N}}+\|\bv\|_1\|\bv\|_\infty Ks_{\Omega}\sqrt{\frac{\log p}{n+N}}\\
		&=O_p\left\{ \|\bv\|_2^2(n^{-1/2}+b_n^2)+\|\bv\|_1\|\bv\|_\infty Ks_{\Omega}\sqrt{\frac{\log p}{n+N}}\right\}.
	\end{align*}
	This implies
	\begin{align*}
		&\left|\bv\trans\left(\wh\sigma^2\wh\bOmega+\frac{n}{n+N}\wh\bGamma\right)\bv-\bv\trans\left(\sigma^2\bOmega+\frac{n}{n+N}\bGamma\right)\bv\right|\\
		&=O_p\left\{ \|\bv\|_2^2(n^{-1/2}+b_n^2)+\|\bv\|_1\|\bv\|_\infty Ks_{\Omega}\sqrt{\frac{\log p}{n+N}}\right.\\
		&\left.~~~~+K^2\|\bv\|_1^2\left(\frac{nb_n}{n+N}+K\sqrt{\frac{n}{n+N}}\sqrt{\frac{s\log p}{n+N}}+\frac{K^2s\log p}{n+N}+Ks_{\Omega}\sqrt{\frac{\log p}{n+N}}\frac{n}{n+N}\right)\right\}.\\
	\end{align*}
	By applying the condition $K \{s\log p/(n+N)\}^{1/2}=o(1)$ and $n/(n+N)\leq 1$, we can further simplify the above rate and derive  (\ref{eq_prop_variance1}). Note that $\bv\trans(\sigma^2\bOmega+\frac{n}{n+N}\bGamma)\bv\geq C\|\bv\|_2^2$ which together with (\ref{eq_prop_variance1}) leads to
	$$
	\left|\bv\trans\left(\wh\sigma^2\wh\bOmega+\frac{n}{n+N}\wh\bGamma\right)\bv/\bv\trans\left(\sigma^2\bOmega+\frac{n}{n+N}\bGamma\right)\bv-1\right|=O_p(n^{-1/2}+b_n^2+Rem_N/\|\bv\|_2^2).
	$$
	Finally, (\ref{eq_prop_variance3}) holds by Theorem \ref{inf1} and the Slutsky Theorem. This completes the proof.
\end{proof}

\subsection{Proof of Theorem \ref{thm_inf2}}

We first state several propositions and lemmas which are used in the proof.

\begin{lemma}\label{bdant}
	Assume that  Assumption \ref{assumption_est} holds. Consider the Dantzig selector $\wh \bt_D$ in (\ref{dantini}) with $\lambda_D\asymp K_1\sqrt{\frac{(\sigma^2+\Phi^2)\log p}{n}}$. We have
	\begin{equation}\label{eq_bdant1}
		\norm{\wh \bt_D-\bt^*}_1=O_p(s\lambda_D),~~\textrm{and}~~\frac{1}{n}\sum_{i=1}^n\{\X_i\trans(\wh \bt_D-\bt^*)\}^2=O_p(s\lambda_D^2).
	\end{equation}
	Moreover, we have
	$$
	\|\wh\bSigma_n(\wh \bt_D-\bt^*)\|_{\infty}=O_p(\lambda_D),
	$$
	where $\wh\bSigma_n=\frac{1}{n}\sum_{i=1}^n\X_i^{\otimes 2}$.
\end{lemma}

\begin{proof}
	The proof of the convergence rate of $\wh\bt_D$ in (\ref{eq_bdant1}) is similar to Theorem 7.1 in \cite{bickel2009}. The key step is to derive
	$$
	\norm{\frac{1}{n}\sum_{i=1}^n \X_i(Y_i-\X_i\trans\bt^*)}_\infty\lesssim K_1(\sigma^2+\Phi^2)^{1/2}\sqrt{\frac{\log p}{n}},
	$$
	which is implied by Lemma \ref{repeated_lemma} together with $\E(Y_i-\X_i\trans\bt^*)^2=\sigma^2+\Phi^2$ and $\|\X_i\|_\infty\leq K_1$. The rest of the proof is omitted. To show the rate of $\|\wh\bSigma_n(\wh \bt_D-\bt^*)\|_{\infty}$, we note that,  with $\lambda_D=C K_1\sqrt{\frac{(\sigma^2+\Phi^2)\log p}{n}}$ for some sufficiently large $C$, we have
	\begin{align*}
		\|\wh\bSigma_n(\wh \bt_D-\bt^*)\|_{\infty}&\leq \left\|\frac{1}{n}\sum_{i=1}^n \X_i(Y_i-\X_i\trans\wh\bt_D)\right\|_{\infty}+\left\|\frac{1}{n}\sum_{i=1}^n \X_i(Y_i-\X_i\trans\bt^*)\right\|_{\infty}\leq 2\bar\lambda,
	\end{align*}
	where we invoke the KKT condition of $\wh\bt_D$ in the last step.
\end{proof}

\begin{pro} \label{newB}
	Under the same conditions in Theorem \ref{thm_inf2}, for any $\wt\lambda_k=\wt\lambda\geq CK_1(c_n+\sqrt{\log p/n})$, we obtain that
	$$
	\norm{\wh \bB_{\cdot k}- \bB_{\cdot k}}_1\lesssim s_B\wt\lambda+\wt\lambda^{-1}\left\{K_1L_B\left(c_n+\sqrt{\frac{\log p}{n}}\right)+K_1^2\sqrt{\frac{s(\sigma^2+\Phi^2)\log p}{n}}\right\}^2.
	$$
	Define
	$$
	\wt\lambda_{opt}=\argmin_{\wt\lambda\geq CK_1(c_n+\sqrt{\log p/n})}s_B\wt\lambda+\wt\lambda^{-1}\left\{K_1L_B\left(c_n+\sqrt{\frac{\log p}{n}}\right)+K_1^2\sqrt{\frac{s(\sigma^2+\Phi^2)\log p}{n}}\right\}^2.
	$$
	By choosing $\wt\lambda_k=\wt\lambda_{opt}$, we obtain
	$$
	\norm{\wh \bB_{\cdot k}- \bB_{\cdot k}}_1\lesssim K_1(s_B+s_B^{1/2}L_B)\left(\sqrt{\frac{\log p}{n}}+c_n\right)+K_1^2\sqrt{\frac{ss_B(\sigma^2+\Phi^2)\log p}{n}},
	$$
	which holds uniformly over $1\leq k \leq p$
\end{pro}

\begin{proof}
	The proof is deferred to the Supplement~\ref{app_newB}.
\end{proof}

Now we are ready to prove Theorem \ref{thm_inf2}. For notational simplicity,  we use $\wh \bt^d_S$ for $\wh \bt^d_{S,\psi}$, $\wh\bxi_S$ for $\wh\bxi_{S,\psi}$ and $\wh\bt$ for $\wh\bt_D$. We can rewrite
\begin{align}
	\bv\trans(\wh \bt^d_S-\bt^*)&=\bv\trans\{\wh\bt-\bt^*-\wh\bOmega\wh\bSigma_n(\wh \bt-\bt^*)+\wh \bOmega(\wh\bxi_S-\wh\bSigma_n \bt^*)\}\nonumber\\
	&=\bv\trans\Big\{(\bI_p-\wh\bOmega\wh\bSigma_n)(\wh\bt-\bt^*)+\J\Big\} \nonumber\\
	&=\bv\trans\Big\{(\bI_p-\bOmega\wh\bSigma_n)(\wh\bt-\bt^*)+(\bOmega-\wh \bOmega)\wh \bSigma_n(\wh\bt-\bt^*)+\J\Big\},\label{eq_pf_thm_inf2_1}
\end{align}
where
$$
\J=\wh \bOmega\left[\frac{\bX\trans (\bY-\bX\bt^*)}{n}-\frac{\psi}{2}\wh \bB\trans\sum_{j=1}^2\left\{\frac{\sum_{i\in D^*_{j}}\X_i \wh m^{-j}(\X_i)}{n_j}-\frac{\sum_{i\in D_j} \X_i \wh m^{-j}(\X_i)}{n_j+N_j}\right\}\right].
$$
We notice that $\|\bOmega \X_i\|_\infty\leq \|\bOmega\|_\infty\|\X_i\|_\infty\leq L_{\Omega}K_1$, applying Hoeffding inequality,\\ $\norm{\bI_p-\bOmega\wh\bSigma_n}_{\max}=O_p\left(K_1^2L_{\Omega}\sqrt{\frac{\log p}{n}}\right)$.
Hence,
\begin{equation}\label{eq_pf_thm_inf2_1_1}
	\norm{(\bI_p-\bOmega\wh\bSigma_n)(\wh\bt-\bt^*)}_\infty\leq \norm{\bI_p-\bOmega\wh\bSigma_n}_{\max}\norm{\wh\bt-\bt^*}_1\lesssim K_1^3L_{\Omega}(\sigma^2+\Phi^2)^{1/2}\frac{s\log p}{n},
\end{equation}
where we use the convergence rate of $\wh\bt$ in Lemma \ref{bdant}.  Moreover, we know  \begin{equation}\label{eq_pf_thm_inf2_1_2}
	\norm{(\bOmega-\wh \bOmega)\wh \bSigma_n(\wh\bt-\bt^*)}_\infty\leq \norm{\bOmega-\wh \bOmega}_\infty\norm{\wh\bSigma_n(\wh\bt-\bt^*)}_\infty\lesssim K_1^2L_{\Omega}s_\Omega\frac{(\sigma^2+\Phi^2)^{1/2}\log p}{\sqrt{(n+N)n}},
\end{equation}
followed by (\ref{eq_rate_nodewise}) (we replace $K$ with $K_1L_{\Omega}$) and again Lemma \ref{bdant}. Now, we focus on the term $\J$. We rewrite $\J$ as
\begin{align*}
	\J&=\bOmega\left[\frac{\bX\trans (\bY-\bX\bt^*)}{n}-\psi\bB\trans\left\{\frac{\sum_{i=1}^{n} \X_im(\X_i)}{n} -\frac{\sum_{i=1}^{n+N} \X_i m(\X_i)}{n+N}\right\}\right]\\
	&+\customunderbrace{(\wh\bOmega-\bOmega)\left[\frac{\bX\trans (\bY-\bX\bt^*)}{n}-\psi\bB\trans\left\{\frac{\sum_{i=1}^{n} \X_i  m(\X_i)}{n}-\frac{\sum_{i=1}^{n+N} \X_i m(\X_i)}{n+N}\right\}\right]}{$\J_1$}\\
	&+\customunderbrace{(\wh\bOmega-\bOmega)(\wh\bxi_S-\bxi_0)}{$\J_2$}+\customunderbrace{\bOmega(\wh \bxi_S-\bxi_0)}{$\J_3$},
\end{align*}
where
$$
\bxi_0=\frac{\bX\trans \bY}{n}- \psi\bB\trans\left\{\frac{\sum_{i=1}^{n} \X_i m(\X_i)}{n}-\frac{\sum_{i=1}^{n+N} \X_i m(\X_i)}{n+N}\right\}.
$$
In the following, we will show that the three terms $\|\J_k\|_\infty$ for $k=1,2,3$ are sufficiently small. We first recall that
$$
\norm{\frac{1}{n}\sum_{i=1}^n \X_im(\X_i)-\bmu}_\infty\lesssim K_1\sqrt{\frac{\log p}{n}},~~~\norm{\frac{1}{n}\sum_{i=1}^n \X_i(Y_i-\X_i\trans\bt^*)}_\infty\lesssim K_1(\sigma^2+\Phi^2)^{1/2}\sqrt{\frac{\log p}{n}},
$$
by Hoeffding inequality and Lemma \ref{repeated_lemma}. For $\J_1$, it holds that
\begin{align}\label{eq_pf_thm_inf2_2}
	\|\J_1\|_\infty  &\leq \norm{\wh\bOmega-\bOmega}_\infty \norm{\frac{\bX\trans (\bY-\bX\bt^*)}{n}-\psi\bB\trans\left\{\frac{\sum_{i=1}^{n} \X_i  m(\X_i)}{n}-\frac{\sum_{i=1}^{n+N} \X_i m(\X_i)}{n+N}\right\}}_\infty\nonumber\\
	&\leq\norm{\wh\bOmega-\bOmega}_\infty  \left\{\norm{\frac{1}{n}\sum_{i=1}^n \X_i(Y_i-\X_i\trans\bt^*)}_\infty+|\psi| L_B\norm{\frac{1}{n}\sum_{i=1}^n \X_im(\X_i)-\bmu}_\infty\right.\nonumber\\
	&\left.~~~~~~~+|\psi| L_B\norm{\frac{1}{n+N}\sum_{i=1}^{n+N} \X_im(\X_i)-\bmu}_\infty\right\}\nonumber\\
	&\lesssim K_1^2L_{\Omega}\left\{(\sigma^2+\Phi^2)^{1/2}+L_B\right\}s_\Omega\frac{\log p}{\sqrt{(n+N)n}}.
\end{align}
To upper bound the supnorm of $\J_2$ and $\J_3$, we need the following bounds. First, by Proposition \ref{newB}, we have
\begin{align*}
	&\norm{(\wh \bB- \bB)\trans\left\{\frac{\sum_{i=1}^{n} \X_i  m(\X_i)}{n}-\frac{\sum_{i=1}^{n+N} \X_i m(\X_i)}{n+N}\right\}}_\infty\\ \nonumber
	&\leq \max_k\norm{\wh\bB_k- \bB_k}_1 \left\{\norm{\frac{1}{n}\sum_{i=1}^n \X_im(\X_i)-\bmu}_\infty+\norm{\frac{1}{n+N}\sum_{i=1}^{n+N} \X_im(\X_i)-\bmu}_\infty\right\}\\ \nonumber
	&\lesssim K_1^2(s_B+s_B^{1/2}L_B)\left(\frac{\log p}{n}+c_n\sqrt{\frac{\log p}{n}}\right)+K_1^3{\frac{\sqrt{ss_B(\sigma^2+\Phi^2)}\log p}{n}}.
\end{align*}
Furthermore,  using a similar argument in the proof of Theorem 3.2 of \cite{deng2023optimal}, we have
\begin{align*}
	&\norm{ \bB\trans\frac{1}{2}\sum_{j=1}^2\left[\frac{\sum_{i\in D^*_{j}} \X_i \left\{\wh m^{-j}(\X_i)-m(\X_i)\right\}}{n_j}-\frac{\sum_{i\in D_j}\X_i \left\{\wh m^{-j}(\X_i)-m(\X_i)\right\}}{n_j+N_j}\right]}_\infty\\
	&\leq L_B\max_{j=1,2}\norm{\frac{\sum_{i\in D^*_{j}} \X_i \left\{\wh m^{-j}(\X_i)-m(\X_i)\right\}}{n_j}-\frac{\sum_{i\in D_j}\X_i \left\{\wh m^{-j}(\X_i)-m(\X_i)\right\}}{n_j+N_j}}_\infty\\
	&\lesssim K_1L_Bc_n\sqrt{\frac{\log p}{n}}.
\end{align*}
Under the condition $s_BK_1^2\left(c_n+\sqrt{\frac{\log p}{n}}\right)=o(1)$, one can easily verify that \\$\max_k\norm{\wh\bB_k- \bB_k}_1\lesssim L_B+K_1^2\sqrt{\frac{ss_B(\sigma^2+\Phi^2)\log p}{n}}$, and therefore,
\begin{align*}
	&\norm{(\wh \bB- \bB)\trans\frac{1}{2}\sum_{j=1}^2\left[\frac{\sum_{i\in D^*_{j}} \X_i \left\{\wh m^{-j}(\X_i)-m(\X_i)\right\}}{n_j}-\frac{\sum_{i\in D_j}\X_i \left\{\wh m^{-j}(\X_i)-m(\X_i)\right\}}{n_j+N_j}\right]}_\infty\\
	&\lesssim K_1L_Bc_n\sqrt{\frac{\log p}{n}}+c_nK_1^3{\frac{\sqrt{ss_B(\sigma^2+\Phi^2)}\log p}{n}},
\end{align*}
which is of smaller order than the previous two terms.  Given these bounds, we can  decompose and bound $\norm{\bxi_0- \wh \bxi_S}_\infty$ as
{\small{
		\begin{align*}
			&\norm{\bxi_0- \wh \bxi_S}_\infty\\
			=&|\psi| \norm{\frac{1}{2}\wh \bB\trans\sum_{j=1}^2\left\{\frac{\sum_{i\in D^*_{j}}\X_i \wh m^{-j}(X_i)}{n_j}-\frac{\sum_{i\in D_j} \X_i \wh m^{-j}(\X_i)}{n_j+N_j}\right\}
				-\bB\trans\left\{\frac{\sum_{i=1}^{n} \X_i m(\X_i)}{n}-\frac{\sum_{i=1}^{n+N} \X_i m(\X_i)}{n+N}\right\}}_\infty\\ \nonumber
			&\leq\norm{(\wh \bB- \bB)\trans\left\{\frac{\sum_{i=1}^{n} \X_i m(\X_i)}{n}-\frac{\sum_{k=1}^{n+N} \X_i m(\X_i)}{n+N}\right\}}_\infty\\ \nonumber
			&~~~+\norm{(\wh \bB- \bB)\trans\frac{1}{2}\sum_{j=1}^2\left[\frac{\sum_{i\in D^*_{j}}\X_i \left\{\wh m^{-j}(\X_i)-m(\X_i)\right\}}{n_j}-\frac{\sum_{i\in D_j} \X_i \left\{\wh m^{-j}(\X_i)-m(\X_i)\right\}}{n_j+N_j}\right]}_\infty\\ \nonumber
			&~~~+\norm{\bB\trans\frac{1}{2}\sum_{j=1}^2\left[\frac{\sum_{i\in D^*_{j}} \X_i \left\{\wh m^{-j}(\X_i)-m(\X_i)\right\}}{n_j}-\frac{\sum_{i\in D_j}\X_i \left\{\wh m^{-j}(\X_i)-m(\X_i)\right\}}{n_j+N_j}\right]}_\infty\\ \nonumber
			&\lesssim K_1^2(s_B+s_B^{1/2}L_B)\left(\frac{\log p}{n}+c_n\sqrt{\frac{\log p}{n}}\right)+K_1^3{\frac{\sqrt{ss_B(\sigma^2+\Phi^2)}\log p}{n}}.
		\end{align*}
}}

Thus, for $\J_2$ and $\J_3$ we have
\begin{align}\label{eq_pf_thm_inf2_3}
	\|J_3\|_\infty\lesssim L_{\Omega} \left\{K_1^2(s_B+s_B^{1/2}L_B)\left(\frac{\log p}{n}+c_n\sqrt{\frac{\log p}{n}}\right)+K_1^3{\frac{\sqrt{ss_B(\sigma^2+\Phi^2)}\log p}{n}}\right\}.
\end{align}
Since we have $K_1L_{\Omega}s_{\Omega}\sqrt{\frac{\log p}{n+N}}=o(1)$, $\|\J_2\|_\infty$ is of smaller order than that of $\|\J_3\|_\infty$. Thus, from (\ref{eq_pf_thm_inf2_2}) and (\ref{eq_pf_thm_inf2_3}),  we have
$$
\J= \bOmega\left[\frac{\bX\trans (\bY-\bX\bt^*)}{n}-\psi\bB\trans\left\{\frac{\sum_{i=1}^{n} \X_im(\X_i)}{n} -\frac{\sum_{i=1}^{n+N} \X_i m(\X_i)}{n+N}\right\}\right]+Rem,
$$
where
\begin{align*}
	\|Rem\|_\infty&\lesssim L_{\Omega} K_1^2\left[\{(\sigma^2+\Phi^2)^{1/2}+L_B\}s_\Omega\frac{\log p}{\sqrt{(n+N)n}}\right.\\
	&\left.~~~~~+(s_B+s_B^{1/2}L_B)\left(\frac{\log p}{n}+c_n\sqrt{\frac{\log p}{n}}\right)+K_1{\frac{\sqrt{ss_B(\sigma^2+\Phi^2)}\log p}{n}}\right].
\end{align*}
Finally, combining with (\ref{eq_pf_thm_inf2_1_1}) and (\ref{eq_pf_thm_inf2_1_2}), we obtain from (\ref{eq_pf_thm_inf2_1}) that
\begin{align}\label{eq_pf_thm_inf2_4}
	\bv\trans (\bt^d_S-\bt^*)&=\bv\trans\bOmega\left[\frac{\bX\trans (\bY-\bX\bt^*)}{n}-\psi \bB\trans\left\{\frac{\sum_{i=1}^{n} \X_i m(\X_i)}{n} -\frac{\sum_{i=1}^{n+N} \X_i m(\X_i)}{n+N}\right\}\right]+O_p(\bar\delta_n),
\end{align}
where
\begin{align*}
	\bar\delta_n&=\|\bv\|_1 L_{\Omega} K_1^2\left[\left\{(\sigma^2+\Phi^2)^{1/2}+L_B\right\}s_\Omega\frac{\log p}{\sqrt{(n+N)n}}\right.\\
	&\left.~~~~~~~~~~~+(s_B+s_B^{1/2}L_B)\left(\frac{\log p}{n}+c_n\sqrt{\frac{\log p}{n}}\right)+K_1(s\vee s_B){\frac{\sqrt{(\sigma^2+\Phi^2)}\log p}{n}}\right].
\end{align*}
To show the asymptotic normality of $\bv\trans (\bt^d_S-\bt^*)$, we denote
$$
\xi_i=\begin{cases}
	\frac{1}{n}\bv\trans\bOmega(\T_{i1}-\frac{N\psi}{n+N}\bB\trans\T_{i2}) & \mbox{for}~ 1\leq i\leq n, \\
	\frac{\psi}{n+N}\bv\trans\bOmega\bB\trans\T_{i2}& \mbox{for}~ n+1\leq i\leq n+N,
\end{cases}
$$
where $\T_{i1}=\X_i(Y_i-\X_i\trans\bt^*)$ and $\T_{i2}=\X_im(\X_i)-\bmu$.
One can rewrite (\ref{eq_pf_thm_inf2_4}) as
$$
\bv\trans (\bt^d_S-\bt^*)=\sum_{i=1}^{n+N}\xi_i+O_p(\bar\delta_n).
$$
To apply the Lindeberg-Feller Central Limit Theorem, we first note that $\E(\xi_i)=0$. Furthermore,
\begin{align*}
	\sum_{i=1}^{n+N}\E(\xi_i^2)&=\frac{1}{n}\bv\trans\bOmega\left\{\E(\T_{i1}^{\otimes 2})-\frac{2N\psi}{n+N}\bB\trans\E(\T_{i2}\T_{i1}\trans)+\frac{N^2\psi^2}{(n+N)^2}\bB\trans\E(\T_{i2}^{\otimes 2})\bB\right\}\bOmega\bv\\
	&~~~~~~~~~~+\frac{\psi^2N}{(n+N)^2}\bv\trans\bOmega\bB\trans\E(\T_{i2}^{\otimes 2})\bB\bOmega\bv\\
	&=\frac{1}{n}\bv\trans\bOmega\left\{\E(\T_{i1}^{\otimes 2})-\frac{2N\psi}{n+N}\bB\trans\E(\T_{i2}\T_{i1}\trans)+\frac{N\psi^2}{n+N}\bB\trans\E(\T_{i2}^{\otimes 2})\bB\right\}\bOmega\bv.
\end{align*}
Recall that $\bB=\{\E(\T_{i2}^{\otimes 2})\}^{-1}\E(\T_{i2}\T_{i1}\trans)$. Thus, we have
$$
\sum_{i=1}^{n+N}\E(\xi_i^2)=\frac{1}{n}\bv\trans\bOmega\left[\E(\T_{i1}^{\otimes 2})-\frac{N(2\psi-\psi^2)}{n+N}\E(\T_{i2}\T_{i1}\trans)\trans\{\E(\T_{i2}^{\otimes 2})\}^{-1}\E(\T_{i2}\T_{i1}\trans)\right]\bOmega\bv:=t_n^2.
$$
Note that $\E\|\T_{i1}\|_\infty^{2+\delta}\leq K_1^{2+\delta}\E|\epsilon_i+\eta_i|^{2+\delta}\lesssim K_1^{2+\delta}$ and $\E\|\T_{i2}\|_\infty^{2+\delta}\lesssim K_1^{2+\delta}$. In addition, our assumption implies $t_n^{2+\delta}\geq C\|\bv\|_2^{2+\delta}/n^{1+\delta/2}$. Finally, we can verify that the Lyapunov condition holds
\begin{align*}
	\frac{\sum_{i=1}^{n+N}\E|\xi_i|^{2+\delta}}{t_n^{2+\delta}}&\leq \frac{\|\bv\|_1^{2+\delta} L_{\Omega}^{2+\delta}}{t_n^{2+\delta}}\left\{\frac{1}{n^{1+\delta}}\E\left\|\T_{i1}-\frac{N\psi}{n+N}\bB\trans\T_{i2}\right\|^{2+\delta}_\infty
	+\frac{\psi^{2+\delta}N}{(n+N)^{2+\delta}}\E\left\|\bB\trans\T_{i2}\right\|_\infty^{2+\delta}\right\}\\
	&\lesssim \left(\frac{\|\bv\|_1L_\Omega K_1}{\|\bv\|_2}\right)^{2+\delta} \frac{1}{n^{\delta/2}}\left(1+\frac{L_B^{2+\delta}N}{n+N}\right)\rightarrow 0.
\end{align*}
Therefore, we obtain the desired result by applying the Lindeberg-Feller Central Limit Theorem and Slutsky Theorem. This completes the proof.

\subsubsection{Proof of Proposition \ref{prop_variance2}}

\begin{proof}
	We first establish the rate of $\wh\bM_1$ and $\wh\bM_2$ in the elementwise supnorm. Define $\bM_1=\E\{(Y_i-\X_i\trans\bt^*)^2\X_i^{\otimes 2}\}$. We note that
	\begin{align*}
		\|\wh\bM_1-\bM_1\|_{\max}&\leq \Big\|\frac{1}{n}\sum_{i=1}^n(Y_i-\X_i\trans\bt^*)^2\X_i^{\otimes 2}-\bM_1\Big\|_{\max}+\Big\|\frac{2}{n}\sum_{i=1}^n(Y_i-\X_i\trans\bt^*)\{\X_i\trans(\wh\bt_D-\bt^*)\}\X_i^{\otimes 2}\Big\|_{\max}\\
		&~~~~~+\Big\|\frac{1}{n}\sum_{i=1}^n\{\X_i\trans(\wh\bt_D-\bt^*)\}^2\X_i^{\otimes 2}\Big\|_{\max}\lesssim K_1^3\sqrt{\frac{s\log p}{n}},
	\end{align*}
	where we use the moment inequality in Lemma \ref{repeated_lemma} and Lemma \ref{bdant}. Similarly, define $\bM_2=\E(\T_{i2}\T_{i1}\trans)$. Since $\wh\bM_2$ is constructed by sample splitting, it suffices to derive the rate of $\wh\bM_2^j-\bM_2$. Following the same type of argument, we can show that
	\begin{align*}
		&\|\wh\bM_2^j-\bM_2\|_{\max}\\
		\leq& \Big\|\frac{1}{n_j}\sum_{i\in D_j^*}(Y_i-\X_i\trans\wh\bt_D)\{\wh m^{-j}(\X_i)-m(\X_i)\}\X_i^{\otimes 2}\Big\|_{\max}+\Big\|\frac{1}{n_j}\sum_{i\in D_j^*}\X_i\trans(\wh\bt_D-\bt^*)m(\X_i)\X_i^{\otimes 2}\Big\|_{\max}\\
		&~~~~~~+\Big\|\frac{1}{n_j}\sum_{i\in D_j^*}(Y_i-\X_i\trans\bt^*)m(\X_i)\X_i^{\otimes 2}-\bM_2\Big\|_{\max}\\
		\lesssim& K_1^2 c_n+K_1^3\sqrt{\frac{s\log p}{n}}.
	\end{align*}
	Together with Proposition \ref{newB}, we have
	\begin{align*}
		&\|\wh\bB\trans\wh\bM_2-\bB\trans\bM_2\|_{\max}\\
		&\leq \|\wh\bB-\bB\|_\infty \|\wh\bM_2-\bM_2\|_{\max}+\|\wh\bB-\bB\|_\infty \|\bM_2\|_{\max}+\|\bB\|_\infty \|\wh\bM_2-\bM_2\|_{\max}\\
		&\lesssim K_1^2 Rem,
	\end{align*}
	where
	$$
	Rem=K_1(s_B+s_B^{1/2}L_B)\left(\sqrt{\frac{\log p}{n}}+c_n\right)+K_1^2\sqrt{\frac{ss_B\log p}{n}}+K_1L_B\sqrt{\frac{s\log p}{n}}
	$$
	As a result,
	$$
	\|\wh\bGamma_\psi-\bGamma_\psi\|_{\max}\lesssim K_1^3\sqrt{\frac{s\log p}{n}}+\frac{NK_1^2}{n+N}Rem.
	$$
	One can easily show that $\|\bGamma_\psi\|_{\max}\lesssim K_1^2+\frac{N}{n+N}L_BK_1^2$, which implies
	$$
	\|\wh\bGamma_\psi\|_{\max}\leq \|\bGamma_\psi\|_{\max}+ \|\wh\bGamma_\psi-\bGamma_\psi\|_{\max}\lesssim K_1^2+\frac{N}{n+N}L_BK_1^2
	$$
	under the assumption that $Rem=o(1)$ and $K_1\sqrt{\frac{s\log p}{n}}=o(1)$. Similarly, $\|\wh\bOmega\|_\infty\lesssim L_{\Omega}$ since $\|\wh\bOmega-\bOmega\|_\infty=o(1)$. Finally, we can establish the rate of convergence of the estimated variance
	\begin{align*}
		&|\bv\trans\wh\bOmega\wh\bGamma_\psi\wh\bOmega\bv-\bv\trans\bOmega\bGamma_\psi\bOmega\bv|\\
		&\leq |\bv\trans(\wh\bOmega-\bOmega)\wh\bGamma_\psi\wh\bOmega\bv|+|\bv\trans\bOmega\wh\bGamma_\psi(\wh\bOmega-\bOmega)\bv|+|\bv\trans\bOmega(\wh\bGamma_\psi-\bGamma_\psi)\bOmega\bv|\\
		&\lesssim \|\bv\|_1^2(L_{\Omega}\|\wh\bOmega-\bOmega\|_\infty\|\wh\bGamma_\psi\|_{\max}+L_{\Omega}^2 \|\wh\bGamma_\psi-\bGamma_\psi\|_{\max})\\
		&\lesssim \|\bv\|_1^2\left(K_1L_{\Omega}^2s_{\Omega}\sqrt{\frac{\log p}{n+N}} \|\bGamma_\psi\|_{\max}+K_1^3L_{\Omega}^2\sqrt{\frac{s\log p}{n}}+\frac{NK_1^2L_{\Omega}^2}{n+N}Rem\right),
	\end{align*}
	which proves (\ref{eq_prop_variance2_1}). The proof of (\ref{eq_prop_variance2_2}) is immediate by the Slutsky Theorem.
\end{proof}

\subsubsection{Proof of Proposition \ref{newB}}\label{app_newB}

\begin{proof}
	To show Proposition \ref{newB}, it suffices to show that the same rate of convergence holds for $\wh \bB^j_{\cdot k}$.  For simplicity of presentation, we use the notation $\wh\bt_D=\wh\bt$, $\wh \bB_{\cdot k}=\wh\bbeta$, $\bB_{\cdot k}=\bbeta$, $\wh\bDelta=\wh\bbeta-\bbeta$ and $\lambda=\wt\lambda_k$, $\wh \bZ_i=\X_i\wh m^{-j}(\X_i) -\wh \bmu^j$ and $ \bZ_i=\X_i m(\X_i) - \bmu$,
	$\wh \Fb=\frac{1}{n_j}\sum_{i\in D_j^*}\wh \bZ_i \wh \bZ_i\trans$ We start from the inequality
	\begin{align*}
		&\frac{1}{n_j}\sum_{i\in D_j^*} \left[X_{ik} (Y_i- \X_i\trans\wh \bt)-\wh\bbeta\trans \left\{\X_i\wh m^{-j}(\X_i) -\wh \bmu^j\right\}\right]^2+\lambda\norm{\wh\bbeta}_1\\
		&\leq \frac{1}{n_j}\sum_{i\in D_j^*} \left[X_{ik} (Y_i- \X_i\trans\wh \bt)-\bbeta\trans \left\{\X_i\wh m^{-j}(\X_i) -\wh \bmu^j\right\}\right]^2+\lambda\norm{\bbeta}_1.
	\end{align*}
	Following the standard argument in the analysis of Lasso (e.g., the proof of Theorem 7.1 in \cite{bickel2009}), the above inequality reduces to
	\begin{align}\label{eq_proof_newB_0}
		\wh\bDelta\trans\wh \Fb \wh\bDelta &\leq \lambda\norm{\bbeta}_1-\lambda\norm{\wh\bbeta}_1+\frac{2}{n_j}\sum_{i\in D_j^*} \left\{X_{ik} (Y_i- \X_i\trans\wh \bt)-\bbeta\trans\wh \bZ_i\right\}\wh \bZ_i\trans\wh\bDelta\\ \nonumber
		&=\lambda\norm{\bbeta}_1-\lambda\norm{\wh\bbeta}_1+I_1+I_2+I_3+I_4,
	\end{align}
	where
	\begin{align*}
		I_1&=\frac{2}{n_j}\sum_{i\in D_j^*} \left\{X_{ik} (Y_i- \X_i\trans \bt^*)-\bbeta\trans \bZ_i\right\} \bZ_i\trans\wh\bDelta,\\
		I_2&=\frac{2}{n_j}\sum_{i\in D_j^*} \left\{X_{ik} (Y_i- \X_i\trans \bt^*)-\bbeta\trans \bZ_i\right\} (\wh\bZ_i-\bZ_i)\trans\wh\bDelta,\\
		I_3&=-\frac{2}{n_j}\sum_{i\in D_j^*} X_{ik} \X_i\trans (\wh\bt-\bt^*) \wh\bZ_i\trans\wh\bDelta,~~\textrm{and}~~~I_4=-\frac{2}{n_j}\sum_{i\in D_j^*} \bbeta\trans(\wh\bZ_i-\bZ_i)\wh\bZ_i\trans\wh\bDelta.
	\end{align*}
	To bound $I_1$, we note that $\E\left[\{X_{ik} (Y_i- \X_i\trans \bt^*)-\bbeta\trans \bZ_i\} \bZ_i\right]=\0$ by the definition of $\bbeta$. To invoke Lemma \ref{repeated_lemma}, we control the second moment as
	\begin{align*}
		\E\Big[\max_{1\leq j\leq p}\{X_{ik} (Y_i- \X_i\trans \bt^*)-\bbeta\trans \bZ_i\}^2 Z_{ij}^2\Big]&\lesssim K_1^2\E\Big[\{X_{ik} (Y_i- \X_i\trans \bt^*)-\bbeta\trans \bZ_i\}^2 \Big]\lesssim K_1^2,
	\end{align*}
	where first step $|Z_{ij}|=|X_{ij} m(\X_i) - \mu_{ij}|\lesssim K_1$ holds as $|X_{ij}|\leq K_1$ and $|m(\X_i)|\leq C$ and second step relies on the assumption that the second moment of $X_{ik} (Y_i- \X_i\trans \bt^*)-\bbeta\trans \bZ_i$ is bounded. Therefore, applying Holder inequality and Lemma \ref{repeated_lemma} with $m=2$, we have
	\begin{align}\label{eq_proof_newB_1}
		|I_1|\lesssim \Big\|\frac{1}{n_j}\sum_{i\in D_j^*} \{X_{ik} (Y_i- \X_i\trans \bt^*)-\bbeta\trans \bZ_i\} \bZ_i\Big\|_\infty\|\wh\bDelta\|_1\lesssim K_1\sqrt{\frac{\log p}{n}}\|\wh\bDelta\|_1.
	\end{align}
	Similarly, we have $|I_2|\lesssim \|\frac{1}{n_j}\sum_{i\in D_j^*} \bE_{ik} (\wh\bZ_i-\bZ_i)\|_\infty\|\wh\bDelta\|_1$, where $\bE_{ik}=X_{ik} (Y_i- \X_i\trans \bt^*)-\bbeta\trans \bZ_i$. Furthermore, we notice that
	\begin{align*}
		\frac{1}{n_j}\sum_{i\in D_j^*} \bE_{ik} (\wh\bZ_i-\bZ_i)&=\frac{1}{n_j}\sum_{i\in D_j^*} \bE_{ik} \X_i\left\{\wh m^{-j}(\X_i)-m(\X_i)\right\}\\
		&~~~-\left(\frac{1}{n_j}\sum_{i\in D_j^*} \bE_{ik}\right) \frac{1}{n_j}\sum_{i\in D_j^*} \Big[\X_i\wh m^{-j}(\X_i)-\E\{\X_i m(\X_i)\}\Big].
	\end{align*}
	For the first term, we can apply  Cauchy–Schwarz inequality to show that
	\begin{align}\label{eq_proof_newB_2}
		&\max_{1\leq l\leq p}\Big|\frac{1}{n_j}\sum_{i\in D_j^*} \bE_{ik} X_{il}\{\wh m^{-j}(\X_i)-m(\X_i)\}\Big|\nonumber\\
		&\leq \max_{1\leq l\leq p}\Big|\frac{1}{n_j}\sum_{i\in D_j^*} \bE^2_{ik} X^2_{il}\Big|^{1/2}\Big|\frac{1}{n_j}\sum_{i\in D_j^*} \{\wh m^{-j}(\X_i)-m(\X_i)\}^2\Big|^{1/2}\lesssim K_1c_n,
	\end{align}
	where in the last step we invoke Lemma \ref{repeated_lemma} again to show
	$\max_{1\leq l\leq p}|\frac{1}{n_j}\sum_{i\in D_j^*} \bE^2_{ik} X^2_{il}-\E(\bE^2_{ik} X^2_{il})|\lesssim K_1^2\sqrt{\frac{\log p}{n}}$ and $\E(\bE^2_{ik} X^2_{il})\lesssim K_1^2$ together with triangle inequality imply the desired bound. For the second term, we first notice that $\frac{1}{n_j}\sum_{i\in D_j^*} \bE_{ik}=O_p(1)$, and then a similar argument leads to
	\begin{align}\label{eq_proof_newB_3}
		&\max_{1\leq l\leq p} \left|\frac{1}{n_j}\sum_{i\in D_j^*} \left[X_{il}\wh m^{-j}(\X_i)-\E\left\{X_{il}m(\X_i)\right\}\right]\right|\nonumber\\
		&\leq\max_{1\leq l\leq p} \Big|\frac{1}{n_j}\sum_{i\in D_j^*} X_{il}\left\{\wh m^{-j}(\X_i) -m(\X_i)\right\}\Big|+\max_{1\leq l\leq p} \Big|\frac{1}{n_j}\sum_{i\in D_j^*} [X_{il}m(\X_i)-\E\{X_{il}m(\X_i)\}]\Big|\nonumber\\
		&\lesssim K_1c_n+K_1\sqrt{\frac{\log p}{n}},
	\end{align}
	where we apply the Hoeffding inequality as $|X_{il}m(\X_i)|\leq CK_1$. Combining (\ref{eq_proof_newB_2}) and (\ref{eq_proof_newB_3}), we have
	\begin{align}\label{eq_proof_newB_4}
		|I_2|\lesssim \Big\|\frac{1}{n_j}\sum_{i\in D_j^*} \bE_{ik} (\wh\bZ_i-\bZ_i)\Big\|_\infty\|\wh\bDelta\|_1\lesssim \left(K_1c_n+K_1\sqrt{\frac{\log p}{n}}\right)\|\wh\bDelta\|_1.
	\end{align}
	We now consider $I_3$. Recall that Lemma \ref{bdant} implies $\frac{1}{n}\sum_{i=1}^n\left\{\X_i\trans(\wh \bt-\bt^*)\right\}^2=O_p\left\{K_1^2s\frac{(\sigma^2+\Phi^2)\log p}{n}\right\}$. Thus,
	\begin{align}\label{eq_proof_newB_5}
		|I_3|&\lesssim \Big|\frac{1}{n_j}\sum_{i\in D_j^*} \{\X_i\trans(\wh \bt-\bt^*)\}^2 \X^2_{ik}\Big|^{1/2}
		(\wh\bDelta\trans\wh \Fb \wh\bDelta)^{1/2}\lesssim K_1^2\sqrt{\frac{s(\sigma^2+\Phi^2)\log p}{n}}(\wh\bDelta\trans\wh \Fb \wh\bDelta)^{1/2}.
	\end{align}
	Similarly, the Cauchy–Schwarz inequality yields $|I_4|\lesssim \left[\frac{1}{n_j}\sum_{i\in D_j^*} \left\{\bbeta\trans(\wh\bZ_i-\bZ_i)\right\}^2\right]^{1/2}(\wh\bDelta\trans\wh \Fb \wh\bDelta)^{1/2}$. Moreover,
	\begin{align*}
		\frac{1}{n_j}\sum_{i\in D_j^*} \{\bbeta\trans (\wh\bZ_i-\bZ_i)\}^2&\leq\frac{2}{n_j}\sum_{i\in D_j^*} (\bbeta\trans \X_i)^2\{\wh m^{-j}(\X_i)-m(\X_i)\}^2\\
		&~~~+2\left(\frac{1}{n_j}\sum_{i\in D_j^*} \left[\bbeta\trans \X_i m(\X_i)-\E\{\bbeta\trans \X_i m(\X_i)\}\right]\right)^2\\
		&\lesssim K_1^2L_B^2c_n^2+\left(K_1L_Bc_n+K_1L_B\sqrt{\frac{\log p}{n}}\right)^2
	\end{align*}
	where the last step holds by using a similar argument as in (\ref{eq_proof_newB_2}) and (\ref{eq_proof_newB_3}) together with the fact that $|\bbeta\trans \X_i|\leq \|\bbeta\|_1\|\X_i\|_\infty\leq K_1L_B$. As a result, we have
	$$
	|I_4|\lesssim \left(K_1L_Bc_n+K_1L_B\sqrt{\frac{\log p}{n}}\right)(\wh\bDelta\trans\wh \Fb \wh\bDelta)^{1/2}.
	$$
	Collecting the bounds in (\ref{eq_proof_newB_1}), (\ref{eq_proof_newB_4}) and (\ref{eq_proof_newB_5}), we obtain from (\ref{eq_proof_newB_0}) that
	\begin{align*}
		\wh\bDelta\trans\wh \Fb \wh\bDelta
		&\leq \lambda\norm{\bbeta}_1-\lambda\norm{\wh\bbeta}_1+t_1\|\wh\bDelta\|_1+t_2(\wh\bDelta\trans\wh \Fb \wh\bDelta )^{1/2}\\
		&\leq \lambda\norm{\wh\bDelta_S}_1-\lambda\norm{\wh\bDelta_{S^c}}_1+t_1\|\wh\bDelta\|_1+t_2(\wh\bDelta\trans\wh \Fb \wh\bDelta )^{1/2},
	\end{align*}
	where  $t_1=C\left(K_1c_n+K_1\sqrt{\frac{\log p}{n}}\right)$ and $t_2=C\left\{K_1L_B\left(c_n+\sqrt{\frac{\log p}{n}}\right)+K_1^2\sqrt{\frac{s(\sigma^2+\Phi^2)\log p}{n}}\right\}$ for some sufficiently large constant $C$. By taking $\lambda\geq 2t_1$, we have
	\begin{align}\label{eq_proof_newB_6}
		\wh\bDelta\trans\wh \Fb \wh\bDelta \leq \frac{3}{2}\lambda\norm{\wh\bDelta_S}_1-\frac{1}{2}\lambda\norm{\wh\bDelta_{S^c}}_1+t_2(\wh\bDelta\trans\wh \Fb \wh\bDelta )^{1/2}.
	\end{align}
	In the following, we consider two cases. In case (1): $t_2(\wh\bDelta\trans\wh \Fb \wh\bDelta )^{1/2}\leq \lambda\norm{\wh\bDelta_S}_1$,  (\ref{eq_proof_newB_6}) further implies
	$$
	\wh\bDelta\trans\wh \Fb \wh\bDelta \leq \frac{5}{2}\lambda\norm{\wh\bDelta_S}_1-\frac{1}{2}\lambda\norm{\wh\bDelta_{S^c}}_1,
	$$
	which leads to the standard cone condition $\norm{\wh\bDelta_{S^c}}_1\leq 5\norm{\wh\bDelta_S}_1$. With lemma \ref{RE1}, we can show that $\|\wh\bDelta\|_2^2\lesssim \lambda\norm{\wh\bDelta_S}_1\leq \lambda s_B^{1/2}\norm{\wh\bDelta_S}_2$ and therefore $\|\wh\bDelta\|_2\lesssim \lambda s_B^{1/2}$. Similarly, we can derive $\|\wh\bDelta\|_1\lesssim \lambda s_B$. In case (2): $t_2(\wh\bDelta\trans\wh \Fb \wh\bDelta )^{1/2}> \lambda\norm{\wh\bDelta_S}_1$,  (\ref{eq_proof_newB_6})  implies
	\begin{align*}
		\wh\bDelta\trans\wh \Fb \wh\bDelta &\leq \frac{3}{2}\lambda\norm{\wh\bDelta_S}_1+t_2(\wh\bDelta\trans\wh \Fb \wh\bDelta )^{1/2}\leq \frac{5}{2}t_2(\wh\bDelta\trans\wh \Fb \wh\bDelta )^{1/2},
	\end{align*}
	and therefore $\wh\bDelta\trans\wh \Fb \wh\bDelta \leq \frac{25}{4}t_2^2$. Since $t_2(\wh\bDelta\trans\wh \Fb \wh\bDelta )^{1/2}> \lambda\norm{\wh\bDelta_S}_1$ holds in case (2), we immediately obtain $\norm{\wh\bDelta_S}_1\leq \frac{5t_2^2}{2\lambda}$. To control $\norm{\wh\bDelta_{S^c}}_1$, we rely on  (\ref{eq_proof_newB_6}) again, which is
	$$
	\frac{1}{2}\lambda\norm{\wh\bDelta_{S^c}}_1\leq \frac{3}{2}\lambda\norm{\wh\bDelta_S}_1+t_2(\wh\bDelta\trans\wh \Fb \wh\bDelta )^{1/2}.
	$$
	This leads to $\norm{\wh\bDelta_{S^c}}_1\leq 3\norm{\wh\bDelta_S}_1+\frac{5t_2^2}{\lambda}\leq \frac{25t_2^2}{2\lambda}$, such that $\|\wh\bDelta\|_1\lesssim t_2^2/\lambda$.  Combining the bounds in these two cases, we derive
	$$
	\|\wh\bDelta\|_1\lesssim \lambda s_B+t_2^2/\lambda,
	$$
	where $\lambda$ is subject to the constraint that $\lambda\geq 2t_1$. To establish a sharp rate of $\|\wh\bDelta\|_1$, we can further minimize $f(\lambda)=\lambda s_B+t_2^2/\lambda$ subject to the constraint $\lambda\geq 2t_1$. Define $\lambda_{opt}=t_2/s_B^{1/2}$. When $\lambda_{opt}\geq 2t_1$, the minimizer of $f(\lambda)$ is $\lambda_{opt}$ and the resulting minimal value is $f(\lambda_{opt})=s_B^{1/2}t_2$. However, when $\lambda_{opt}<2t_1$, by the monotonicity of $f(\lambda)$ the minimal is given by $f(2t_1)\asymp t_1s_B$. Combining these two cases, finally, we obtain the desired rate
	$$
	\|\wh\bDelta\|_1\lesssim t_1s_B+s_B^{1/2}t_2.
	$$
	With a slight modification of the proof (e.g., $|I_1|\lesssim K_1\sqrt{\frac{\log p}{n}}\|\wh\bDelta\|_1$ still holds uniformly over $1\leq k\leq p$), we obtain the same rate for $\|\wh \bB_{\cdot k}-\bB_{\cdot k}\|_1$ uniformly over $1\leq k\leq p$. This concludes the proof.
	
\end{proof}

Recall that  $\wh \bZ_i=\X_i\wh m^{-j}(\X_i) -\wh \bmu^j$, $ \bZ_i=\X_i m(\X_i) - \bmu$ and  $\wh\Fb=\frac{1}{n_j}\sum_{i\in D_j^*} \wh \bZ_i^{\otimes 2}$

\begin{lemma}[RE condition for $\wh\bB_k$]\label{RE1}
	Assume that the same conditions in Theorem \ref{thm_inf2} hold. Then with probability tending to 1,
	$$
	\inf_{\bv\in \mathcal{C},\bv\neq 0}\frac{\bv\trans\wh\Fb\bv}{\|\bv\|_2^2}\geq C,
	$$
	where $\mathcal{C}=\{\bv\in \RR^p:\exists S\subseteq \{1,...,p\},|S|=s_B, \norm{\bv_{S^c}}_1\leq \xi\norm{\bv_S}_1\}$ for some constants $C,\xi>0$.
\end{lemma}

\begin{proof}
	We define $\Fb=\frac{1}{n_j}\sum_{i\in D_j^*} \bZ_i^{\otimes 2}$. It holds that
	\begin{align*}
		\bv\trans\wh\Fb\bv&=\bv\trans(\wh\Fb-\Fb)\bv+\bv\trans\{\Fb-\E(\Fb)\}\bv+\bv\trans\E(\Fb)\bv\\
		&\geq \bv\trans\E(\Fb)\bv-|\bv\trans(\wh\Fb-\Fb)\bv|-|\bv\trans\{\Fb-\E(\Fb)\}\bv|.
	\end{align*}
	In what follows, we will bound the three terms in the last line one by one. Clearly,
	\begin{align}\label{eq_pf_RE1_00}
		\bv\trans\E(\Fb)\bv\geq C\|\bv\|_2^2
	\end{align}
	uniformly over $\bv$, as $\E(\Fb)$ has bounded smallest eigenvalues.  For the last term,
	$$
	|\bv\trans\{\Fb-\E(\Fb)\}\bv|\leq \|\bv\|_1^2\|\Fb-\E(\Fb)\|_{\max}\leq s_B(\xi+1)^2\|\bv\|_2^2\|\Fb-\E(\Fb)\|_{\max}\lesssim s_B\|\bv\|_2^2K_1^2\sqrt{\frac{\log p}{n}},
	$$
	where the second step holds as $\|\bv\|_1\leq (\xi+1)\|\bv_S\|_1\leq (\xi+1)s_B^{1/2}\|\bv_S\|_2\leq (\xi+1)s_B^{1/2}\|\bv\|_2$ and the last step is obtained by the Hoeffding inequality together with the bound $\|\X_im(\X_i)\|_\infty\leq CK_1$. Under the condition $s_BK_1^2\sqrt{\frac{\log p}{n}}=o(1)$, we have
	\begin{align}\label{eq_pf_RE1_0}
		\sup_{\bv\in \mathcal{C},\bv\neq 0}\frac{|\bv\trans\{\Fb-\E(\Fb)\}\bv|}{\|\bv\|_2^2}=o(1).
	\end{align}
	Now, we focus on the second term $|\bv\trans(\wh\Fb-\Fb)\bv|$. To this end, we first note that
	\begin{align}\label{eq_pf_RE1_1}
		|\bv\trans(\wh\bmu^j-\bmu)|&\leq \Big|\frac{1}{n_j}\sum_{i\in D_j^*} \bv\trans \X_i\{\wh m^{-j}(\X_i)-m(\X_i)\}\Big|+\Big|\frac{1}{n_j}\sum_{i\in D_j^*} \bv\trans [\X_im(\X_i)-\E\{\X_im(\X_i)\}]\Big|\nonumber\\
		&\leq \Big|\frac{1}{n_j}\sum_{i\in D_j^*} (\bv\trans \X_i)^2\Big|^{1/2}\Big|\frac{1}{n_j}\sum_{i\in D_j^*} \{\wh m^{-j}(\X_i)-m(\X_i)\}^2\Big|^{1/2}\nonumber\\
		&~~~~~~+\|\bv\|_1\Big\|\frac{1}{n_j}\sum_{i\in D_j^*} [\X_im(\X_i)-\E\{\X_im(\X_i)\}]\Big\|_\infty\nonumber\\
		&\lesssim \|\bv\|_1K_1\left(c_n+\sqrt{\frac{\log p}{n}}\right),
	\end{align}
	which is implied by the Hoeffding inequality in the last step and $\|\wh m^{-j}-m\|_2\lesssim c_n$. In addition,
	$$
	\frac{1}{n_j}\sum_{i\in D_j^*} (\bv\trans \X_i)^2\{\wh m^{-j}(\X_i)-m(\X_i)\}^2\lesssim \|\bv\|_1^2K_1^2c_n^2.
	$$
	Combined with (\ref{eq_pf_RE1_1}), we have
	\begin{align}\label{eq_pf_RE1_2}
		&\frac{1}{n_j}\sum_{i\in D_j^*} [\bv\trans \X_i \{\wh m^{-j}(\X_i)-m(\X_i)\}-\bv\trans(\wh\bmu^j-\bmu)]^2\nonumber\\
		&\leq \frac{2}{n_j}\sum_{i\in D_j^*} [\bv\trans \X_i \{\wh m^{-j}(\X_i)-m(\X_i)\}]^2+\frac{2}{n_j}\sum_{i\in D_j^*} \{\bv\trans(\wh\bmu^j-\bmu)\}^2\nonumber\\
		&\lesssim \|\bv\|_1^2K_1^2\left(c_n+\sqrt{\frac{\log p}{n}}\right)^2.
	\end{align}
	An implication of (\ref{eq_pf_RE1_2}) is  the following inequality
	\begin{align}\label{eq_pf_RE1_22}
		&\frac{1}{n_j}\sum_{i\in D_j^*} [\bv\trans \X_i \{\wh m^{-j}(\X_i)+m(\X_i)\}-\bv\trans(\wh\bmu^j+\bmu)]^2\nonumber\\
		&\leq \frac{2}{n_j}\sum_{i\in D_j^*} [\bv\trans \X_i \{\wh m^{-j}(\X_i)-m(\X_i)\}-\bv\trans(\wh\bmu^j-\bmu)]^2+\frac{2}{n_j}\sum_{i\in D_j^*} \{2\bv\trans \X_i m(\X_i)-2\bv\trans\bmu\}^2\nonumber\\
		&\lesssim \|\bv\|_1^2K_1^2\left(c_n+\sqrt{\frac{\log p}{n}}\right)^2+\|\bv\|_1^2K_1^2 \lesssim \|\bv\|_1^2K_1^2.
	\end{align}
	Finally, applying Cauchy–Schwarz inequality we can show that
	\begin{align}\label{eq_pf_RE1_3}
		&|\bv\trans(\wh\Fb-\Fb)\bv|\nonumber\\
		&= \Big|\frac{1}{n_j}\sum_{i\in D_j^*} \Big[\left\{\bv\trans \X_i \wh m^{-j}(\X_i)-\bv\trans\wh\bmu\right\}^2- \left\{\bv\trans \X_i  m(\X_i)-\bv\trans\bmu\right\}^2\Big]\Big|\nonumber\\
		&=\Big|\frac{1}{n_j}\sum_{i\in D_j^*} [\bv\trans \X_i \{\wh m^{-j}(\X_i)-m(\X_i)\}-\bv\trans(\wh\bmu-\bmu)][\bv\trans X_i \{\wh m^{-j}(\X_i)+m(\X_i)\}-\bv\trans(\wh\bmu+\bmu)]\Big|\nonumber\\
		&\leq \Big|\frac{1}{n_j}\sum_{i\in D_j^*} [\bv\trans \X_i \{\wh m^{-j}(\X_i)-m(\X_i)\}-\bv\trans(\wh\bmu^j-\bmu)]^2\Big|^{1/2}\nonumber\\
		&~~~~~~~~~~~~~~\times\Big|\frac{1}{n_j}\sum_{i\in D_j^*} [\bv\trans \X_i \{\wh m^{-j}(\X_i)+m(\X_i)\}-\bv\trans(\wh\bmu^j+\bmu)]^2\Big|^{1/2}\nonumber\\
		&\lesssim \|\bv\|_1^2K_1^2\left(c_n+\sqrt{\frac{\log p}{n}}\right)\lesssim \|\bv\|_2^2s_BK_1^2\left(c_n+\sqrt{\frac{\log p}{n}}\right),
	\end{align}
	where we use (\ref{eq_pf_RE1_2}) and (\ref{eq_pf_RE1_22}). Therefore, from (\ref{eq_pf_RE1_00}), (\ref{eq_pf_RE1_0}) and (\ref{eq_pf_RE1_3}), we obtain
	$$
	\inf_{\bv\in \mathcal{C},\bv\neq 0}\frac{\bv\trans\wh\Fb\bv}{\|\bv\|_2^2}\geq C-o(1).
	$$
\end{proof}

\subsection{Some Technical Details}\label{app_sup}

\subsubsection{Sparsity assumption on $\bB$}\label{app_sparseB}

Here, we first comprehensively illustrate the meaning and the implication of the sparsity assumption on the coefficient matrix $\bB$, then we investigate one scenario where this sparsity assumption is satisfied, together with a concrete example.

\paragraph{Meaning and implications of the sparsity of $\bB$.}
Recall that the matrix $\bB\in\RR^{p\times p}$ is defined as the coefficient matrix where $\X(Y-\X\trans\bt^*)$ is the response and $\X m(\X)-\bmu$ is the covariate; i.e.,
\begin{equation*}
	\X(Y-\X\trans\bt^*)=\bB\trans\{\X m(\X)-\bmu\}+\bE.
\end{equation*}
For each coordinate $j$, one can write 
\bse
X_j(Y-\X\trans\bt^*) = \sum_{k=1}^p B_{k j}\,\{X_k m(\X)-\mu_k\}+E_j.
\ese
Thus, the $j$th column of $\bB$ is sparse; i.e., $|\{k: B_{k j}\neq 0\}|\ll p$, indicates that, among all the $p$ variables $\{X_k m(\X)-\mu_k\}$, $k=1,\ldots,p$, only a small portion of them are truly contributing to the outcome $X_j(Y-\X\trans\bt^*)$.

The matrix $\bB$ has the closed-form expression 
\begin{align*}
	\bB=(\E[\{\X m(\X)-\bmu\}^{\otimes 2}])^{-1}\E\{\X^{\otimes 2}m(\X)(Y-\X\trans\bt^*)\}.
\end{align*}
From this expression, whether the sparsity assumption can be satisfied is driven by a few factors, such as the choice of the function $m(\x)$, and the structure of covariate $\X$.
Below, we first provide general justification for a scenario in which this assumption is naturally satisfied. Then, we present a concrete example.

\paragraph{General justification: when covariate $\X$ has the blockwise independence structure.}
We denote the support of function $m$ by $S_m\subseteq \{1, . . . , p\}$ which is the index set of all the variables present in $m(\cdot)$. Assume that the predictor variables exhibit block
independence with blocks corresponding to the block-diagonal covariance matrix $\bSigma$, and the maximal block-size is equal to $b_{\max}$. Under this assumption, we firstly know that for $1\leq k\leq p$, if $k \notin S_m $,
\begin{equation*}
	\|\E[\{\X m(\X)-\bmu\}X_k m(\X)]\|_0\leq b_{\max},
\end{equation*}
where we use $X_k$ to denote the $k$th component of $\X$. Otherwise,
\begin{equation*}
	\|\E[\{\X m(\X)-\bmu\}X_k m(\X)]\|_0\leq b_{\max}|S_m|.
\end{equation*}
Therefore, we can view the covariance matrix of $\X m(\X)$ as a different block-diagonal matrix with the same blocks as in $\bSigma$ for those variables not in working model $m$. Moreover, we can claim that $(\E[\{\X m(\X)-\bmu\}^{\otimes 2}])^{-1}$ presents the same block structure as in covariance matrix of $\X m(\X)$.

On the other hand, for $k \notin S_m\cup S_\eta $, where $\eta(\X)=f(\X)-\X\trans\bt^*$, $$\|\E[\{\X X_k m(\X)\eta(\X)\}]\|_0\leq b_{\max},$$ and the non-zero elements are within the dependence block of $X_k$.
Therefore, given that
\begin{equation}
	\label{app:B}
	\bB_{\cdot k}=(\E[\{\X m(\X)-\bmu\}^{\otimes 2}])^{-1}\E[\{\X X_k m(\X)\eta(\X)\}],
\end{equation}
we know $\|\bB_{\cdot k}\|_0\leq b_{\max}$ since $B_{jk}$ is nonzero only when $j$ is in the corresponding dependence block of $X^k$.
For $k \in S_m\cup S_\eta$, $$\|\E[\{\X X_k m(\X)\eta(\X)\}]\|_0\leq b_{\max}|S_m\cup S_\eta|,$$ and by equation (\ref{app:B}) and the block structure of $(\E[\{\X m(\X)-\bmu\}^{\otimes 2}])^{-1}$, we know $$\|\bB_{\cdot k}\|_0\leq b_{\max}|S_m\cup S_\eta|.$$
This investigation indicates that, if the supports of working model $m(\cdot)$ and $\eta(\cdot)$, and the nonlinear part in $f(\cdot)$ are sparse, the blockwise independence structure of $\X$ assumption is sufficient to guarantee the sparsity of $\bB$.

The following example even indicates, in some special cases, only the blockwise independence structure of $\X$ can guarantee the sparsity of $\bB$.
\paragraph{Concrete example.}
Let $p=2M$ with $M\ge 2$ and partition the features into independent bivariate blocks
\begin{align*}
	\X=\left(
	\customunderbrace{X_1,X_2}{$G_1$}
	,\ \ldots,\ \customunderbrace{X_{p-1},X_p}{$G_M$}
	\right)\trans,
	\qquad G_r\sim N \left(\0,\begin{bmatrix}1&\rho\\ \rho&1\end{bmatrix}\right),
\end{align*}
where $G_r$ independent across $r$, and $|\rho|<1$.
Now, consider
\bse
f(\X)=\bt\trans\X + \sum_{r=1}^M \gamma_r \left\{(X_{2r-1})^2-1\right\}.
\ese
By some simple algebra, one can compute that
$\bt^* = \bt$, hence $\eta(\X)=\sum_{r=1}^M \gamma_r\left\{(X_{2r-1})^2-1\right\}$.
In the following derivations, we take $m(\X):=\sum_{r=1}^M \left\{(X_{2r-1})^2-1\right\}$, adopt the simpler notation that $\Z_m(\X):=\X m(\X)$, then it is clear that $\bmu=\E\{\X m(\X)\}=\0$.
By block independence and symmetry, both $\E(\Z_m\Z_m\trans)$ and $\E\{\X\X\trans m(\X)\eta(\X)\}$ are block‑diagonal with identical $2\times 2$ blocks across $r$:
\bse
\E\left(\Z_m\Z_m\trans\right)=\mathrm{diag}\left(\A,\ldots,\A\right),\qquad
\E\left\{\X\X\trans m(\X)\eta(\X)\right\}=\mathrm{diag}\left(\C_1,\ldots,\C_M\right),
\ese
where,
\bse
\mathbf{A}
=\begin{bmatrix}
	2M+8 & \rho(2M+8)\\
	\rho(2M+8) & 2M+8\rho^2
\end{bmatrix},
\qquad
\mathbf{C}_r
=\begin{bmatrix}
	2G+8\gamma_r & \rho(2G+8\gamma_r)\\
	\rho(2G+8\gamma_r) & 2G+8\rho^2\gamma_r
\end{bmatrix},
\ese
where $G:=\sum_{r=1}^M \gamma_r$.
Therefore, each block of $\bB$ is
\bse
\mathbf{B}_r \;=\; \mathbf{A}^{-1}\mathbf{C}_r
\;=\;
\begin{bmatrix}
	\dfrac{G+4\gamma_r}{M+4}
	&
	\dfrac{4\rho\,(M\gamma_r-G)}{M(M+4)}
	\\[10pt]
	0
	&
	\dfrac{G}{M}
\end{bmatrix},
\qquad r=1,\ldots,M.
\ese
Hence $\bB=\mathrm{diag}(\mathbf{B}_1,\ldots,\mathbf{B}_M)$ is block‑diagonal, hence sparse.
An interesting special case is when $\gamma_r\equiv 1$, then $\mathbf{B}_r=\I_2$ for all $r$, then $\bB$ becomes the identity matrix $\I_p$.

\subsubsection{The formal proof that the condition $0<\psi<2$ is necessary and sufficient}\label{sec:functional_class}

Recall that
\bse
\bGamma_{\psi}
= \E(\T_{i1}^{\otimes 2})
- \frac{N(2\psi-\psi^{2})}{n+N}\,
\left\{\E(\T_{i2}\T_{i1}\trans)\right\}\trans
\left\{\E(\T_{i2}^{\otimes 2})\right\}^{-1}
\E(\T_{i2}\T_{i1}\trans),
\qquad
\Kb=\E(\T_{i1}^{\otimes 2}).
\ese
Define
\bse
c_\psi := \frac{N(2\psi-\psi^{2})}{n+N},\quad
\A := \E(\T_{i2}\T_{i1}\trans),\quad
\C := \E(\T_{i2}^{\otimes 2}) (\succ \0).
\ese
Then $\bGamma_{\psi} = \Kb - c_\psi\A\trans\C^{-1}\A$.
For any $\bv\neq \0$,
\bse
\bv\trans \bOmega \bGamma_{\psi}\bOmega \bv
= \bv\trans \bOmega \Kb \bOmega \bv - c_\psi\,\bv\trans \bOmega \A\trans\C^{-1}\A \bOmega \bv
= \bv\trans \bOmega \Kb \bOmega \bv - c_\psi\,\big\|\,\C^{-1/2}\A \bOmega \bv\,\big\|_2^2.
\ese

If $2\psi-\psi^{2}>0$ (equivalently $0<\psi<2$), then $c_\psi>0$ and
\bse
\bv\trans \bOmega \bGamma_{\psi}\bOmega \bv \le \bv\trans \bOmega \Kb \bOmega \bv,
\ese
with equality if and only if $\A\bOmega\bv=\0$. In Remark~\ref{rem_efficiency2}, we assume $\A$ has full rank (equivalently $\A\trans\C^{-1}\A\succ \0$), hence the inequality is strict for all $\bv\neq \0$:
\bse
\bv\trans \bOmega \bGamma_{\psi}\bOmega \bv < \bv\trans \bOmega \Kb \bOmega \bv.
\ese

On the other hand, if $\bv\trans \bOmega \bGamma_{\psi}\bOmega \bv < \bv\trans \bOmega \Kb \bOmega \bv$, then it follows that $c_\psi\,\big\|\,\C^{-1/2}\A \bOmega \bv\,\big\|_2^2 > 0$. Because $\A$ is of full rank which implies that $\|\C^{-1/2}\A \bOmega \bv\|_2^2 >0$ for any $\bv \neq \0$, it comes that $c_\psi >0$ which implies $0 < \psi < 2$.
This completes the proof.

\subsubsection{Efficiency gain compared to the supervised estimator}\label{sec:supp_efficiency}

We are able to see the efficiency gain by computing the relative efficiency between the supervised and the semi-supervised estimators.
Note that the proposed semi-supervised estimator admits the asymptotic variance $\bv\trans\bOmega \bGamma_{\psi} \bOmega\bv$, 
while the supervised estimator has the asymptotic variance $\bv\trans\bOmega \Kb \bOmega\bv$,
with
\begin{align*}
	\bGamma_\psi = \Kb - (1-\rho)(2\psi-\psi^2)\bM, \qquad \Kb=\E(\T_{i1}^{\otimes 2}),
\end{align*}
and $\bM:=\{\E(\T_{i2}\T_{i1}\trans)\}\trans\{\E(\T_{i2}^{\otimes 2})\}^{-1}\E(\T_{i2}\T_{i1}\trans)$.
Thus, the relative efficiency equals
\begin{align*}
	\frac{\bv\trans\bOmega \bGamma_{\psi} \bOmega\bv}{\bv\trans\bOmega \Kb \bOmega\bv} = 1 - (1-\rho)(2\psi-\psi^2) \frac{\bv\trans\bOmega \bM \bOmega\bv}{\bv\trans\bOmega \Kb \bOmega\bv}.
\end{align*}
Clearly, if we have more unlabeled data, $\rho$ gets smaller, then we have a smaller relative efficiency that indicates a greater efficiency gain.
Also, it is clear that the optimal choice of $\psi$ is 1.
The term $\bv\trans\bOmega \bM \bOmega\bv/\bv\trans\bOmega \Kb \bOmega\bv$ is always bounded between 0 and 1, since $\Kb-\bM$ is positive semi-definite.
In the special situation that $\T_{i1}$ is a linear function of $\T_{i2}$, $\Kb$=$\bM$ and this terms becomes 1,
where $\T_{i1}=\X_i(Y_i-\X_i\trans\bt^*)$ and $\T_{i2}=\X_i m(\X_i)-\bmu$.

\subsubsection{Choice of functional class $\cG$ for estimating $m(\x)$}\label{sec:supp_cG}

The main assumption regarding the estimator $\wh m(\x)$ in Assumption~\ref{assumptionE} pertains to the deterministic sequence $c_n$ where
$\|\wh m^{-j}-m\|_2=O_p(c_n)$.
It is required that $s_BK_1^2\Big(c_n+\sqrt{\frac{\log p}{n}}\Big)=o(1)$.

In practice, we recommend that users of our method choose the function class $\cG$ from commonly used ones, such as linear functions, additive functions, interaction models \citep{zhao2016analysis}, single-index models \citep{radchenko2015high, yang2017high, eftekhari2021inference}, or multi-index models \citep{yang2017learning}.

If the function class $\cG$ is chosen to be sparse linear, Theorem 7.20 in \cite{wainwright2019high} showed that $c_n \asymp \sqrt{s\log p/n}$.
Then the requirement becomes to $s_B K_1^2 \sqrt{s\log p/n} =o(1)$.
If the function class $\cG$ is chosen to be sparse additive, from some oracle inequalities for sparse additive estimator \citep{koltchinskii2010sparsity} as well as classical nonparametric rate \citep{tsy}, it implies that $c_n$ can be chosen as
\bse
\sqrt{s}\,n^{-\alpha/(2\alpha+1)} + \sqrt{s\log p/n},
\ese
where it is assumed that each active univariate component has smoothness parameter $\alpha > 0$.
For more complex choices of $\cG$, relevant theoretical results can be found in the respective literature, such as those for interaction models \citep{zhao2016analysis}, single-index models \citep{radchenko2015high, yang2017high, eftekhari2021inference}, and multi-index models \citep{yang2017learning}.

\subsubsection{Comparison with related work in Section~\ref{sec_dependable}}\label{sec_compare}

When the dimension $p$ is fixed and small, \cite{azriel} and \cite{chakrabortty2018} investigated how to  incorporate the unlabeled data to improve the estimation efficiency for regression coefficients in a working linear regression. In addition to the technical challenges arise from the high dimensionality (e.g., regularization and one-step update), a key difference from the previous works is that our dependable semi-supervised approach leads to a more efficient estimator for any linear combination of $\bt^*$. In the following, we briefly summarize their methodologies and explain the differences.

To improve the estimation efficiency for $\theta_j^*$, \cite{azriel} considered the following adjusted linear regression, for any $j\in [p]$
$$
\wt Y_{ij}=\theta^*_j+\bphi_j\trans \bU_{ij}+\wt\delta_{ij},
$$
where $\wt Y_{ij}=Y_i\wt X_{ij}$, $\bU_{ij}=(U_{ij1},...,U_{ijp})\trans$ with $U_{ijk}=X_{ik}\wt X_{ij}$ for $k\neq j$ and $U_{ijj}=X_{ij}\wt X_{ij}-1$ and $\wt\delta_{ij}$ is a mean 0 random variable. We use the notation $\wt X_{ij}=(X_{ij}-\bgamma_j\trans \X_{i,-j})/\E\{(X_{ij}-\bgamma_j\trans \X_{i,-j})^2\}$ where $\bgamma_j$ is  the estimand for the nodewise lasso (\ref{eq_hatgamma}). One interesting property of the adjusted linear regression is that the parameter of interest $\theta_j^*$ becomes the intercept parameter, because $\E(\bU_{ij})=0$ and $\theta_j^*=\E(Y_i\wt X_{ij})=\E(\wt Y_{ij})$ by the definition of $\wt Y_{ij}$. Thus, when $p$ is fixed and small, $\theta^*_j$ can be estimated by $\wh\theta_j^{A}$ the LSE from the adjusted linear regression, where the unlabeled data can help the estimation of $\bgamma_j$ and $\E\{(X_{ij}-\bgamma_j\trans \X_{i,-j})^2\}$. Thanks to the orthogonality of $\wt\delta_{ij}$ and $\bU_{ij}$, the asymptotic variance of $\wh\theta_j^{A}$ is shown to be no greater than $\wh\theta_j^{LSE}$, the $j$th component of the standard LSE $\wh\bt^{LSE}=(\bX\trans\bX)^{-1}\bX\trans\bY$; see their Theorem 2. As a result, if the parameter of interest is any component of $\bt^*$, their estimator provides the dependable semi-supervised inference.

However, since the adjusted linear regression is estimated for each $j\in [p]$ separately, their procedure does not guarantee the orthogonality of $\wt \delta_{ij}$ and $\bU_{ij'}$ for any $j'\neq j$ when the true regression function $f(\X)$ is nonlinear. Therefore, the linear combination of their estimators such as $\wh\theta_j^{A}+\wh\theta_{j'}^{A}$ may not be more efficient than the standard LSE $\wh\theta_j^{LSE}+\wh\theta_{j'}^{LSE}$. Unlike their approach, our estimator is constructed based on the geometric interpretation of estimating functions. The projection theory from estimating functions motivates us to consider the working regression model (\ref{eq_regression2}), which is different from the adjusted linear regression in \cite{azriel}.


\cite{chakrabortty2018} proposed a class of Efficient and Adaptive Semi-Supervised Estimators (EASE) which exploit the unlabeled data based on a semi-non-parametric smoothing and refitting
estimate of a target imputation function $\mu(\X)$. They mainly focused on the context that $N$ is much larger than $n$ and $p$ is fixed. With an estimated imputation function $\wh \mu(\X)$, they derived an initial semi-supervised estimator $\wh \bt_r$ through the estimating equation
\begin{equation}
	\frac{1}{N}\sum_{i=n+1}^{n+N} \X_i\left\{\wh \mu(\X_i)-\X_i\trans\bt\right\}=\0.
\end{equation}
The estimator $\wh \bt_r$ attains the semi-parametric efficiency bound when the imputation is sufficient (i.e., the imputation function equals the conditional mean function $\mu(\X)=f(\X)$) or the conditional mean function $f(\X)$ is linear $f(\X)=\X\trans\bt^*$. As seen in Remark \ref{rem_efficiency}, these properties also hold for our efficient semi-supervised estimator $\wh \bt^d$ in (\ref{eq_debias1}). To ensure the improved efficiency of EASE, they considered a further step of calibration which searched an optimal linear combination of $\wh \bt_r$ and the LSE $\wh\bt^{LSE}$. Their adaptive estimator is defined as $\wh \bt^E=\wh\bt^{LSE}+\bDelta(\wh \bt_r-\wh\bt^{LSE})$, where $\bDelta$ is a diagonal matrix that minimizes the asymptotic variance of $\wh \bt^E_j$ for each $j\in[p]$. When $\bDelta$ is consistently estimated, $\wh \theta^E_j$ is always no less efficient than the LSE no matter whether the imputation is sufficient or $f(\X)$ is linear. However, by the construction of $\wh \bt^E$, the efficiency improvement is not guaranteed if a linear combination of $\bt^*$ is considered.



\subsubsection{The difference between our framework and variable/feature importance}\label{sec:supp_importance}

The setting we adopt in this paper is the so-called assumption-lean framework.
This framework is suitable when one in interested in some simple but interpretable parameter of interest, such as the association between a certain phenotype and SNPs in the genome-wide association study.
This framework does not have to be confined in the linear working model.
In fact, in Section 4, we extend the methodology to a more general M-estimation framework, and 
we have clarified that the key ideas carry over to this broader setting without substantial difficulty.

Variable importance is a relevant, but distinct, concept.
In recent years, owing to its interpretability and generality, variable importance has attracted significant interest in the literature, particularly when estimated with flexible machine learning methods; see, e.g. \cite{williamson2021nonparametric, williamson2023general, verdinelli2024decorrelated}, as well as applications in survival analysis \citep{wolock2025assessing} and causal inference \citep{hines2025variable}.
Many variable importance measures exist, such as the one based on the Shapley value \citep{verdinelli2024feature}.
Below, we briefly review one of such measures.
Using the notation we use in our paper, decompose the covariate $\X$ as $\X=(\X_1\trans, \X_2\trans)\trans$ and denote $f(\X)=\E(Y\mid \X)$, $h(\X) = \E(Y\mid \X_2)$, then the importance of covariate $\X_1$ can be defined as
\begin{align*}
	\psi = \E\left[ \{m(\X) - h(\X)\}^2 \right].
\end{align*}
By construction, the parameter $\psi$ quantifies the improvement in predictive performance achieved by incorporating $\X_1$, relative to solely using $\X_2$.

While variable importance measures are inherently model-agnostic, their primary objective is to quantify the predictive value of specific features. 
In contrast, our goal in this paper is to estimate and conduct inference on a generally defined parameter (e.g., an association measure derived via a linear working model or a general M-estimator) within an assumption-lean framework. 
Therefore, the two objectives are fundamentally distinct.

\subsection{Extra Numerical Results}\label{sec:sim_supp}

\begin{table}[!htbp] 
	\caption{
		Simulation results for \underline{Model 1} with $p=500$:
		Bias, SD and RMSE stand for empirical bias, standard deviation, and root mean squared error, respectively, len represents the length of 95\% confidence interval.
		The estimators D-Lasso1 (that only uses labeled data with sample size $n$) and D-Lasso2 are $\wh\bt^d_1$ and $\wh \bt^d_2$, defined in (24).
		The straightforward debiased estimator D-SSL is defined in (5).
		The proposed dependable semi-supervised estimator S-SSL is defined in (14).
		The best performance is \textbf{bolded} during the comparison.
	}\label{sim:500main}
	\center
	\scalebox{0.70}{
		\begin{tabular}{cccrrrrrrrrrrrr}
			\hline
			&		&	&\multicolumn{4}{c}{$n=100$}  &\multicolumn{4}{c}{$n=300$}&\multicolumn{4}{c}{$n=500$}		\\
			&	$N$	& &\text{Bias}& \text{SD}& \text{RMSE}  &\text{len/2}  &\text{Bias}&\text{SD}  & \text{RMSE}& \text{len/2} &\text{Bias}&\text{SD}  & \text{RMSE}& \text{len/2}\\
			\hline
			\multirow{10}{*}{$\theta_1$}&&	{D-Lasso1}
			& 0.077     & 1.182     & 1.179     & 1.976     &  \bf{-0.072}& 0.296     & 0.303     & 0.548     & -0.038     & 0.223     & 0.225     & 0.434     \\
			\cline{3-15}
			&\multirow{3}{*}{$n$}&	{D-Lasso2}
			& \bf{0.008}& 0.695     & 0.692     & 1.142     &	 -0.078     & 0.293     & 0.302     & 0.549     & \bf{-0.038}& 0.222     & 0.224     & 0.433     \\
			&		&	{D-SSL}
			&-0.188     & 0.618     & 0.643     & 2.192     &	 -0.134     & 0.270     & 0.300     & 0.565     & -0.074     & \bf{0.196}& \bf{0.208}& \bf{0.351}\\
			&		&	{S-SSL}
			&-0.198     & \bf{0.562}& \bf{0.594}& \bf{0.963}&	 -0.112     & \bf{0.260}& \bf{0.282}& \bf{0.466}& -0.062     & 0.204     & 0.212     & 0.368     \\
			\cline{3-15}
			&\multirow{3}{*}{$4n$}&	{D-Lasso2}
			&\bf{-0.009}& 0.677     & 0.674     & 1.113     &	 -0.077     & 0.290     & 0.298     & 0.546     & \bf{-0.038}& 0.222     & 0.225     & 0.432     \\
			&		&	{D-SSL}
			&-0.184     & 0.591     & 0.616     & 1.203     &	 -0.148     & 0.280     & 0.316     & \bf{0.401}& -0.085     & 0.192     & 0.209     & \bf{0.279}\\
			&		&	{S-SSL}
			&-0.284     & \bf{0.461}& \bf{0.539}& \bf{0.851}&	 -0.140     & \bf{0.245}& \bf{0.281}& 0.408     & -0.086     & \bf{0.185}& \bf{0.203}& 0.322     \\
			\cline{3-15}
			&\multirow{3}{*}{$8n$}&	{D-Lasso2}
			&\bf{-0.008}& 0.678     & 0.674     & 1.119     &	 -0.079     & 0.290     & 0.299     & 0.546     & \bf{-0.039}& 0.222     & 0.225     & 0.432     \\
			&		&	{D-SSL}
			&-0.200     & 0.575     & 0.606     & 0.906     &	 -0.149     & 0.280     & 0.316     & \bf{0.357}& -0.089     & 0.190     & 0.209     & \bf{0.259}\\
			&		&	{S-SSL}
			&-0.318     & \bf{0.439}& \bf{0.540}& \bf{0.798}&	 -0.146     & \bf{0.243}& \bf{0.283}& 0.390     & -0.093     & \bf{0.177}& \bf{0.199}& 0.307     \\
			\hline
			\multirow{10}{*}{$\theta_2$}&&	{D-Lasso1}
			&-0.386     & 1.867     & 1.898     & 2.504     &     0.017     & 0.162     & 0.162     & 0.330     & -0.002     & 0.136     & 0.136     & 0.261     \\
			\cline{3-15}
			&\multirow{3}{*}{$n$}&	{D-Lasso2}
			& \bf{0.031}& 0.363     & 0.362     & 0.648     &	  0.016     & 0.162     & 0.162     & 0.330     & \bf{-0.003}& 0.135     & 0.134     & 0.262     \\
			&		&	{D-SSL}
			&-0.079     & 0.311     & 0.320     & 1.260     &	 -0.017     & \bf{0.141}& \bf{0.141}& 0.328     & -0.014     & 0.124     & 0.124     & \bf{0.212}\\
			&		&	{S-SSL}
			&-0.047     & \bf{0.309}& \bf{0.311}& \bf{0.551}&	 \bf{-0.003}& 0.142     & \bf{0.141}& \bf{0.288}& -0.012     & \bf{0.119}& \bf{0.119}& 0.228     \\
			\cline{3-15}
			&\multirow{3}{*}{$4n$}&	{D-Lasso2}
			& \bf{0.026}& 0.347     & 0.347     & 0.641     &	  \bf{0.013}& 0.164     & 0.163     & 0.330     & \bf{-0.006}& 0.136     & 0.135     & 0.262     \\
			&		&	{D-SSL}
			&-0.104     & 0.311     & 0.326     & 0.695     &	 -0.025     & 0.139     & 0.141     & \bf{0.245}& -0.019     & \bf{0.113}& \bf{0.114}& \bf{0.174}\\
			&		&	{S-SSL}
			&-0.118     & \bf{0.285}& \bf{0.307}& \bf{0.484}&	 -0.024     & \bf{0.138}& \bf{0.139}& 0.261     & -0.021     & \bf{0.113}& 0.115     & 0.205     \\
			\cline{3-15}
			&\multirow{3}{*}{$8n$}&	{D-Lasso2}
			& \bf{0.014}& 0.348     & 0.346     & 0.641     &	  \bf{0.012}& 0.164     & 0.163     & 0.331     & \bf{-0.006}& 0.137     & 0.136     & 0.263     \\
			&		&	{D-SSL}
			&-0.109     & 0.314     & 0.331     & 0.542     &	 -0.024     & 0.139     & 0.140     & \bf{0.222}& -0.021     & \bf{0.109}& \bf{0.111}& \bf{0.163}\\
			&		&	{S-SSL}
			&-0.139     & \bf{0.281}& \bf{0.312}& \bf{0.459}&	 -0.028     & \bf{0.135}& \bf{0.138}& 0.252     & -0.025     & 0.111     & 0.113     & 0.199     \\
			\hline
			\multirow{10}{*}{$\theta_4$}&&	{D-Lasso1}
			&-0.284     & 2.374     & 2.379     & 3.906     &    -0.064     & 0.189     & 0.199     & 0.377     & -0.032     & 0.142     & 0.145     & 0.300     \\
			\cline{3-15}
			&\multirow{3}{*}{$n$}&	{D-Lasso2}
			&\bf{-0.101}& 0.358     & 0.370     & 0.743     &	 \bf{-0.059}& 0.194     & 0.202     & 0.377     & \bf{-0.023}& 0.138     & 0.139     & 0.302     \\
			&		&	{D-SSL}
			&-0.233     & \bf{0.297}& 0.376     & 1.352     &	 -0.083     & 0.180     & 0.197     & 0.345     & -0.048     & 0.120     & \bf{0.129}& \bf{0.222}\\
			&		&	{S-SSL}
			&-0.203     & 0.309     & \bf{0.368}& \bf{0.604}&	 -0.098     & \bf{0.161}& \bf{0.187}& \bf{0.315}& -0.053     & \bf{0.118}& \bf{0.129}& 0.251     \\
			\cline{3-15}
			&\multirow{3}{*}{$4n$}&	{D-Lasso2}
			&\bf{-0.079}& 0.329     & \bf{0.337}& 0.722     &	 \bf{-0.046}& 0.195     & 0.199     & 0.378     & \bf{-0.010}& 0.139     & 0.138     & 0.303     \\
			&		&	{D-SSL}
			&-0.234     & \bf{0.269}& 0.356     & 0.724     &	 -0.085     & 0.167     & 0.187     & \bf{0.256}& -0.049     & 0.111     & \bf{0.121}& \bf{0.182}\\
			&		&	{S-SSL}
			&-0.251     & 0.273     & 0.370     & \bf{0.510}&	 -0.108     & \bf{0.150}& \bf{0.184}& 0.269     & -0.058     & \bf{0.108}& 0.122     & 0.213     \\
			\cline{3-15}
			&\multirow{3}{*}{$8n$}&	{D-Lasso2}
			&\bf{-0.071}& 0.340     & \bf{0.346}& 0.735     &	 \bf{-0.039}& 0.195     & 0.198     & 0.379     & \bf{-0.006}& 0.140     & 0.139     & 0.304     \\
			&		&	{D-SSL}
			&-0.230     & \bf{0.265}& 0.350     & 0.563     &	 -0.086     & 0.162     & 0.183     & \bf{0.231}& -0.047     & 0.116     & 0.125     & \bf{0.170}\\
			&		&	{S-SSL}
			&-0.264     & 0.268     & 0.375     & \bf{0.485}&	 -0.107     & \bf{0.147}& \bf{0.181}& 0.253     & -0.056     & \bf{0.108}& \bf{0.121}& 0.201     \\
			\hline
			\multirow{10}{*}{$\theta_5$}&&	{D-Lasso1}
			& 0.481     & 1.909     & 1.959     & 1.912     &     0.139     & 0.135     & 0.194     & 0.251     &  0.080     & 0.092     & 0.122     & 0.191     \\
			\cline{3-15}
			&\multirow{3}{*}{$n$}&	{D-Lasso2}
			& \bf{0.370}& 0.254     & 0.448     & 0.519     &	  0.112     & 0.131     & 0.172     & 0.251     &  0.057     & 0.089     & 0.105     & 0.192     \\
			&		&	{D-SSL}
			& 0.390     & \bf{0.189}& \bf{0.433}& 1.134     &	  \bf{0.095}& 0.123     & \bf{0.155}& 0.323     &  \bf{0.050}& \bf{0.073}& \bf{0.088}& 0.213     \\
			&		&	{S-SSL}
			& 0.400     & 0.204     & 0.449     & \bf{0.454}&	  0.125     & \bf{0.122}& 0.174     & \bf{0.232}&  0.069     & 0.076     & 0.102     & \bf{0.176}\\
			\cline{3-15}
			&\multirow{3}{*}{$4n$}&	{D-Lasso2}
			& \bf{0.268}& 0.256     & \bf{0.370}& 0.521     &	  \bf{0.078}& 0.133     & 0.154     & 0.255     &  \bf{0.033}& 0.089     & 0.095     & 0.194     \\
			&		&	{D-SSL}
			& 0.351     & \bf{0.180}& 0.394     & 0.673     &	  \bf{0.078}& 0.126     & 0.148     & 0.251     &  0.037     & 0.073     & \bf{0.082}& 0.178     \\
			&		&	{S-SSL}
			& 0.346     & 0.183     & 0.391     & \bf{0.414}&	  0.089     & \bf{0.115}& \bf{0.145}& \bf{0.218}&  0.047     & \bf{0.072}& 0.085     & \bf{0.164}\\
			\cline{3-15}
			&\multirow{3}{*}{$8n$}&	{D-Lasso2}
			& \bf{0.229}& 0.271     & \bf{0.354}& 0.530     &	  \bf{0.056}& 0.136     & 0.147     & 0.257     &  \bf{0.019}& 0.091     & 0.092     & 0.196     \\
			&		&	{D-SSL}
			& 0.328     & 0.190     & 0.379     & 0.544     &	  0.069     & 0.123     & 0.140     & 0.229     &  0.027     & 0.075     & \bf{0.080}& 0.168     \\
			&		&	{S-SSL}
			& 0.310     & \bf{0.181}& 0.359     & \bf{0.402}&	  0.077     & \bf{0.111}& \bf{0.135}& \bf{0.213}&  0.036     & \bf{0.071}& \bf{0.080}& \bf{0.161}\\
			\hline
			\multirow{10}{*}{$\theta_6$}&&	{D-Lasso1}
			&\bf{-0.097}& 0.676     & 0.680     & 1.164     &  -0.057     & 0.135     & 0.146     & 0.245     & -0.040     & 0.097     & 0.105     & 0.190     \\
			\cline{3-15}
			&\multirow{3}{*}{$n$}&	{D-Lasso2}
			&-0.133     & 0.280     & \bf{0.309}& 0.528     &	 \bf{-0.042}& 0.139     & 0.145     & 0.246     & \bf{-0.028}& 0.099     & 0.103     & 0.191     \\
			&		&	{D-SSL}
			&-0.313     & 0.243     & 0.395     & 1.154     &	 -0.056     & \bf{0.120}& \bf{0.132}& 0.327     & -0.030     & \bf{0.086}& \bf{0.091}& 0.217     \\
			&		&	{S-SSL}
			&-0.219     & \bf{0.238}& 0.323     & \bf{0.456}&	 -0.060     & 0.125     & 0.138     & \bf{0.226}& -0.040     & 0.093     & 0.101     & \bf{0.175}\\
			\cline{3-15}
			&\multirow{3}{*}{$4n$}&	{D-Lasso2}
			&\bf{-0.099}& 0.265     & \bf{0.281}& 0.529     &	 \bf{-0.023}& 0.142     & 0.143     & 0.248     & \bf{-0.017}& 0.102     & 0.103     & 0.193     \\
			&		&	{D-SSL}
			&-0.281     & 0.257     & 0.380     & 0.700     &	 -0.048     & \bf{0.118}& \bf{0.127}& 0.252     & -0.030     & \bf{0.080}& \bf{0.085}& 0.180     \\
			&		&	{S-SSL} 
			&-0.218     & \bf{0.248}& 0.329     & \bf{0.436}&	 -0.049     & 0.120     & 0.129     & \bf{0.213}& -0.036     & 0.088     & 0.094     & \bf{0.163}\\
			\cline{3-15}
			&\multirow{3}{*}{$8n$}&	{D-Lasso2}
			&\bf{-0.075}& 0.262     & \bf{0.271}& 0.534     &	 \bf{-0.015}& 0.143     & 0.143     & 0.250     & \bf{-0.010}& 0.103     & 0.103     & 0.194     \\
			&		&	{D-SSL}
			&-0.282     & 0.261     & 0.383     & 0.555     &	 -0.045     & \bf{0.117}& 0.125     & 0.230     & -0.026     & \bf{0.078}& \bf{0.082}& 0.169     \\
			&		&	{S-SSL}
			&-0.218     & \bf{0.244}& 0.326     & \bf{0.421}&	 -0.042     & 0.118     & \bf{0.124}& \bf{0.210}& -0.030     & 0.086     & 0.091     & \bf{0.160}\\
			\hline
	\end{tabular}}
\end{table}

\begin{figure}[!htbp] 
	\centering
	\includegraphics[width=\textwidth,height=1.15\textwidth]{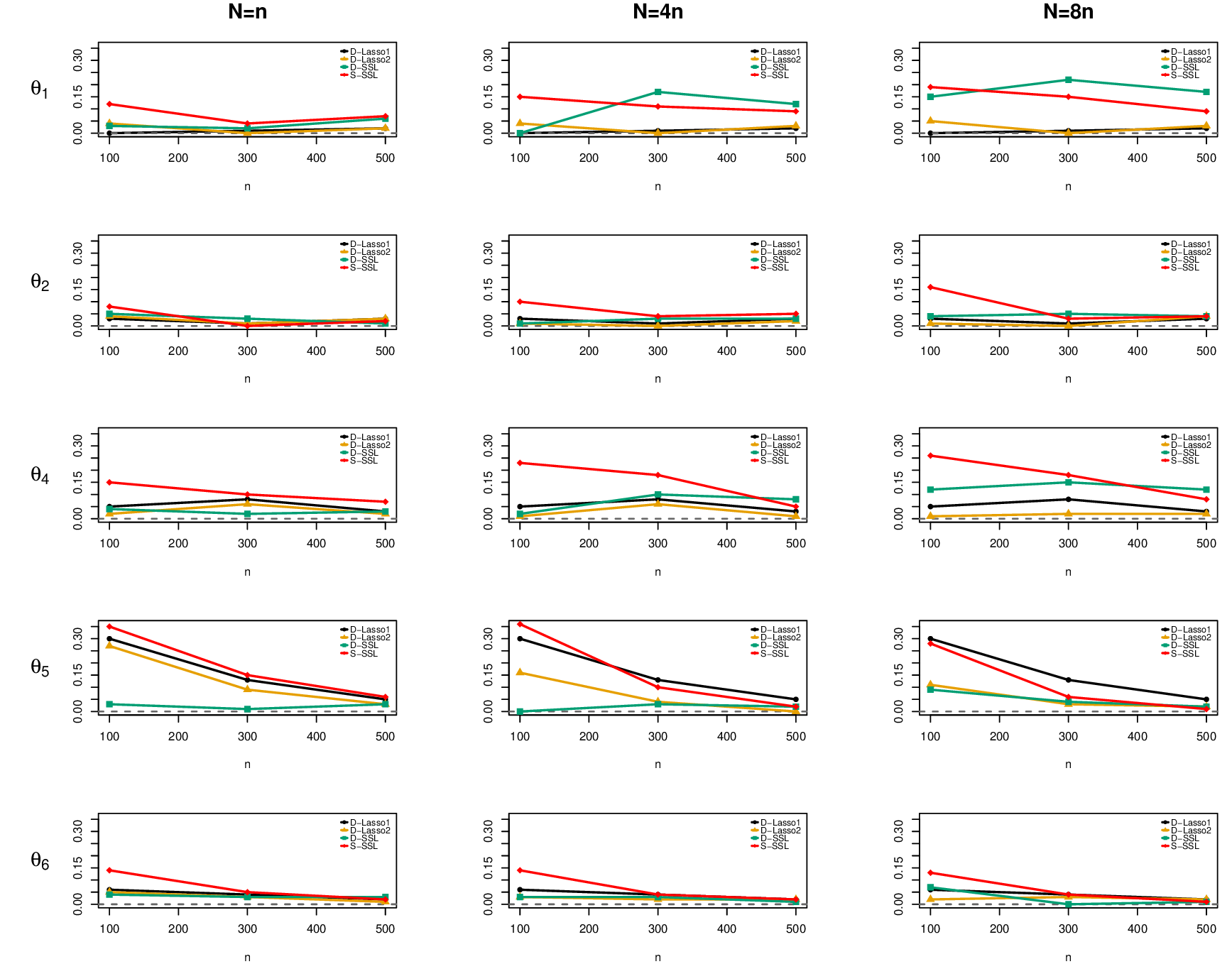}
	\caption{Simulation results for \underline{Model 1} with $p=500$: absolute difference between the empirical 95\% coverage probability and the nominal level 0.95.
		In all panels, rows represent different parameters, columns represent different $N/n$ ratios, and each panel plots the trend over the sample size $n$.
	}
	\label{fig:500}
\end{figure}

We consider the following additive model for $Y$, that 
$Y=0.5X_1^2+0.8X_3^3-(X_4-2)^2+2(X_5+1)^2+2X_6+\epsilon$,
where $\epsilon \sim \mathcal{N}(0,1)$.
Similar to Section~\ref{sec_simu}, we first generate a $p$-dimensional multivariate normal random vector $\bU \sim \mathcal{N}(0,\bSigma)$ with  $\Sigma_{jk}=0.3^{|j-k|}$. We set the covariate $\X=(X_1,...,X_p)\trans$ to be $X_1=|U_1|$ and $X_j=U_j ~\mathrm{for}~ 1<j\leq p$. The reason we take $X_1=|U_1|$ is that this transformation implies $\E(X_1^k X_j)=0$ for $j\neq 1$ but the parameter $\theta^*_1$ for centered $X_1$ is nonzero. 
To calculate the corresponding regression parameter $\bt^*$ under the working linear model, we first center $Y$ and $X_1$ so that their means are 0. By Proposition 4 in \cite{mann2015}, we know that the support of $\bt^*$ is $S=\{1,3,4,5,6\}$ and $\theta_j^*$ for any $j\in S$ is given by the $L_2(\PP)$ projection in the sub-model only with the variable $X_j$ (e.g, $\theta_3^*=\arg\min \E(0.8X_3^2-\theta_3 X_3)^2$). 
After some calculation, we obtain $\bt^*=(1.1,0,2.4,4,4,2,0,...,0)\trans$, which is sparse.

With sample size $n\in\{100,300\}$, the ratio $N/n\in\{1,4,8\}$ and the dimension $p\in\{200,500\}$, we compare the performance of the four methods: $\wh\bt^d_1$ (D-Lasso1, that only uses labeled data with sample size $n$), $\wh \bt^d_2$ (D-Lasso2), both defined in (\ref{est_Ld}), the straightforward debiased estimator D-SSL $\wh\bt^d$ defined in (\ref{eq_debias1}), and the proposed dependable semi-supervised estimator S-SSL $\wh \bt^d_{S,\psi=1}$ defined in (\ref{eq_debias2}).

Based on 100 simulation replicates, we report the empirical bias (Bias), standard deviation (SD), root mean squared error (RMSE), the half length of 95\% confidence interval (len/2) and the coverage probability (CVP) for each of the single parameters $\theta_1$, $\theta_3$, $\theta_4$, $\theta_5$ and $\theta_6$, for $p=200$ and $p=500$, in Table~\ref{sim:t1} and Table~\ref{sim:t3}, respectively.
In Table~\ref{sim:time_supp}, we also report the computation time (in seconds) of one simulation replication of these four methods.

It is seen that the coverage rates of all the methods are very close to the desired level $0.95$, especially when the sample size $n=300$.
In the majority of the scenarios we consider, the D-SSL method produces the shortest CIs among all four methods, and the CIs from the S-SSL method are always shorter than those from D-Lasso1 and D-Lasso2.
As expected, the CIs from both D-SSL and S-SSL become shorter as the size of unlabeled data $N$ increases.
Finally, we note that the length of CIs from D-Lasso1 and D-Lasso2 is very similar, which shows that the way of estimating $\bOmega$ has little effect on debiased lasso estimators.

In addition, in Table~\ref{sim:t9}, we compare the performance of the proposed method S-SSL with different methods of estimating the conditional mean function $f(\cdot)$. Other than the method in \cite{huang2010} (huang), we also implement the random forest method (randomforest) and the neural network method (nnet). 
Under the setting we consider, all three methods perform very similar and it is hard tell which one might be better.
In Table~\ref{sim:t12}, we compare the performance of the proposed dependable semi-supervised estimator S-SSL with different criteria of choosing the tuning parameters $(\lambda_{\Omega},\lambda_B)$ in estimating $(\bOmega, \bB)$.
Here, $\lambda_{\min}$ refers to the value of the tuning parameter $\lambda$ that yields the minimum mean cross-validated error, and $\lambda_{1se}$ refers to the largest (most regularized, or "simplest") value of the tuning parameter $\lambda$ such that the cross-validation error is within one standard error of the minimum error achieved by $\lambda_{\min}$.
Across the four scenarios we compare, their performances are very similar which indicates that the proposed estimator is not sensitive to the tuning parameter selection.
Further, in Table~\ref{sim:t10}, we assess the performance of the proposed method S-SSL for different linear combinations of the parameter of interest: $-\theta_1+2\theta_6$ (case 1), $-\theta_1+\theta_3+2\theta_6$ (case 2), $-\theta_1+\theta_3+\theta_5+\theta_6$ (case 3), and $-\theta_1+\theta_3-\theta_4+\theta_5$ (case 4).
In addition, in Table~\ref{sim:t11}, we assess the performance of the proposed method S-SSL for different estimands where the value $\|\bv\|_1/\|\bv\|_2$ ranges from 1 to $\sqrt{5}$.

\begin{table}[!htbp] 
	\caption{Simulation results for \underline{Model 2} with $p=200$:
		Bias, SD and RMSE stand for empirical bias, standard deviation, and root mean squared error, respectively,
		len represents the length of 95\% confidence interval, and CVP is the coverage probability.
		The estimators D-Lasso1 (that only uses labeled data with sample size $n$) and D-Lasso2 are $\wh\bt^d_1$ and $\wh \bt^d_2$, defined in (24).
		The straightforward debiased estimator D-SSL is defined in (5).
		The proposed dependable semi-supervised estimator S-SSL is defined in (14).
		The best performance is \textbf{bolded} during the comparison.
	}\label{sim:t1}
	\center
	\scalebox{0.70}{
		\begin{tabular}{cccrccccrcccc}
			\hline
			&		&	&\multicolumn{5}{c}{$n=100$}  &\multicolumn{5}{c}{$n=300$}		\\
			&	$N$	& &\text{Bias}& \text{SD}& \text{RMSE}  &\text{len/2}&\text{CVP}  &\text{Bias}&\text{SD}  & \text{RMSE}& \text{len/2} &\text{CVP}\\
			\hline
			\multirow{10}{*}{$\theta_1$}&&	{D-Lasso1}
			&-0.135     & 1.229     & 1.230     & 2.003     & \bf{0.96}  &  \bf{-0.059} & 0.420     & 0.422     & 0.837     & 0.97\\
			\cline{3-13}
			&\multirow{3}{*}{$n$}&	{D-Lasso2}
			& \bf{0.028}& 0.653     & 0.650     & 1.483     & \bf{0.96}  &	 -0.062     & 0.419     & 0.421     & 0.829     & \bf{0.96}\\
			&		&	{D-SSL}
			&-0.069     & \bf{0.531}& \bf{0.533}& 1.768     & 1.00       &	 -0.086     & 0.379     & 0.387     & 0.748     & 0.91\\
			&		&	{S-SSL}  
			&-0.049     & 0.607     & 0.606     & \bf{1.298}& 0.97       &	 -0.070     & \bf{0.338}& \bf{0.343}& \bf{0.739}& 0.97\\
			\cline{3-13}
			&\multirow{3}{*}{$4n$}&	{D-Lasso2}
			& 0.064     & 0.662     & 0.662     & 1.486     & 0.97       &	 \bf{-0.057}& 0.423     & 0.425     & 0.827     & 0.96\\
			&		&	{D-SSL}
			&-0.105     & \bf{0.465}& \bf{0.475}& 1.285     & 1.00       &	 -0.064     & \bf{0.304}& \bf{0.310}& \bf{0.612}& \bf{0.95}\\
			&		&	{S-SSL}
			&\bf{-0.013}& 0.561     & 0.558     & \bf{1.189}& 0.97       &	 -0.094     & 0.339     & 0.350     & 0.691     & 0.93\\
			\cline{3-13}
			&\multirow{3}{*}{$8n$}&	{D-Lasso2}
			& 0.036     & 0.661     & 0.658     & 1.469     & 0.97       &	 \bf{-0.059}& 0.425     & 0.427     & 0.831     & \bf{0.96}\\
			&		&	{D-SSL}
			&-0.086     & \bf{0.443}& \bf{0.449}& \bf{1.150}& 0.99       &	 -0.061     & \bf{0.302}& \bf{0.306}& \bf{0.572}& \bf{0.96}\\
			&		&	{S-SSL}
			&\bf{-0.013}& 0.560     & 0.558     & 1.155     & \bf{0.96}  &	 -0.095     & 0.336     & 0.347     & 0.675     & 0.94\\
			\hline
			\multirow{10}{*}{$\theta_3$}&&	{D-Lasso1}
			& 0.195     & 0.917     & 1.384     & 1.606     & \bf{0.91}       &     0.039    & 0.398     & 0.398     & 0.771     & 0.96\\
			\cline{3-13}
			&\multirow{3}{*}{$n$}&	{D-Lasso2}
			& \bf{0.021}& 0.701     & 1.328     & 1.237     & 0.90       &	  \bf{0.010}& 0.399     & 0.397     & 0.772     & \bf{0.95}\\
			&		&	{D-SSL}
			&-0.173     & \bf{0.551}& 1.312     & 1.219     & \bf{0.97}  &	 -0.076     & \bf{0.325}& 0.333     & \bf{0.493}& \bf{0.95}\\
			&		&	{S-SSL}
			&-0.087     & 0.583     & \bf{1.295}& \bf{1.069}& 0.91       &	 -0.015     & 0.333     & \bf{0.332}& 0.644     & \bf{0.95}\\
			\cline{3-13}
			&\multirow{3}{*}{$4n$}&	{D-Lasso2}
			&\bf{-0.031}& 0.687     & 1.328     & 1.248     & 0.90       &	 \bf{-0.004}& 0.401     & 0.399     & 0.775     & \bf{0.95}\\
			&		&	{D-SSL}
			&-0.253     & \bf{0.519}& 1.356     & \bf{0.844}& 0.82       &	 -0.108     & 0.300     & 0.317     & \bf{0.405}& 0.92\\
			&		&	{S-SSL}
			&-0.187     & 0.525     & \bf{1.319}& 0.932     & 0.88       &	 -0.082     & \bf{0.285}& \bf{0.295}& 0.549     & 0.92\\
			\cline{3-13}
			&\multirow{3}{*}{$8n$}&	{D-Lasso2}
			&\bf{-0.057}& 0.689     & 0.687     & 1.240     & \bf{0.91}  &	 \bf{-0.015}& 0.401     & 0.399     & 0.782     & \bf{0.96}\\
			&		&	{D-SSL}
			&-0.283     & \bf{0.511}& 0.582     & \bf{0.750}& 0.84       &	 -0.119     & 0.288     & 0.311     & \bf{0.376}& 0.89\\
			&		&	{S-SSL}
			&-0.225     & 0.518     & \bf{0.563}& 0.884     & 0.86       &	 -0.096     & \bf{0.265}& \bf{0.281}& 0.519     & 0.91\\
			\hline
			\multirow{10}{*}{$\theta_4$}&&	{D-Lasso1}
			& 0.398     & 0.659     & 0.767     & 1.230     & 0.88       &     0.158    & 0.314     & 0.350     & 0.578     & 0.92\\
			\cline{3-13}
			&\multirow{3}{*}{$n$}&	{D-Lasso2}
			& 0.315     & 0.546     & 0.628     & 1.033     & 0.88       &	  0.108     & 0.314     & 0.330     & 0.577     & 0.92\\
			&		&	{D-SSL}
			& \bf{0.037}& \bf{0.440}& \bf{0.440}& 1.138     & \bf{1.00}  &	  \bf{0.023}& \bf{0.247}& \bf{0.247}& \bf{0.498}& \bf{0.95}\\
			&		&	{S-SSL}
			& 0.310     & 0.506     & 0.591     & \bf{0.926}& 0.87       &	  0.103     & 0.263     & 0.281     & 0.528     & \bf{0.95}\\
			\cline{3-13}
			&\multirow{3}{*}{$4n$}&	{D-Lasso2}
			& 0.203     & 0.550     & 0.584     & 1.026     & \bf{0.93}  &	  0.075     & 0.313     & 0.321     & 0.580     & 0.94\\
			&		&	{D-SSL}
			&\bf{-0.051}& \bf{0.408}& \bf{0.409}& 0.835     & \bf{0.97}  &	  \bf{0.010}& \bf{0.206}& \bf{0.205}& \bf{0.411}& 0.96\\
			&		&	{S-SSL}
			& 0.170     & 0.480     & 0.507     & \bf{0.832}& 0.87       &	  0.052     & 0.237     & 0.242     & 0.492     & \bf{0.95}\\
			\cline{3-13}
			&\multirow{3}{*}{$8n$}&	{D-Lasso2}
			& 0.135     & 0.539     & 0.553     & 1.019     & \bf{0.97}  &	  0.052     & 0.316     & 0.319     & 0.585     & \bf{0.94}\\
			&		&	{D-SSL}
			&\bf{-0.093}& \bf{0.400}& \bf{0.408}& \bf{0.750}& \bf{0.93}  &	 \bf{-0.003}& \bf{0.197}& \bf{0.196}& \bf{0.381}& \bf{0.94}\\
			&		&	{S-SSL}
			& 0.111     & 0.477     & 0.488     & 0.810     & 0.87       &	  0.024     & 0.236     & 0.236     & 0.482     & \bf{0.94}\\
			\hline
			\multirow{10}{*}{$\theta_5$}&&	{D-Lasso1}
			& 0.319     & 0.985     & 1.122     & 1.837     & 0.91       &   0.229      & 0.431     & 0.486     & 0.787     & 0.90\\
			\cline{3-13}
			&\multirow{3}{*}{$n$}&	{D-Lasso2}
			& 0.213     & 0.712     & 0.740     & 1.235     & 0.90       &	  0.177     & 0.430     & 0.463     & 0.788     & \bf{0.94}\\
			&		&	{D-SSL}
			& \bf{0.044}& \bf{0.594}& \bf{0.593}& \bf{1.180}& \bf{0.94}  &	  \bf{0.069}& \bf{0.315}& \bf{0.321}& \bf{0.520}& 0.88\\
			&		&	{S-SSL}
			& 0.195     & 0.646     & 0.672     & 1.137     & 0.90       &	  0.142     & 0.359     & 0.384     & 0.702     & 0.91\\
			\cline{3-13}
			&\multirow{3}{*}{$4n$}&	{D-Lasso2}
			& 0.122     & 0.709     & 0.716     & 1.253     & 0.90       &	  0.129     & 0.429     & 0.446     & 0.791     & \bf{0.94}\\
			&		&	{D-SSL}
			&\bf{-0.019}& \bf{0.530}& \bf{0.528}& \bf{0.852}& 0.87       &	  \bf{0.028}& \bf{0.253}& \bf{0.254}& \bf{0.431}& 0.90\\
			&		&	{S-SSL}
			& 0.061     & 0.586     & 0.587     & 1.072     & \bf{0.91}  &	  0.084     & 0.329     & 0.338     & 0.636     & 0.92\\
			\cline{3-13}
			&\multirow{3}{*}{$8n$}&	{D-Lasso2}
			& 0.072     & 0.711     & 0.711     & 1.247     & 0.89       &	  0.111     & 0.434     & 0.446     & 0.796     & \bf{0.95}\\
			&		&	{D-SSL}
			&-0.048     & \bf{0.508}& \bf{0.508}& \bf{0.758}& 0.87       &	  \bf{0.019}& \bf{0.232}& \bf{0.232}& \bf{0.395}& 0.91\\
			&		&	{S-SSL}
			& \bf{0.004}& 0.580     & 0.577     & 1.038     & \bf{0.92}  &	  0.054     & 0.322     & 0.325     & 0.621     & 0.92\\
			\hline
			\multirow{10}{*}{$\theta_6$}&&	{D-Lasso1}
			& 0.168     & 0.943     & 0.953     & 1.912     & 0.98       &     0.057     & 0.254     & 0.259     & 0.492    & \bf{0.96}\\
			\cline{3-13}
			&\multirow{3}{*}{$n$}&	{D-Lasso2}
			& 0.147     & 0.478     & 0.498     & 0.889     & \bf{0.95}  &	  0.031     & 0.237     & 0.238     & 0.488     & 0.97\\
			&		&	{D-SSL}
			& \bf{0.002}& \bf{0.377}& \bf{0.375}& 1.117     & 0.99       &	  0.052     & 0.217     & 0.222     & \bf{0.450}& 0.97\\
			&		&	{S-SSL}
			& 0.113     & 0.412     & 0.425     & \bf{0.815}& 0.94       &	 \bf{-0.001}& \bf{0.189}& \bf{0.188}& 0.475     & 0.99\\
			\cline{3-13}
			&\multirow{3}{*}{$4n$}&	{D-Lasso2}
			& 0.069     & 0.480     & 0.482     & 0.884     & 0.97       &	  \bf{0.011}& 0.238     & 0.237     & 0.492     & 0.98\\
			&		&	{D-SSL}
			&-0.045     & \bf{0.395}& \bf{0.396}& 0.826     & \bf{0.95}  &	 -0.014     & \bf{0.187}& \bf{0.186}& \bf{0.398}& 0.97\\
			&		&	{S-SSL}
			& \bf{0.043}& 0.401     & 0.401     & \bf{0.734}& 0.93       &	  0.027     & 0.209     & 0.210     & 0.422     & \bf{0.96}\\
			\cline{3-13}
			&\multirow{3}{*}{$8n$}&	{D-Lasso2} 
			& 0.039     & 0.472     & 0.471     & 0.881     & \bf{0.96}  &	  \bf{0.000}& 0.236     & 0.235     & 0.497     & 0.98\\
			&		&	{D-SSL}
			&-0.065     & \bf{0.373}& \bf{0.377}& 0.739     & 0.93       &	 -0.017     & \bf{0.166}& \bf{0.166}& \bf{0.372}& 0.98\\
			&		&	{S-SSL}
			& \bf{0.022}& 0.406     & 0.404     & \bf{0.709}& 0.91       &	  0.013     & 0.203     & 0.202     & 0.414     & \bf{0.96}\\
			\hline
	\end{tabular}}
\end{table}


\begin{table}[!htbp] 
	\caption{Simulation results for \underline{Model 2} with $p=500$:
		Bias, SD and RMSE stand for empirical bias, standard deviation, and root mean squared error, respectively,
		len represents the length of 95\% confidence interval, and CVP is the coverage probability.
		The estimators D-Lasso1 (that only uses labeled data with sample size $n$) and D-Lasso2 are $\wh\bt^d_1$ and $\wh \bt^d_2$, defined in (24).
		The straightforward debiased estimator D-SSL is defined in (5).
		The proposed dependable semi-supervised estimator S-SSL is defined in (14).
		The best performance is \textbf{bolded} during the comparison.
	}\label{sim:t3}
	\center
	\scalebox{0.70}{
		\begin{tabular}{cccrccccrcccc}
			\hline
			&		&	&\multicolumn{5}{c}{$n=100$}  &\multicolumn{5}{c}{$n=300$}		\\
			&	$N$	& &\text{Bias}& \text{SD}& \text{RMSE}  &\text{len/2}&\text{CVP}  &\text{Bias}&\text{SD}  & \text{RMSE}& \text{len/2} &\text{CVP}\\
			\hline
			\multirow{10}{*}{$\theta_1$}&&	{D-Lasso1}
			& \bf{0.104}& 0.885     & 0.887     & 1.555     & 0.97       &    -0.146     & 0.448     & 0.469     & 0.842    & 0.93\\
			\cline{3-13}
			&\multirow{3}{*}{$n$}&	{D-Lasso2}
			&-0.162     & 0.789     & 0.801     & 1.512     & \bf{0.94}  &	 -0.149     & 0.441     & 0.463     & 0.837     & \bf{0.94}\\
			&		&	{D-SSL}
			&-0.292     & \bf{0.579}& 0.646     & 3.043     & 1.00       &	 -0.162     & \bf{0.349}& \bf{0.383}& 0.923     & \bf{0.96}\\
			&		&	{S-SSL}
			&-0.200     & 0.599     & \bf{0.629}& \bf{1.229}& 0.91       &	 \bf{-0.125}& 0.387     & 0.405     & \bf{0.740}& 0.92\\
			\cline{3-13}
			&\multirow{3}{*}{$4n$}&	{D-Lasso2}
			&-0.193     & 0.749     & 0.770     & 1.462     & 0.91       &	 -0.152     & 0.448     & 0.471     & 0.832     & 0.93\\
			&		&	{D-SSL}
			&-0.319     & 0.544     & 0.628     & 1.706     & 0.99       &	 -0.151     & \bf{0.319}& \bf{0.351}& 0.680     & \bf{0.96}\\
			&		&	{S-SSL}
			&-0.197     & \bf{0.540}& \bf{0.572}& \bf{1.120}& \bf{0.93}  &	 \bf{-0.119}& 0.350     & 0.368     & \bf{0.679}& 0.91\\
			\cline{3-13}
			&\multirow{3}{*}{$8n$}&	{D-Lasso2}
			&-0.181     & 0.758     & 0.776     & 1.467     & 0.90       &	 -0.147     & 0.448     & 0.469     & 0.835     & \bf{0.94}\\
			&		&	{D-SSL}
			&-0.304     & 0.542     & 0.620     & 1.349     & \bf{0.96}  &	 -0.134     & \bf{0.297}& \bf{0.325}& \bf{0.610}& \bf{0.96}\\
			&		&	{S-SSL}
			&-0.196     & \bf{0.536}& \bf{0.568}& \bf{1.084}& 0.91       &	 \bf{-0.105}& 0.339     & 0.353     & 0.661     & 0.92\\
			\hline
			\multirow{10}{*}{$\theta_3$}&&	{D-Lasso1}
			& 0.177     & 0.689     & 1.243     & 1.385     & 0.96       &   0.116       & 0.332     & 0.350     & 0.808    & 0.98\\
			\cline{3-13}
			&\multirow{3}{*}{$n$}&	{D-Lasso2}
			& 0.137     & 0.601     & 0.613     & 1.349     & 0.94       &	  0.084     & 0.335     & 0.344     & 0.805     & 0.98\\
			&		&	{D-SSL}
			&-0.125     & \bf{0.553}& 0.565     & 2.383     & 1.00       &	 \bf{-0.008}& 0.277     & 0.275     & \bf{0.645}& \bf{0.97}\\
			&		&	{S-SSL}
			&\bf{-0.046}& 0.558     & \bf{0.557}& \bf{1.120}& \bf{0.95}  &	  0.046     & \bf{0.268}& \bf{0.270}& 0.673     & \bf{0.97}\\
			\cline{3-13}
			&\multirow{3}{*}{$4n$}&	{D-Lasso2}
			& \bf{0.076}& 0.630     & 0.631     & 1.321     & 0.94       &	  0.054     & 0.339     & 0.342     & 0.810     & 0.98\\
			&		&	{D-SSL}
			&-0.239     & 0.526     & 0.575     & 1.181     & \bf{0.95}  &	 -0.046     & 0.271     & 0.274     & \bf{0.454}& 0.92\\
			&		&	{S-SSL}
			&-0.126     & \bf{0.455}& \bf{0.470}& \bf{0.950}& 0.91       &	 \bf{-0.001}& \bf{0.266}& \bf{0.264}& 0.568     & \bf{0.94}\\
			\cline{3-13}
			&\multirow{3}{*}{$8n$}&	{D-Lasso2}
			& \bf{0.051}& 0.644     & 0.643     & 1.336     & \bf{0.93}  &	  0.049     & 0.341     & 0.343     & 0.813     & 0.98\\
			&		&	{D-SSL}
			&-0.245     & 0.516     & 0.569     & 0.898     & 0.85       &	 -0.055     & 0.263     & 0.267     & \bf{0.403}& 0.85\\
			&		&	{S-SSL}
			&-0.177     & \bf{0.447}& \bf{0.478}& \bf{0.897}& 0.87       &	 \bf{-0.029}& \bf{0.258}& \bf{0.259}& 0.534     & \bf{0.94}\\
			\hline
			\multirow{10}{*}{$\theta_4$}&&	{D-Lasso1}
			& 0.550     & 0.552     & 0.777     & 1.079     & 0.87       &     0.182     & 0.304     & 0.353     & 0.572    & 0.87\\
			\cline{3-13}
			&\multirow{3}{*}{$n$}&	{D-Lasso2}
			& 0.388     & 0.596     & 0.709     & 1.049     & \bf{0.90}  &	  0.123     & 0.295     & 0.318     & 0.570     & \bf{0.95}\\
			&		&	{D-SSL}
			& \bf{0.089}& \bf{0.453}& \bf{0.460}& 2.006     & \bf{1.00}  &	  \bf{0.029}& \bf{0.227}& \bf{0.228}& 0.599     & 0.99\\
			&		&	{S-SSL}
			& 0.360     & 0.541     & 0.647     & \bf{0.917}& 0.85       &	  0.126     & 0.260     & 0.288     & \bf{0.520}& 0.88\\
			\cline{3-13}
			&\multirow{3}{*}{$4n$}&	{D-Lasso2}
			& 0.278     & 0.600     & 0.659     & 1.025     & 0.89       &	  0.069     & 0.296     & 0.302     & 0.574     & \bf{0.95}\\
			&		&	{D-SSL}
			& \bf{0.010}& \bf{0.398}& \bf{0.396}& 1.096     & \bf{1.00}  &	 \bf{-0.001}& \bf{0.217}& \bf{0.216}& \bf{0.450}& 0.97\\
			&		&	{S-SSL}
			& 0.209     & 0.512     & 0.551     & \bf{0.848}& 0.88       &	  0.053     & 0.254     & 0.258     & 0.491     & 0.91\\
			\cline{3-13}
			&\multirow{3}{*}{$8n$}&	{D-Lasso2}
			& 0.249     & 0.616     & 0.661     & 1.032     & \bf{0.90}  &	  0.044     & 0.298     & 0.300     & 0.577     & \bf{0.95}\\
			&		&	{D-SSL}
			&\bf{-0.009}& \bf{0.398}& \bf{0.396}& 0.876     & \bf{1.00}  &	 \bf{-0.007}& \bf{0.217}& \bf{0.216}& \bf{0.403}& 0.94\\
			&		&	{S-SSL}
			& 0.158     & 0.491     & 0.513     & \bf{0.838}& \bf{0.90}  &	  0.028     & 0.250     & 0.251     & 0.481     & 0.92\\
			\hline
			\multirow{10}{*}{$\theta_5$}&&	{D-Lasso1}
			& 0.605     & 0.696     & 0.919     & 1.306     & 0.84       &     0.177     & 0.419     & 0.453     & 0.777    & 0.91\\
			\cline{3-13}
			&\multirow{3}{*}{$n$}&	{D-Lasso2}
			& 0.307     & 0.785     & 0.839     & 1.302     & 0.89       &	  0.128     & 0.417     & 0.434     & 0.779     & 0.92\\
			&		&	{D-SSL}
			& \bf{0.098}& \bf{0.638}& \bf{0.642}& 2.289     & \bf{1.00}  &	  \bf{0.017}& \bf{0.330}& \bf{0.329}& \bf{0.611}& 0.91\\
			&		&	{S-SSL}
			& 0.291     & 0.700     & 0.755     & \bf{1.160}& 0.85       &	  0.106     & 0.356     & 0.370     & 0.695     & \bf{0.94}\\
			\cline{3-13}
			&\multirow{3}{*}{$4n$}&	{D-Lasso2}
			& 0.192     & 0.772     & 0.791     & 1.288     & 0.87       &	  0.077     & 0.426     & 0.430     & 0.784     & \bf{0.94}\\
			&		&	{D-SSL}
			& \bf{0.026}& \bf{0.545}& \bf{0.543}& 1.147     & \bf{0.97}  &	  \bf{0.004}& \bf{0.284}& \bf{0.283}& \bf{0.459}& 0.87\\
			&		&	{S-SSL}
			& 0.102     & 0.672     & 0.677     & \bf{1.073}& 0.86       &	  0.026     & 0.332     & 0.331     & 0.644     & 0.93\\
			\cline{3-13}
			&\multirow{3}{*}{$8n$}&	{D-Lasso2}
			& 0.139     & 0.781     & 0.789     & 1.305     & 0.88       &	  0.044     & 0.428     & 0.428     & 0.789     & \bf{0.94}\\
			&		&	{D-SSL}
			&\bf{-0.020}& \bf{0.530}& \bf{0.528}& \bf{0.890}& \bf{0.90}  &	 -0.017     & \bf{0.265}& \bf{0.264}& \bf{0.413}& 0.93\\
			&		&	{S-SSL}
			& 0.029     & 0.657     & 0.655     & 1.049     & 0.85       &	  \bf{0.002}& 0.325     & 0.324     & 0.626     & \bf{0.94}\\
			\hline
			\multirow{10}{*}{$\theta_6$}&&	{D-Lasso1}
			& 0.177     & 0.450     & 0.482     & 0.909     & 0.96       &   0.130       & 0.232     & 0.265     & 0.505     & 0.93\\
			\cline{3-13}
			&\multirow{3}{*}{$n$}&	{D-Lasso2}
			& 0.125     & 0.497     & 0.510     & 0.929     & \bf{0.94}  &	  0.112     & 0.225     & 0.251     & 0.504     & \bf{0.96}\\
			&		&	{D-SSL}
			&\bf{-0.008}& 0.404     & \bf{0.402}& 1.920     & 1.00       &	  \bf{0.028}& \bf{0.197}& \bf{0.198}& 0.586     & 1.00\\
			&		&	{S-SSL}
			& 0.124     & \bf{0.387}& 0.404     & \bf{0.836}& 0.93       &	  0.103     & 0.207     & 0.230     & \bf{0.458}& \bf{0.94}\\
			\cline{3-13}
			&\multirow{3}{*}{$4n$}&	{D-Lasso2}
			& 0.072     & 0.479     & 0.482     & 0.902     & 0.93       &	  0.082     & 0.224     & 0.238     & 0.504     & \bf{0.95}\\
			&		&	{D-SSL}
			&\bf{-0.049}& 0.396     & 0.397     & 1.085     & 1.00       &	  \bf{0.014}& \bf{0.172}& \bf{0.171}& 0.439     & 0.99\\
			&		&	{S-SSL}
			& 0.052     & \bf{0.395}& \bf{0.396}& \bf{0.766}& \bf{0.96}  &	  0.065     & 0.191     & 0.201     & \bf{0.423}& \bf{0.95}\\
			\cline{3-13}
			&\multirow{3}{*}{$8n$}&	{D-Lasso2}
			& 0.047     & 0.495     & 0.495     & 0.912     & 0.91       &	  0.067     & 0.232     & 0.240     & 0.508     & \bf{0.95}\\
			&		&	{D-SSL}
			&-0.059     & \bf{0.388}& 0.391     & 0.865     & 0.98       &	  \bf{0.005}& \bf{0.171}& \bf{0.170}& \bf{0.396}& 0.98\\
			&		&	{S-SSL}
			& \bf{0.013}& 0.390     & \bf{0.388}& \bf{0.741}& \bf{0.93}  &	  0.046     & 0.183     & 0.188     & 0.413     & \bf{0.95}\\
			\hline
	\end{tabular}}
\end{table}


\begin{table}[!htbp] 
	\caption{Simulation results for \underline{Model 2}: computational time (in seconds) of one simulation replication. 
		The estimates of $\bB$ and $\bOmega$ are implemented in parallel, with each utilizing 11 cores.}\label{sim:time_supp}
	\center
	\begin{tabular}{cccrrrr}
		\hline
		& & & \multicolumn{2}{c}{$p=200$} & \multicolumn{2}{c}{$p=500$} \\
		\cline{4-7}
		&		&	&$n=100$&$n=300$ &$n=100$&$n=300$ 	\\
		\cline{3-7}
		&&	{D-Lasso1}
		& 3.897&     6.563  & 10.360&  28.994   \\
		\hline
		\multirow{9}{*}{$N$}&\multirow{3}{*}{$n$}&	{D-Lasso2}
		& 8.020&	  5.201	  & 19.085& 137.224 \\
		&		&	{D-SSL}
		& 7.939&	  5.167	  & 18.326& 136.795   \\
		&		&	{S-SSL}
		& 9.141&	  13.526  & 22.040& 145.803    \\
		\cline{2-7}
		&\multirow{3}{*}{$4n$}&	{D-Lasso2}
		& 5.290&	  6.533  & 66.312&  63.046 \\
		&		&	{D-SSL}
		& 5.255&	  6.706	   & 65.746&  63.373    \\
		&		&	{S-SSL}
		& 6.425&	  14.975  & 69.298&  71.910   \\
		\cline{2-7}
		&\multirow{3}{*}{$8n$}&	{D-Lasso2}
		& 5.558&	  8.724	 & 58.987& 115.264   \\
		&		&	{D-SSL}
		& 5.601&	  9.358	  & 58.736& 117.025  \\
		&		&	{S-SSL}
		& 6.727&	 17.474    & 62.066& 124.901  \\
		\hline
	\end{tabular}
\end{table}

\begin{table}[!htbp] 
	\caption{Simulation results for \underline{Model 2} with $p=200$  and $n=100$:
		Bias, SD and RMSE stand for empirical bias, standard deviation, and root mean squared error, respectively,
		len represents the length of 95\% confidence interval, and CVP is the coverage probability.
		The performance comparison of the proposed dependable semi-supervised estimator S-SSL with different methods to estimate $\wh f(\cdot)$.
	}\label{sim:t9}
	\scalebox{0.70}{
		\begin{tabular}{ccrccccrccccrcccc}
			\hline
			&		&	&&\multicolumn{1}{c}{huang}&&  &&&\multicolumn{1}{c}{randomforest}&&		&&&\multicolumn{1}{c}{nnet}&&		\\
			&	$N$	 &\text{Bias}& \text{SD}& \text{RMSE}  &\text{len/2}& \text{CVP}  &\text{Bias}&\text{SD}  & \text{RMSE}& \text{len/2}& \text{CVP} &\text{Bias}&\text{SD}  & \text{RMSE}& \text{len/2}& \text{CVP}\\
			\hline
			\multirow{3}{*}{$\theta_1$}
			&$n$		& 0.022& 0.655& 0.656& 1.282& 0.93&	 -0.068& 0.635& 0.639& 1.264& 0.92	  & -0.071& 0.749& 0.752& 1.356&  0.92	    \\
			&$4n$		&-0.005& 0.625& 0.625& 1.162& 0.90&	 -0.164& 0.594& 0.616& 1.128& 0.87	  & -0.142& 0.724& 0.737& 1.293&  0.90	    \\
			&$8n$		&-0.020& 0.610& 0.612& 1.127& 0.90&	 -0.183& 0.586& 0.613& 1.086& 0.86    & -0.153& 0.731& 2.210& 1.275&  0.90\\
			\hline
			\multirow{3}{*}{$\theta_3$}
			&$n$		& 0.113& 0.472& 0.485& 1.135& 0.98&	 -0.179& 0.469& 0.502& 1.138& 0.91	  &  0.075& 0.574& 0.578& 1.276&  0.95	    \\
			&$4n$		&-0.006& 0.435& 0.436& 0.979& 0.95&	 -0.158& 0.420& 0.448& 0.987& 0.80	  & -0.032& 0.572& 0.572& 1.243&  0.93	    \\
			&$8n$		&-0.033& 0.429& 0.430& 0.906& 0.94&	 -0.245& 0.408& 0.476& 0.923& 0.71    & -0.075& 0.564& 0.569& 1.209&  0.93\\
			\hline
			\multirow{3}{*}{$\theta_4$}
			&$n$		& 0.311& 0.568& 0.647& 0.911& 0.84&	  0.143& 0.564& 0.581& 0.913& 0.90	  &  0.185& 0.610& 0.637& 0.983&  0.90	    \\
			&$4n$		& 0.156& 0.513& 0.536& 0.825& 0.89&	 -0.060& 0.542& 0.545& 0.829& 0.82	  & -0.016& 0.604& 0.604& 0.950&  0.88	    \\
			&$8n$		& 0.106& 0.507& 0.518& 0.808& 0.89&	 -0.116& 0.551& 0.563& 0.813& 0.82    & -0.071& 0.612& 0.616& 0.952&  0.85\\
			\hline
			\multirow{3}{*}{$\theta_5$}
			&$n$		& 0.274& 0.625& 0.682& 1.156& 0.88&	  0.065& 0.633& 0.636& 1.155& 0.90	  &  0.103& 0.686& 0.693& 1.246&  0.91	    \\
			&$4n$		& 0.113& 0.587& 0.598& 1.053& 0.90&	 -0.191& 0.625& 0.653& 1.052& 0.86	  & -0.143& 0.685& 0.699& 1.206&  0.91	    \\
			&$8n$		& 0.072& 0.581& 0.585& 1.024& 0.92&	 -0.255& 0.627& 0.676& 1.024& 0.82    & -0.194& 0.694& 0.720& 1.199&  0.88\\
			\hline
			\multirow{3}{*}{$\theta_6$}
			&$n$		& 0.071& 0.411& 0.417& 0.826& 0.98&	 -0.087& 0.413& 0.422& 0.822& 0.95	  & -0.012& 0.468& 0.468& 0.873&  0.95	    \\
			&$4n$		&-0.018& 0.386& 0.388& 0.747& 0.96&	 -0.245& 0.403& 0.471& 0.740& 0.88	  & -0.121& 0.462& 0.477& 0.832&  0.92	    \\
			&$8n$		&-0.048& 0.379& 0.382& 0.723& 0.94&	 -0.294& 0.407& 0.502& 0.716& 0.83    & -0.143& 0.459& 0.480& 0.821&  0.91\\
			\hline
	\end{tabular}}
\end{table}

\begin{table}[!htbp] 
	\caption{Simulation results for \underline{Model 2} with $p=200$  and $n=100$:
		Bias, SD and RMSE stand for empirical bias, standard deviation, and root mean squared error, respectively,
		len represents the length of 95\% confidence interval, and CVP is the coverage probability.
		The performance comparison of the proposed dependable semi-supervised estimator S-SSL with different criteria of choosing the tuning parameters $(\lambda_{\Omega},\lambda_B)$ in estimating $(\bOmega, \bB)$.
	}\label{sim:t12}
	\scalebox{0.5}{
		\begin{tabular}{ccrccccrccccrccccrcccc}
			\hline
			&		&	&&\multicolumn{1}{c}{$(\lambda_{min},\lambda_{min})$}&&  &&&\multicolumn{1}{c}{$(\lambda_{1se},\lambda_{min})$}&&		&&&\multicolumn{1}{c}{$(\lambda_{min},\lambda_{1se})$}&&		&&&\multicolumn{1}{c}{$(\lambda_{1se},\lambda_{1se})$}&&\\
			&	$N$	 &\text{Bias}& \text{SD}& \text{RMSE}  &\text{len/2}& \text{CVP}  &\text{Bias}&\text{SD}  & \text{RMSE}& \text{len/2}& \text{CVP} &\text{Bias}&\text{SD}  & \text{RMSE}& \text{len/2}& \text{CVP} &\text{Bias}&\text{SD}  & \text{RMSE}& \text{len/2}& \text{CVP}\\
			\hline
			\multirow{3}{*}{$\theta_1$}
			&$n$		&-0.268& 0.609& 0.663& 1.212& 0.92&	 -0.274& 0.612& 0.667& 1.239& 0.92	  & -0.224& 0.681& 0.714& 1.302&  0.93	    & -0.232& 0.684& 0.719& 1.332&  0.92\\
			&$4n$		&-0.217& 0.562& 0.600& 1.138& 0.93&	 -0.218& 0.566& 0.604& 1.144& 0.93	  & -0.179& 0.672& 0.692& 1.301&  0.92	    & -0.181& 0.679& 0.699& 1.308&  0.92\\
			&$8n$		&-0.214& 0.534& 0.573& 1.094& 0.92&	 -0.220& 0.538& 0.579& 1.098& 0.92    & -0.172& 0.647& 0.666& 1.279&  0.93      & -0.179& 0.652& 0.673& 1.284&  0.93\\
			\hline
			\multirow{3}{*}{$\theta_3$}
			&$n$		& 0.006& 0.480& 0.478& 1.070& 0.96&	  0.118& 0.491& 0.502& 1.076& 0.98 	  &  0.023& 0.520& 0.518& 1.161&  0.98	    &  0.141& 0.538& 0.553& 1.170&  0.96\\
			&$4n$		&-0.121& 0.440& 0.454& 0.906& 0.92&	 -0.018& 0.450& 0.448& 0.909& 0.93 	  & -0.087& 0.498& 0.503& 1.069&  0.92	    &  0.022& 0.512& 0.510& 1.078&  0.94\\
			&$8n$		&-0.168& 0.445& 0.473& 0.852& 0.89&	 -0.066& 0.445& 0.447& 0.854& 0.89    & -0.120& 0.495& 0.507& 1.045&  0.91      & -0.012& 0.499& 0.496& 1.052&  0.94\\
			\hline
			\multirow{3}{*}{$\theta_4$}
			&$n$		& 0.281& 0.490& 0.562& 0.916& 0.86&	  0.498& 0.503& 0.706& 0.910& 0.77	  &  0.305& 0.522& 0.602& 0.982&  0.89	    &  0.529& 0.531& 0.748& 0.980&  0.81\\
			&$4n$		& 0.176& 0.464& 0.494& 0.844& 0.84&	  0.358& 0.472& 0.591& 0.835& 0.80	  &  0.213& 0.521& 0.561& 0.951&  0.91	    &  0.408& 0.533& 0.669& 0.948&  0.83\\
			&$8n$		& 0.115& 0.452& 0.464& 0.823& 0.85&	  0.299& 0.456& 0.544& 0.804& 0.84    &  0.157& 0.513& 0.535& 0.945&  0.90      &  0.355& 0.520& 0.627& 0.930&  0.87\\
			\hline
			\multirow{3}{*}{$\theta_5$}
			&$n$		& 0.101& 0.664& 0.668& 1.144& 0.90&	  0.322& 0.689& 0.757& 1.124& 0.88	  &  0.151& 0.734& 0.746& 1.233&  0.91	    &  0.376& 0.758& 0.843& 1.216&  0.90\\
			&$4n$		&-0.040& 0.598& 0.597& 1.056& 0.87&	  0.162& 0.618& 0.636& 1.039& 0.88 	  &  0.029& 0.705& 0.707& 1.208&  0.90	    &  0.236& 0.728& 0.762& 1.195&  0.86\\
			&$8n$		&-0.092& 0.585& 0.589& 1.035& 0.85&	  0.095& 0.600& 0.604& 1.011& 0.88    & -0.020& 0.696& 0.698& 1.210&  0.88      &  0.173& 0.715& 0.732& 1.187&  0.89\\
			\hline
			\multirow{3}{*}{$\theta_6$}
			&$n$		& 0.172& 0.370& 0.406& 0.803& 0.95&	  0.282& 0.369& 0.463& 0.809& 0.93	  &  0.194& 0.400& 0.443& 0.863&  0.98	    &  0.311& 0.391& 0.498& 0.873&  0.95\\
			&$4n$		& 0.078& 0.368& 0.375& 0.731& 0.93&	  0.179& 0.355& 0.396& 0.723& 0.91	  &  0.101& 0.411& 0.421& 0.836&  0.96	    &  0.212& 0.396& 0.447& 0.831&  0.95\\
			&$8n$		& 0.046& 0.355& 0.356& 0.711& 0.95&	  0.145& 0.348& 0.375& 0.696& 0.92    &  0.073& 0.404& 0.408& 0.831&  0.96      &  0.180& 0.399& 0.436& 0.817&  0.96\\
			\hline
	\end{tabular}}
\end{table}

\begin{table}[!htbp] 
	\center
	\caption{Simulation results for \underline{Model 2} with $p=200$  and $n=100$:
		Bias, SD and RMSE stand for empirical bias, standard deviation, and root mean squared error, respectively,
		len represents the length of 95\% confidence interval, and CVP is the coverage probability.
		The performance comparison of the proposed dependable semi-supervised estimator S-SSL with different linear combinations $\bv$: Case 1 with $\bv=(-1,0,0,0,0,2,0,\cdots,0)\trans$, Case 2 with $\bv=(-1,0,1,0,0,2,0,\cdots,0)\trans$, Case 3 with $\bv=(-1,0,1,0,1,1,0,\cdots,0)\trans$, and Case 4 with $\bv=(-1,0,1,-1,1,0,0,\cdots,0)\trans$.
	}\label{sim:t10}
	\begin{tabular}{ccrcrcccc}
		\hline
		&	$N$	 &\text{Bias}& \text{SD}& \text{RMSE}  &\text{len/2}& \text{CVP}  \\
		\hline
		\multirow{3}{*}{Case 1}
		&$n$		& 0.105& 0.956& 0.962& 2.020& 0.97	 	    \\
		&$4n$		&-0.032& 0.915& 0.916& 1.820& 0.96	 	    \\
		&$8n$		&-0.081& 0.892& 0.895& 1.773& 0.97	 \\
		\hline
		\multirow{3}{*}{Case 2}
		&$n$		& 0.217& 1.158& 1.173& 2.355& 0.97	 	    \\
		&$4n$		&-0.040& 1.094& 1.095& 2.074& 0.96	 	    \\
		&$8n$		&-0.114& 1.060& 1.066& 1.986& 0.95	 \\
		\hline
		\multirow{3}{*}{Case 3}
		&$n$		& 0.426& 0.962& 1.048& 2.261& 0.96	 	    \\
		&$4n$		& 0.091& 0.910& 0.914& 1.937& 0.95	 	    \\
		&$8n$		& 0.005& 0.864& 0.864& 1.830& 0.93	 \\
		\hline
		\multirow{3}{*}{Case 4}
		&$n$		& 0.046& 1.052& 1.053& 2.104& 0.94	  	    \\
		&$4n$		&-0.050& 1.008& 1.009& 1.976& 0.93	 	    \\
		&$8n$		&-0.051& 0.961& 0.962& 1.938& 0.95	 \\
		\hline
	\end{tabular}
\end{table}

\begin{table}[!htbp] 
	\center
	\caption{Simulation results for \underline{Model 2} with $p=200$  and $n=100$:
		Bias, SD and RMSE stand for empirical bias, standard deviation, and root mean squared error, respectively,
		len represents the length of 95\% confidence interval, and CVP is the coverage probability.
		The performance comparison of the proposed dependable semi-supervised estimator S-SSL with different values of $\|\bv\|_1/\|\bv\|_2$: Case 1 with $\bv=(-1,0,0,0,0,0,0,\cdots,0)\trans$, Case 2 with $\bv=(-1,0,1,0,0,0,0,\cdots,0)\trans$, Case 3 with $\bv=(-1,0,1,-1,0,0,0,\cdots,0)\trans$, Case 4 with $\bv=(-1,0,1,-1,1,0,0,\cdots,0)\trans$, and Case 5 with $\bv=(-1,0,1,-1,1,-1,0,\cdots,0)\trans$.
	}\label{sim:t11}
	\begin{tabular}{ccrrcccc}
		\hline
		&	$N$	 &\text{Bias}& \text{SD}& \text{RMSE}  &\text{len/2}& \text{CVP}  \\
		\hline
		\multirow{3}{*}{Case 1}
		&$n$		&-0.026& 0.654& 0.651& 1.279& 0.930	  	    \\
		&$4n$		& 0.001& 0.623& 0.620& 1.161& 0.900	 	    \\
		&$8n$		& 0.017& 0.612& 0.609& 1.127& 0.900	 \\
		\hline
		\multirow{3}{*}{Case 2}
		&$n$		& 0.086& 0.769& 0.769& 1.632& 0.970	  	    \\
		&$4n$		&-0.007& 0.728& 0.724& 1.450& 0.950	 	    \\
		&$8n$		&-0.016& 0.703& 0.700& 1.375& 0.930	 \\
		\hline
		\multirow{3}{*}{Case 3}
		&$n$		&-0.228& 0.858& 0.883& 1.778& 0.980	  	    \\
		&$4n$		&-0.164& 0.810& 0.822& 1.651& 0.940	 	    \\
		&$8n$		&-0.122& 0.787& 0.792& 1.599& 0.920	 \\
		\hline
		\multirow{3}{*}{Case 4}
		&$n$		& 0.046& 1.052& 1.048& 2.104& 0.940	  	    \\
		&$4n$		&-0.050& 1.008& 1.004& 1.976& 0.930	 	    \\
		&$8n$		&-0.051& 0.961& 0.958& 1.938& 0.950	 \\
		\hline
		\multirow{3}{*}{Case 5}
		&$n$		&-0.020& 1.140& 1.134& 2.199& 0.930	  	    \\
		&$4n$		&-0.034& 1.089& 1.084& 2.100& 0.940	 	    \\
		&$8n$		&-0.002& 1.051& 1.045& 2.073& 0.940	 \\
		\hline
	\end{tabular}
\end{table}

	\end{document}